\documentclass{article}

\PassOptionsToPackage{numbers}{natbib}

\usepackage[preprint]{neurips_2024}

\usepackage[utf8]{inputenc} % allow utf-8 input
\usepackage[T1]{fontenc}    % use 8-bit T1 fonts
\usepackage{hyperref}       % hyperlinks
\usepackage{url}            % simple URL typesetting
\usepackage{booktabs}       % professional-quality tables
\usepackage{amsfonts}       % blackboard math symbols
\usepackage{nicefrac}       % compact symbols for 1/2, etc.
\usepackage{microtype}      % microtypography

\usepackage{graphicx} 
\usepackage{amsmath}
\usepackage{enumitem}
\usepackage{multicol}
\usepackage{multirow}
\usepackage{makecell}
\usepackage{balance}
\usepackage{xspace}
\usepackage[table,xcdraw]{xcolor}

\usepackage{subfigure}
\usepackage{graphbox}
\usepackage{svg}
\usepackage{nicefrac}
\usepackage{wrapfig}

\usepackage{bold-extra}
\usepackage{adjustbox}
\usepackage{array}
\newcolumntype{P}[1]{>{\centering\arraybackslash}p{#1}}
\newcolumntype{M}[1]{>{\centering\arraybackslash}m{#1}}
\newcolumntype{R}[1]{>{\arraybackslash}m{#1}}

\definecolor{bg-lightgreen}{RGB}{200, 247, 200}
\definecolor{ft-lightgreen}{RGB}{15,157,88}

\definecolor{bg-lightred}{RGB}{255,198,196}
\definecolor{ft-lightred}{RGB}{219,68,55}

\definecolor{bg-lightyellow}{RGB}{253, 249, 205}
\definecolor{ft-lightyellow}{RGB}{207, 161, 13}

\definecolor{bg-lightblue}{RGB}{197, 241, 255}
\definecolor{ft-lightblue}{RGB}{66,133,244}

\definecolor{bg-lightpurple}{RGB}{230,230,250}
\definecolor{ft-lightpurple}{RGB}{138,43,226}

\definecolor{bg-blue}{RGB}{197, 220, 255}
\definecolor{ft-blue}{RGB}{0,0,120}

\definecolor{bg-lightpink}{RGB}{255, 228, 225}
\definecolor{ft-lightpink}{RGB}{255, 105, 180}

\definecolor{bg-lightorange}{RGB}{255, 223, 186}
\definecolor{ft-lightorange}{RGB}{255, 140, 0}

\definecolor{bg-lightviolet}{RGB}{238, 221, 255}
\definecolor{ft-lightviolet}{RGB}{148, 0, 211}

\usepackage{rotate}
\usepackage{capt-of}
\usepackage{tabulary}
\usepackage{tabularx}
\usepackage{amsmath}  % For align environment
\usepackage{setspace}
\usepackage{amssymb}
\usepackage{mathtools}

\DeclareMathOperator*{\argmax}{arg\,max}

\usepackage{hyperref}

\newcommand{\emojisci}{\includegraphics[width=0.02\textwidth]{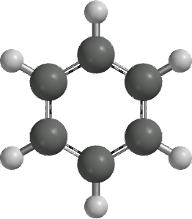}\xspace}
\newcommand{\emojiscititle}{\includegraphics[width=0.03\textwidth]{Figure/logo_sci.png}\xspace}

\newcommand{\emojitab}{\includegraphics[width=0.02\textwidth]{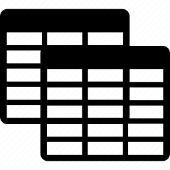}\xspace}
\newcommand{\emojitabtitle}{\includegraphics[width=0.03\textwidth]{Figure/logo_tab.png}\xspace}

\newcommand{\emojiinfra}{\includegraphics[width=0.02\textwidth]{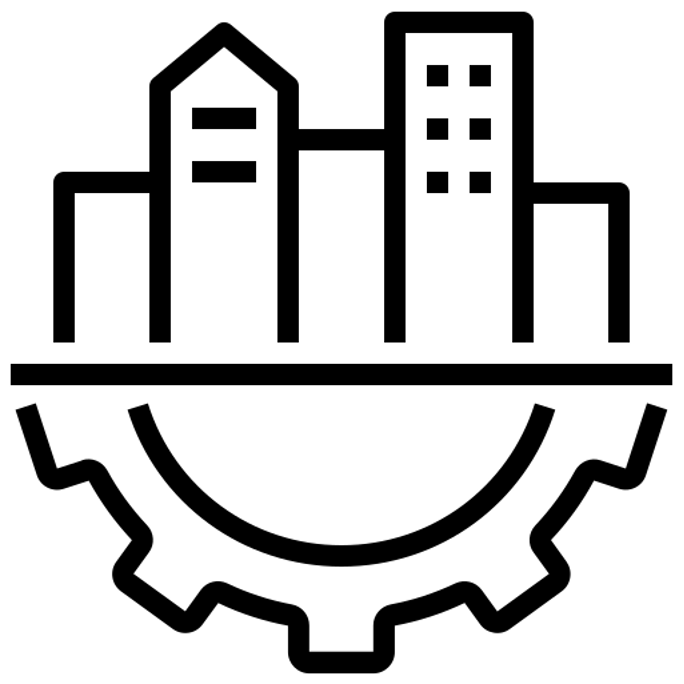}\xspace}
\newcommand{\emojiinfratitle}{\includegraphics[width=0.03\textwidth]{Figure/logo_infra.png}\xspace}

\newcommand{\emojiscene}{\includegraphics[width=0.02\textwidth]{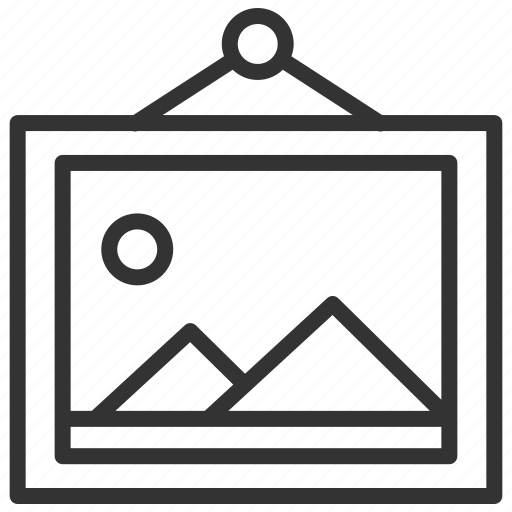}\xspace}
\newcommand{\emojiscenetitle}{\includegraphics[width=0.03\textwidth]
{Figure/logo_scene.png}\xspace}

\newcommand{\emojidoc}{\includegraphics[width=0.02\textwidth]{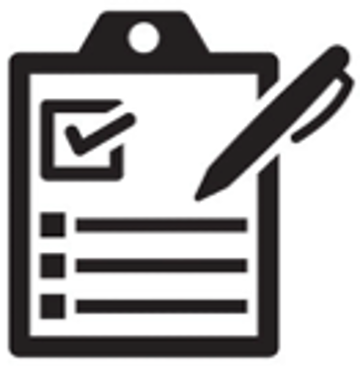}\xspace}
\newcommand{\emojidoctitle}{\includegraphics[width=0.03\textwidth]{Figure/logo_doc.png}\xspace}

\newcommand{\emojisoc}{\includegraphics[width=0.02\textwidth]{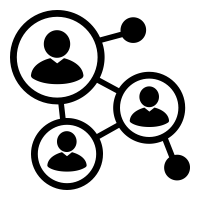}\xspace}
\newcommand{\emojisoctitle}{\includegraphics[width=0.03\textwidth]{Figure/logo_soc.png}\xspace}

\newcommand{\emojikg}{\includegraphics[width=0.02\textwidth]{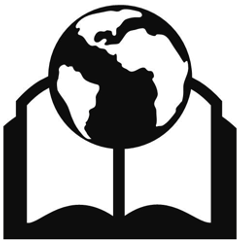}\xspace}
\newcommand{\emojikgtitle}{\includegraphics[width=0.03\textwidth]{Figure/logo_kg.png}\xspace}

\newcommand{\emojiplan}{\includegraphics[width=0.02\textwidth]{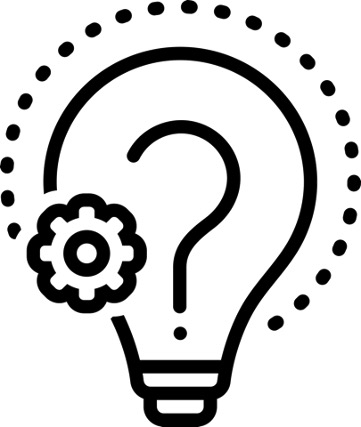}\xspace}
\newcommand{\emojiplantitle}{\includegraphics[width=0.03\textwidth]{Figure/logo_plan.png}\xspace}

\newcommand{\emojirandom}{\includegraphics[width=0.02\textwidth]{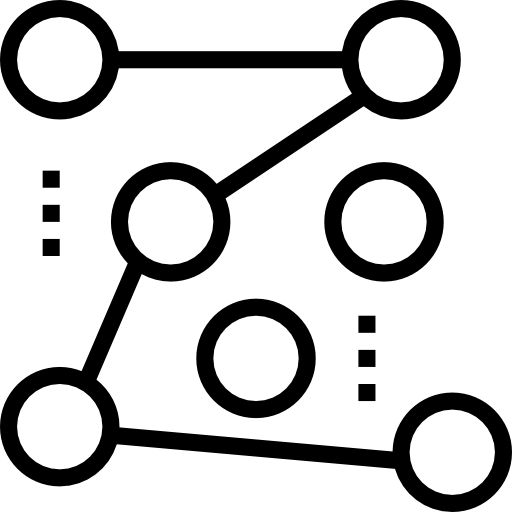}\xspace}
\newcommand{\emojirandomtitle}{\includegraphics[width=0.03\textwidth]{Figure/logo_random.png}\xspace}

\newcommand{\emojibio}{\includegraphics[width=0.02\textwidth]{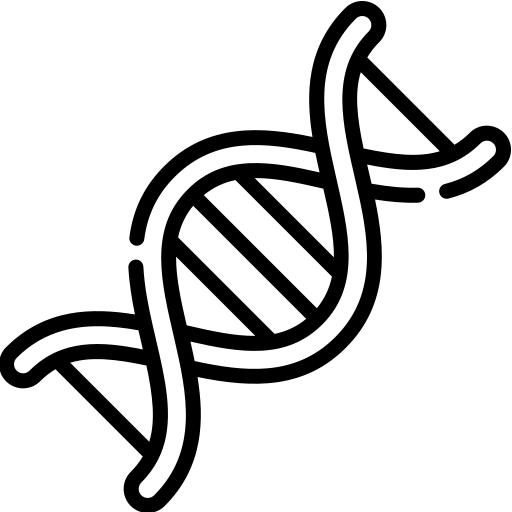}\xspace}
\newcommand{\emojibiotitle}{\includegraphics[width=0.03\textwidth]{Figure/logo_bio.png}\xspace}

\usepackage{hyperref}
\hypersetup{
    colorlinks = true,
    linkcolor = purple,
    citecolor = purple,
}
\title{Retrieval-Augmented Generation with Graphs (GraphRAG)}

\author{%
Haoyu Han\textsuperscript{1}\thanks{Equal contribution.}, Yu Wang\textsuperscript{2}\footnotemark[1], Harry Shomer\textsuperscript{1}, Kai Guo\textsuperscript{1}, Jiayuan Ding\textsuperscript{5}, Yongjia Lei\textsuperscript{2},\\
  \textbf{Mahantesh Halappanavar\textsuperscript{3}, Ryan A. Rossi\textsuperscript{4}, Subhabrata Mukherjee\textsuperscript{5}, Xianfeng Tang\textsuperscript{6}, Qi He\textsuperscript{6}}, \\ 
  \textbf{Zhigang Hua\textsuperscript{7}, Bo Long\textsuperscript{7}, 
  Tong Zhao\textsuperscript{8}, Neil Shah\textsuperscript{8}, Amin Javari\textsuperscript{9}, Yinglong Xia\textsuperscript{7}, Jiliang Tang\textsuperscript{1}} \\
  \textsuperscript{1}Michigan State University, \textsuperscript{2}University of Oregon, \textsuperscript{3}Pacific Northwest National Laboratory\\
  \textsuperscript{4}Adobe Research, \textsuperscript{5}Hippocratic AI, \textsuperscript{6}Amazon, \textsuperscript{7}Meta ,\textsuperscript{8}Snap Inc.,  ,\textsuperscript{9}The Home Depot,\\
  \texttt{\{hanhaoy1, shomerha, guokai1, tangjili\}@msu.edu}, \\
  \texttt{\{yuwang, yongjia\}@uoregon.edu}, \texttt{hala@pnnl.gov}, \texttt{ryarossi@gmail.com}, \\
    \texttt{\{jiayuan, subho\}@hippocraticai.com}, \texttt{\{xianft, qih\}@amazon.com}, \\ \texttt{\{zhua, bolong, yxia\}@meta.com},
  \texttt{\{tong, nshah\}@snap.com}, \texttt{amin\_javari@homedepot.com}
}

\begin{document}

\maketitle

\begin{abstract}
Retrieval-augmented generation (RAG) is a powerful technique that enhances downstream task execution by retrieving additional information, such as knowledge, skills, and tools from external sources. Graph, by its intrinsic "nodes connected by edges" nature, encodes massive heterogeneous and relational information, making it a golden resource for RAG in tremendous real-world applications. As a result, we have recently witnessed increasing attention on equipping RAG with Graph, i.e., GraphRAG. However, unlike conventional RAG, where the retriever, generator, and external data sources can be uniformly designed in the neural-embedding space, the uniqueness of graph-structured data, such as diverse-formatted and domain-specific relational knowledge, poses unique and significant challenges when designing GraphRAG for different domains. Given the broad applicability, the associated design challenges, and the recent surge in GraphRAG, a systematic and up-to-date survey of its key concepts and techniques is urgently desired. Following this motivation, we present a comprehensive and up-to-date survey on GraphRAG. Our survey first proposes a holistic GraphRAG framework by defining its key components, including query processor, retriever, organizer, generator, and data source. Furthermore, recognizing that graphs in different domains exhibit distinct relational patterns and require dedicated designs, we review GraphRAG techniques uniquely tailored to each domain. Finally, we discuss research challenges and brainstorm directions to inspire cross-disciplinary opportunities. Our survey repository is publicly maintained at \url{https://github.com/Graph-RAG/GraphRAG/}.
\end{abstract}

\vspace{-2ex}
\section{Introduction}\label{sec-intro}
\vspace{-1ex}
Retrieval-Augmented Generation (RAG), as a powerful technique to improve downstream tasks by retrieving additional information from external data sources, has been successfully applied to various real-world applications~\cite{ding2024survey, gao2023retrieval, zeng2024towards, zhao2024retrieval}. In RAG frameworks, retrievers search for additional knowledge, skills, and tools based on user-defined queries or task instructions. The retrieved content is then refined by an organizer and seamlessly integrated with the original query or instruction, which is further fed into the generator to produce the final answer. For example, when conducting question-answering (QA) tasks, the classic "Retriever-then-Reader" frameworks~\cite{ju2022grape, karpukhin2020dense, xiong2020answering, zhu2021retrieving} retrieve external factual knowledge to improve the answer faithfulness, which significantly benefit social goodness and mitigate risks in high-stake scenarios (e.g., medical, legal,  financial, and education consultation~\cite{xiong2024benchmarking, xu2024ram, zhang2023enhancing}). Moreover, recent advancements in large language models (LLMs) have further underscored the power of RAG in enhancing the social responsibility of LLMs, such as mitigating hallucinations~\cite{tonmoy2024comprehensive}, enhancing interpretability and transparency~\cite{kim2024re}, enabling dynamic adaptability~\cite{shi2024retrieval, wang2023knowledge}, reducing privacy risks~\cite{zeng2024mitigating, zeng2024good}, ensuring reliability/robust responses~\cite{fang2024enhancing, xiang2024certifiably}, and promoting fair treatment~\cite{shrestha2024fairrag}.

Building on the unprecedented success of RAG and further considering the ubiquity of graphs in real-world applications~\cite{zhang2020deep}, recent research has explored the integration of RAG with graph-structured data. Unlike textual or visual data, graph-structured data encodes heterogeneous and relational information through its intrinsic "nodes connected by edges" nature. For example, individuals connected by social relationships of social networks usually exhibit homophily behaviors~\cite{mcpherson2001birds}, sequential decision-making steps in plans follow casual dependency~\cite{wu2024can}, and atoms belonging to the same functional group within a molecule possess unique structural properties~\cite{ertl2017algorithm, zang2023hierarchical}. Designing the RAG that utilizes relational information requires adapting its core components, such as the retriever and generator, to seamlessly integrate graph-structured data, resulting in GraphRAG. Different from RAG, which predominantly uses semantic/lexical similarity search~\cite{fan2024survey, gao2023retrieval}, GraphRAG offers unique advantages in capturing relational knowledge by leveraging graph-based machine learning (e.g., Graph Neural Networks (GNNs)) and graph/network analysis techniques (e.g., Graph Traversal Search and Community Detection~\cite{edge2024local, wang2024knowledge}). For example, considering the query ``What drugs are used to treat epithelioid sarcoma and also affect the EZH2 gene product?"~\cite{wu2024stark}, blindly executing the existing BM25 or embedding-based search that relies solely on semantic/lexical similarity ignores relational knowledge encoded in graph structure. In contrast, some GraphRAG methods traverse the graph along the relational path ``Disease (Epithelioid Sarcoma) $\rightarrow$ [indication] $\rightarrow$ Drug $\leftarrow$ [target] $\leftarrow$ Gene/Protein (EZH2 gene product)" to retrieve neighbors of Epithelioid Disease following the relation [indication], neighbors of Gene EZH2 following the relation [target], and find their intersected drug~\cite{jin2024graph, luoreasoning, wang2024knowledge}. Moreover, some domains involve entities with extremely complex geometry that require dedicated model design to characterize.
% \aj{incomplete sentence?Moreover, some domains involve entities with extremely complex geometry that require careful handling}. 
For example, 3D structures in molecular graphs~\cite{chen2025graph, wigh2022review} and hierarchical tree structures commonly found in product taxonomies (e.g., on Amazon~\cite{zhang2021deep}), in document sections (e.g., when using Adobe Acrobat~\cite{zhang2024tree}), and social networks (e.g., at Snap~\cite{ma2024harec}) requires carefully designed graph encoders (or, more precisely, geometric encoders) with appropriate expressiveness to capture structural nuances~\cite{ma2024harec, zhang2024hierarchical}. Simply verbalizing node texts and feeding them into LLMs cannot express complex geometric information and becomes infeasible given the exponentially growing textual descriptions as neighborhood layers expand.

\begin{figure*}[t!]
    \centering
    \includegraphics[width=1\linewidth]{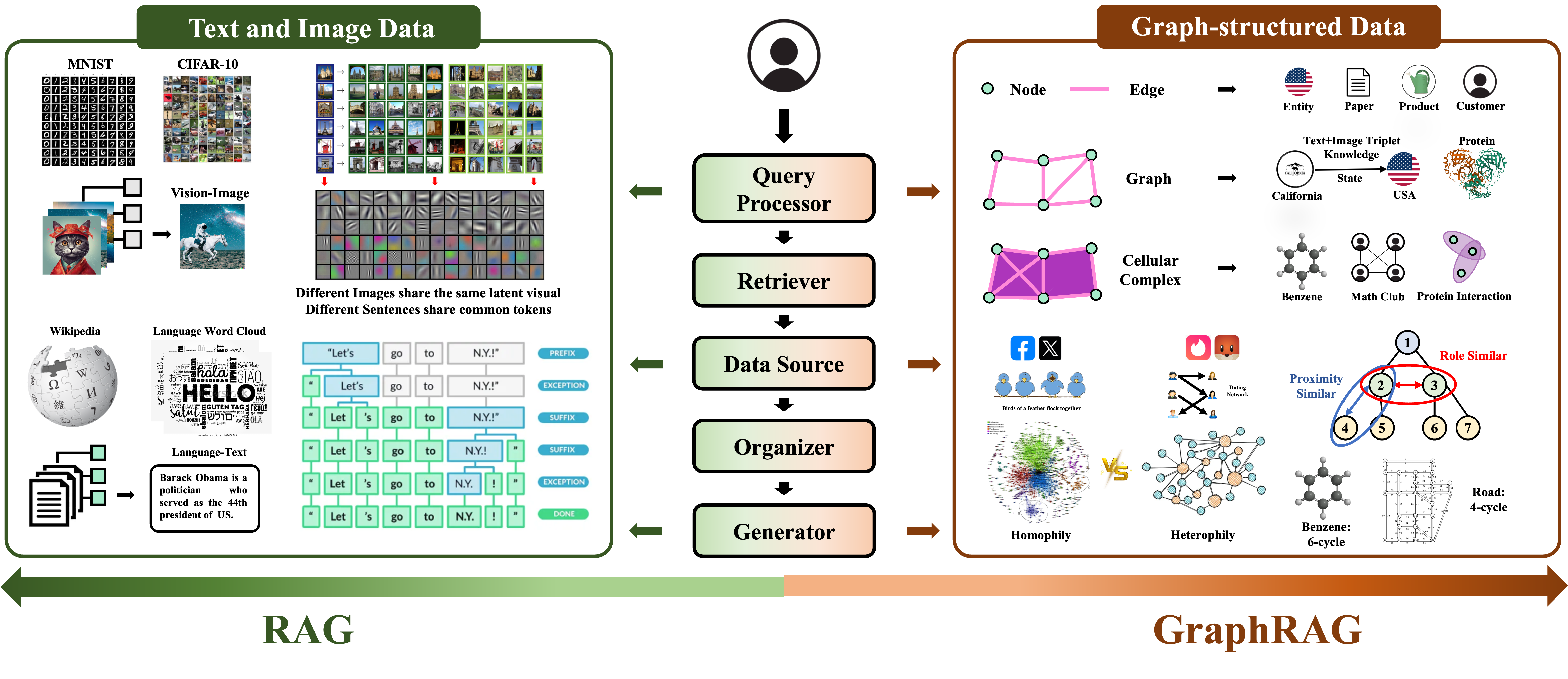}
    \vspace{-3ex}
    \caption{RAG works on text and images, which can be uniformly formatted as 1D sequences or 2D grids with no relational information. In contrast, GraphRAG works on graph-structured data, which encompasses diverse formats and includes domain-specific relational information.}
    \label{fig-challenge}
    \vspace{-4ex}
\end{figure*}

Despite the above advantages of GraphRAGs over RAGs, designing appropriate GraphRAGs faces unprecedented challenges due to the following differences in graph-structured data:

\vspace{-1ex}
\begin{itemize}[leftmargin=*]
    \item \textbf{Difference 1 - Unified versus (vs.) Diverse-Formatted Information}: Unlike conventional RAG, where semantic information can be uniformly represented as a 2D grid of image patches or a 1D sequence of textual corpora, graph-structured data often encompass diverse formats and are stored in heterogeneous sources~\cite{aggarwal2017edge, bodnar2021weisfeiler, wang2022retrieval}. For example, document graphs embed entities as sentence chunks~\cite{edge2024local, wang2024knowledge}, knowledge graphs store graph information as triplets or paths~\cite{chang2024path}, and molecule graphs consist of higher-order structures (e.g., cellular complexes)~\cite{bodnar2021weisfeiler}, as shown in Figure~\ref{fig-challenge}. Some graph data may even be multimodal (e.g., text-attributed graphs include both structural and textual attributes, and scene graphs combine structures and vision). Consequently, this diversity necessitates different RAG designs. For retrievers, conventional RAG assumes the target information is indexed in an image or text corpus, which can be uniformly represented as vector embeddings and enable one-size-fits-all embedding-based retrieval. However, retrievers for GraphRAG must consider the concrete format and source of the desired information, making the one-size-fits-all design impractical. When dealing with knowledge graph question-answering, information of nodes, edges, or subgraphs is usually fetched by graph search before embedding matching-based retrieval~\cite{wang2023knowledge, yasunaga2021qa}. This fetching operation is usually conducted by identifying relevant nodes/edges/subgraphs via entity linking, relational matching, and graph search algorithms (e.g., Breadth-First Search, Depth-First Search, Monte Carlo Tree Search, and A* search)~\cite{tian2024graph, wang2023knowledge, zhuang2023toolchain}, which is unachievable if solely through deep learning-based embedding similarity search. 
    Furthermore, the design of the retriever should ensure sufficient geometric expressiveness to capture structural nuances. For instance, when retrieving APIs from a plan graph to accomplish specific goals~\cite{shen2023taskbench, shen2024hugginggpt, wu2024can}, it is essential to equip the retriever with directional awareness. This enables the execution of APIs with resource dependencies in the correct order, preventing conflicts and avoiding invalid operations. Similarly, designing expressive retrievers capable of differentiating high-order subgraph structures, such as 6-cycle benzene versus vs. 4-star Methane, and 3-star T-junction vs. 4-square road, is essential in drug design for disease treatment~\cite{hamilton2020graph} and road construction for city planning~\cite{kocayusufoglu2022flowgen}. Beyond the retriever, the generator also requires specialized designs. When retrieved content includes complex graph structures with textual attributes, simply verbalizing the text of the subgraph and concatenating it into a prompt may obscure critical structural information. In these cases, encoding the graph with graph encoders such as GNNs before integrating it into generation can help preserve structural nuances~\cite{knowledgenavigator, unioqa, wang2022retrieval, wen2023mindmap, retrieve_rewrite_answer}.

    \item \textbf{Difference 2 - Independent vs. Interdependent Information:} In conventional RAG, information is stored and utilized independently. For example, documents are split into chunks, such as individual sentences, paragraphs, or a fixed number of tokens, based on the document context and downstream task~\cite{barnett2024seven, zhu2021retrieving}. Each chunk is then indexed and stored independently in a vector database. This independence prevents the retrieval from capturing chunk relations, which hinders performance on tasks requiring multi-hop reasoning and long-form planning. However, GraphRAG stores chunks as interconnected nodes with edges denoting their relations, which can benefit retrieval, organization, and generation. For retrieval, these edges could enable multi-hop traversal to capture other chunks that share a logical connection with existing retrieved chunks. Furthermore, the retrieved content can be organized not only by their semantic meaning (e.g., reranking~\cite{chen2023understanding, ivgi2023efficient, liu2024lost}) but also their structural relations (e.g., graph pruning~\cite{su2024pipenet, wang2024reasoning}). During the generation phase, squeezing interdependency (e.g., positional encoding~\cite{shomer2024lpformer, zhao2023gimlet}) to the generator would encode richer structural signals into the generated content.

    \item \textbf{Difference 3 - Domain Invariance vs. Domain-specific Information}: The relations in graph-structured data are domain-specific. Unlike images and texts, where different domains often share transferable semantics~\cite{liu2023towards, mao2024graph}, such as textures and grains in images or vocabulary defined by the tokenizer in texts, graph-structured data lacks explicit transferable units. This shared basis in images and texts lays the foundation for designing encoders with geometric invariance and enables the well-known data-scaling law. However, for graph-structured data, the underlying data generation process governing the generated graphs varies significantly across different domains. This variability makes the relational information highly domain-specific, and it is nearly impossible to design a unified GraphRAG applicable to different domains. For example, when predicting the topic of an academic paper, the widely accepted homophily assumption suggests retrieving references from the paper to inform its topic prediction~\cite{zhu2020beyond}. However, this homophily assumption is not suitable when classifying the role of an airport in a flight network, where hubs are often sparsely distributed across a country with no direct connections~\cite{cui2022positional}. Moreover, even within the same graph from the same domain, different tasks may necessitate distinct GraphRAG designs. For example, when designing an automatic email completion system to optimize communication efficiency in a company, both content relevance and tone coherence should be considered~\cite{wang2024augmenting}. To ensure the content relevance of the generated emails, one might assume that close emails (i.e., emails from the same conversation thread) share similar content and thus should be retrieved for reference. However, to maintain tone coherence, emails from staff with similar roles might be retrieved, even if they do not share close social relations (e.g., between subordinates and superiors) but instead hold similar structural roles within the company (e.g., as managers of different teams).
\end{itemize}

Despite the above differences that have driven extensive research in GraphRAG, the current research landscape in this field remains fragmented, with significant variation in concepts, techniques, and datasets across studies. Moreover, current GraphRAG research primarily focuses on knowledge and document graphs as surveyed in Figure~\ref{fig-imbalance-pub}, often overlooking broader applications in other domains like infrastructure graphs. This imbalance not only hampers the advancement of GraphRAG but also risks creating a "bubble effect" that restricts the scope of future exploration. To address these challenges, we present a comprehensive and up-to-date review of GraphRAG, aiming to unify the GraphRAG framework from the global perspective while also specializing its unique design for each domain from the local perspective. The key contributions of this survey are as follows:

\begin{wrapfigure}{r}{0.35\textwidth}
\vspace{-0.2in}
\centering
    \includegraphics[width=0.35\textwidth]{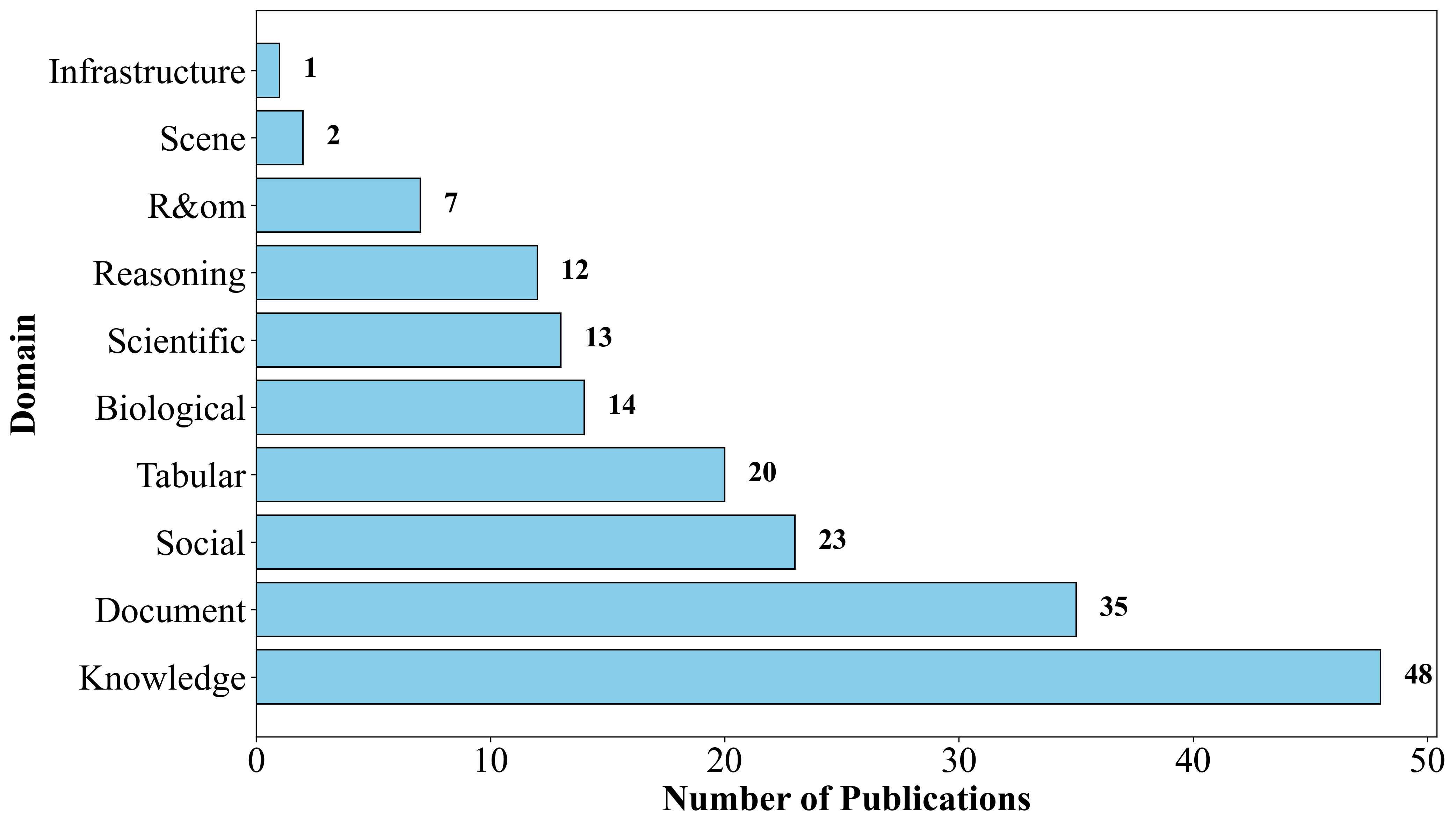}
    \vspace{-0.15in}
     \caption{Publications of GraphRAGs in different domains based on surveyed papers}
    \label{fig-imbalance-pub}
\vspace{-0.1in}
\end{wrapfigure}

\begin{itemize}[leftmargin=*]
    \item \textbf{A Holistic Framework of GraphRAG}: 
    We propose a holistic framework of GraphRAG consisting of five key components: query processor, retriever, organizer, generator, and graph data source. Within each component, we review representative GraphRAG techniques.

    \item \textbf{Specialization of GraphRAG in different domains}: 
    We categorize GraphRAG designs into 10 distinct domains based on their specific applications, including knowledge graph \emojikg, document graph \emojidoc, scientific graph \emojisci, social graph \emojisoc, planning \& reasoning graph \emojiplan, tabular graph \emojitab, infrastructure graph \emojiinfra, biological graph \emojibio, scene graph \emojiscene, and random graph \emojirandom. For each domain, we review their unique applications and specific graph construction methods. We then summarize the distinctive designs of each component within our proposed holistic GraphRAG framework and collect rich benchmark datasets and tool resources.
    
    \item \textbf{Challenges and Future Directions:} We highlight the challenges of current GraphRAG research and pinpoint future opportunities for further advancing GraphRAG into the new frontier.
\end{itemize}

In the following, we highlight the differences between our survey and existing surveys. Despite the urgent need for a systematic overview of GraphRAG, most existing surveys focus on general RAG within the context of i.i.d. data~\cite{asai2023retrieval, gao2023retrieval, li2022survey, zhao2024retrieval, zhou2024trustworthiness}. Before the advent of LLMs, earlier surveys focused on textual RAGs~\cite{asai2023retrieval, li2022survey}. With the recent unprecedented success achieved by foundational models such as LLMs, various surveys have explored foundational-model-powered RAG in different modalities. \citet{gao2023retrieval} group existing RAG approaches into three categories (Naive, Advanced, and Modular RAGs), summarize three core techniques (Retrieval, Generation, and Augmentation), and review evaluation metrics. In parallel, \citet{zhao2024retrieval} review representative RAG systems according to their corresponding application and data modality. \cite{zhou2024trustworthiness} focuses on reviewing trustworthy concerns and techniques of RAG. However, none of them have a dedicated focus on graph-structured data. To the best of our knowledge, only one very recent study~\cite{peng2024graph} has specifically surveyed RAG in the context of graph-structured data. However, this work mainly focuses on reviewing techniques introduced by graphs under the conventional RAG architecture without specializing in reviewing diverse relations and technical designs for graphs across different domains. In contrast to its holistic review philosophy, we recognize the inherent heterogeneity of graph-structured data and specialize our GraphRAG review across different domains. Specifically, we uncover the fundamental task applications (when to retrieve), graph construction methods and relational rationales (what to retrieve), and GraphRAG techniques (how to retrieve) for each domain. In this way, our survey provides a comprehensive overview of GraphRAG for information retrieval, data mining, and machine learning communities and domain-specific insights that facilitate interdisciplinary research and industrial opportunities.

Our survey is structured as follows: Section~\ref{sec-framework} introduces the holistic framework of GraphRAG and introduces representative techniques for its five key components. From Section~\ref{sec-kg} to \ref{sec-other}, we delve into specific domains, reviewing unique task applications, summarizing existing graph construction methods that guide GraphRAG design for that domain, highlighting domain-specific techniques for each of the five components within our proposed holistic framework, and presenting existing GraphRAG resources (e.g., benchmark datasets and tools) used across different domains. Finally, we discuss research challenges and opportunities in Section~\ref{sec-challenge} and conclude our survey in Section~\ref{sec-conclusion}.

\begin{figure*}[t!]
    \centering
    \includegraphics[width=1\linewidth]{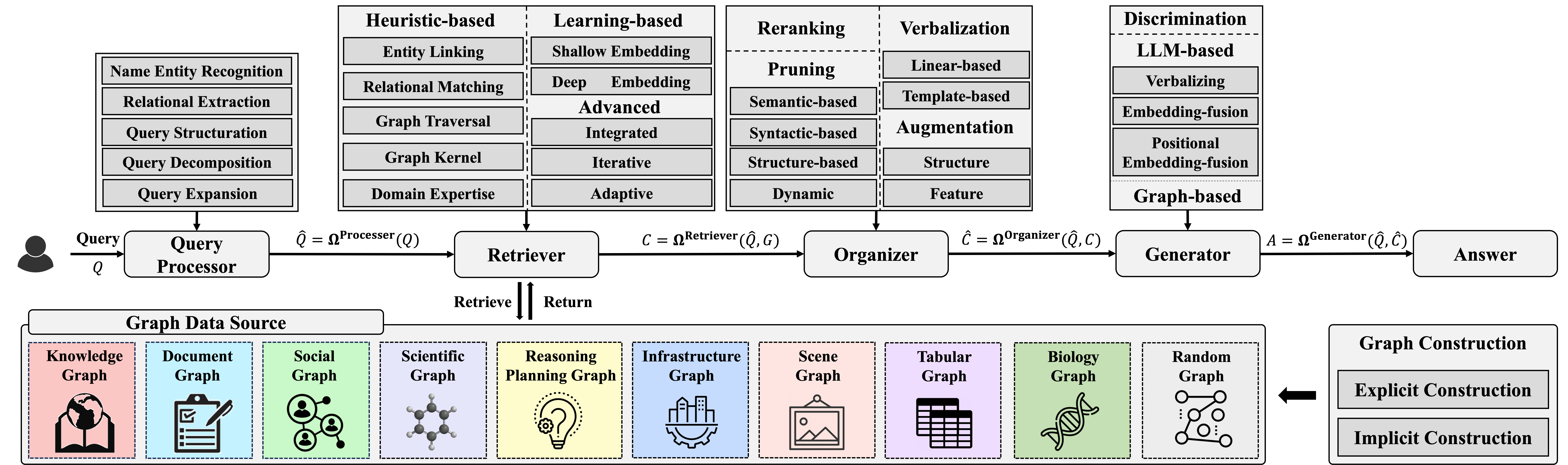}
    \caption{A holistic framework of GraphRAG and representative techniques for its key components.}
    \label{fig-frameworkGraphRAG}
\end{figure*}

\section{A Holistic Framework of GraphRAG}\label{sec-framework}
Based on existing literature on GraphRAG, we present a holistic framework of GraphRAG. Next, we introduce the basic problem setting and notation used throughout the whole framework.

\subsection{Problem Setting and Notations}
Following the general setting of RAG, given a graph-structured data source $G$, the user-defined query $Q$ is further sent to query processor $\boldsymbol{\Omega}^{\textbf{Processor}}$ to obtain the pre-processed query $\hat{Q}$. After that, the retriever $\boldsymbol{\Omega}^{\textbf{Retriever}}$ retrieves the content $C$ from the graph data source $G$ based on $\hat{Q}$. Next, the retrieved content $C$ is refined by the organizer $\boldsymbol{\Omega}^{\textbf{Organizer}}$ to formulate the refined content $\hat{C}$. Finally, the refined content $\hat{C}$ triggers the generator $\boldsymbol{\Omega}^{\textbf{Generator}}$ to generate the final answer $A$. The above five components are summarized as follows:

\begin{itemize}[leftmargin=*]
    \item \textbf{Query Processor $\boldsymbol{\Omega}^{\textbf{Processor}}$}: Preprocessing the given query $\hat{Q} = \boldsymbol{\Omega}^{\textbf{Processor}}(Q)$.

    \item \textbf{Graph Data Source $G$}: Information organized in graph-structured format.

    \item \textbf{Retriever $\boldsymbol{\Omega}^{\textbf{Retriever}}$}: Retrieve the content $C = \boldsymbol{\Omega}^{\textbf{Retriever}}(\hat{Q}, G)$ from $G$ based on the query $\hat{Q}$.

    \item \textbf{Organizer $\boldsymbol{\Omega}^{\textbf{Organizer}}$}: Arrange and refine the retrieved content $\hat{C}=\boldsymbol{\Omega}^{\textbf{Organizer}}(\hat{Q}, C)$. 

    \item \textbf{Generator}: Generate answers $A=\boldsymbol{\Omega}^{\textbf{Generator}}(\hat{Q}, \hat{C})$ to answer query $Q$.
\end{itemize}

Unlike sequential-based textual data and grid-structured image data, graph-structured data encapsulates relational information. To effectively harness this relational information, the above five core components of GraphRAG desire dedicated designs to handle graph-structured input/output and support graph-based operations. For example, in the retriever component, conventional RAG in the Natural Language Processing (NLP) utilizes sparse/dense encoders for index search~\cite{karpukhin2020dense, xiong2020answering}. In contrast, GraphRAG employs graph traversal methods (e.g., entity linking and BFS/DFS) and graph-based encoders (e.g., Graph Neural Networks (GNNs)) to produce embeddings for retrieval. 
This motivates us to summarize key innovations and representative designs of GraphRAG for each of the above five components under the holistic GraphRAG framework in the following.

\subsection{Task Applications and Example Query \texorpdfstring{$Q$}{Q}}
Similar to the general RAG framework where the text-formatted query $Q$ specifies the question context or the task instruction. Query $Q$ in GraphRAG could also be in the format of text. For example, in knowledge graph-based question-answering, the query could be "What is the Capital of China?"~\cite{luo2023reasoning, tian2024graph}. In addition, the query could also be in other formats, such as smile strings for molecular graphs~\cite{guo2023can}, or could even be the combination of multiple formats, such as the scene graph along with the text instruction~\cite{he2024g}. Table~\ref{tab-QA-summary} summarizes the common task applications and exemplary queries used in each domain, as well as their representative references.

% \begin{sidewaystable}
\begin{table*}[ht!]
    % \small
    \centering
    \caption{Summary of Task Applications and Exemplary Queries for GraphRAG in each domain.}
    \resizebox{\textwidth}{!}{
    \begin{adjustbox}{angle=90}
    \begin{tabular}{lp{4.5cm}p{15cm}p{7cm}
    }

    \toprule
    \textbf{Domain} & \textbf{Task Applications} & \textbf{Exemplary Query} & \textbf{References}\\
    \midrule

    %%%%%%%%%%%%%%%%%%%%%%%%%%%%
    \multirow{9}{*}{\textbf{\makecell[l]{Knowledge}}} 
    & \multirow{4}{*}{\textbf{\makecell[l]{KG QA}}} &  \multirow{4}{*}{\makecell[l]{Where is the 16th President of the United States?}} &  \cite{yasunaga2021qa,zhanggreaselm, yasunaga2022deep, sunthink, feng2020scalable, jiang2024hykge, delile2024graph, hu2022empowering, zhang2022subgraph, sanmartin2024kg, luo2023reasoning, knowledgenavigator, retrieve_rewrite_answer, wang2024reasoning, pullnet, pahuja2023retrieve, niu2024mitigating, tian2024graph, jiang2023structgpt, kim2024causal, gao2022graph, mvp_tuning, dai2024counter, luo2023chatkbqa, keqing, mavromatis2024gnn, gnn_llm, fang2024reano, wu2024stark, kim2023kg, wen2023mindmap, li2024dalk, unioqa, kgrank, li2024simple, wang2023knowledgpt, kgagent, zhu2024knowagent} \\    
    &  \textbf{KG Completion} & (Louvre, in, ?) & \cite{choudhary2023complex, pahuja2023retrieve, taunk2023grapeqa, kicgpt} \\
    & \textbf{Temporal QA} & Who was the president of the United States in 2006? & \cite{liao2024gentkg, gao2024two} \\
    & \textbf{Medical Prediction} & How long will the patient be spent in the hospital? & \cite{zhu2024emerge, zhu2024realm} \\
    & \textbf{Fact-Checking} & Who was the founder of Apple? & \cite{kim2023kg} \\
    & \textbf{Cyber Analysis and Defense} & Threat and vulnerability analysis & \cite{rahman2024retrieval} \\
    % & Entity Linking & \harry{todo} & \cite{} \\
    % & Relational Extraction & \harry{todo} & \cite{} \\
    %%%%%%%%%%%%%%%%%%%%%%%%%%%%
    \midrule
    %%%%%%%%%%%%%%%%%%%%%%%%%%%%
    \multirow{7}{*}{\textbf{\makecell[l]{Document}}} & \multirow{2}{*}{\textbf{\makecell[l]{Document Summarization}}} & \multirow{2}{*}{\makecell[l]{Please summarize the following documents.}} & \cite{yasunaga2017graph, wang2020heterogeneous, li2020leveraging, zhang2023contrastive, zhao2024hierarchical, xie2022gretel, zhang2023enhancing1, yang2018integrated, li2023compressed, chen2021sgsum, zhang2024coarse, edge2024local} \\ 
    & \textbf{Text Generation} & Please generate an abstract based on this paper and its references. & \cite{wang2024augmenting, chen2024llaga, kong2024gofa} \\ 
    & \textbf{Document Retrieval} & Find relevant documents related to Houston. & \cite{dong2024don, yu2021graph, zhang2018graph, liu2018matching, guo2024lightrag} \\
    & \textbf{Document Classification} &  What is the category of the following document?  & \cite{zhang2020every, liu2022hierarchical, zhang2020text, xiao2022graph} \\
    & \textbf{Document QA} & What are the main causes of climate change mentioned in the document? & \cite{nie2022capturing, wang2023docgraphlm, he2024g, wang2024knowledge, fang2019hierarchical}. \\
    & \textbf{Relational Extraction} & What is the relation between "Surfers Riverwalk" and "Queensland"? & \cite{wang2020global, nan2020reasoning, sahu2019inter, christopoulou2019connecting, zhou2020global, zhang2021improving} \\
    % & Node Classification &  &  \\
    % & Link Prediction &  &  \\
    %%%%%%%%%%%%%%%%%%%%%%%%%%%%
    \midrule
    %%%%%%%%%%%%%%%%%%%%%%%%%%%%
    \multirow{3}{*}{\textbf{\makecell[l]{Scientific}}} & \textbf{Molecule Generation} & Given an input molecule, retrieve exemplar molecules to guide the generation process. &\cite{wang2022retrieval, huanginteraction} \\
    & \textbf{Molecule Property Prediction} & Lumo is the lowest unoccupied molecular orbital energy. What's the Lumo value of this molecule? &\cite{liu2024moleculargpt} \\
    & \textbf{Scientific QA} & After meals, I feel a bit of stomach reflux. What medication should I take for it? & \cite{yang2024kg, jiang2024hykge, wen2023mindmap, li2024dalk, jeong2024improving,wu2024medical,delile2024graph,pelletier2024explainable,soman2024biomedical,wu2024medical} \\
    %%%%%%%%%%%%%%%%%%%%%%%%%%%%
    \midrule
    %%%%%%%%%%%%%%%%%%%%%%%%%%%%
    \multirow{5}{*}{\textbf{\makecell[l]{Social}}} & \textbf{Entity Property Prediction} & Morality Assessment, Partisanship Detection & \cite{jiang2023social, wang2024knowledge2} \\
    & \textbf{Text Generation} & Review Generation, Social Post Generation & \cite{ahmed2018learning, wang2024augmenting, kim2020retrieval, xie2023factual, sharma2018cyclegen, dong2017learning, park2015retrieval} \\
    & \textbf{Recommendation} & Graph-based/Sequential/Conversational Recommendation & \cite{du2024large, wei2024llmrec, wang2024knowledge2, huang2021graph, zeng2024federated, wang2023zero, qiu2020exploiting, hou2024large, friedman2023leveraging} \\
    & \textbf{Social QA} & What are the best parks for familiar gatherings around Los Gatoks? & \cite{zeng2024large, kemper2024retrieval, wu2024stark} \\
    & \textbf{Fake News Detection} & The COVID-19 vaccine contains microchips for government tracking. & \cite{ram2024credirag, li2024re} \\
    % & Node Classification &  &  \\
    % & Link Prediction &  &  \\
    %%%%%%%%%%%%%%%%%%%%%%%%%%%%
    \midrule
    %%%%%%%%%%%%%%%%%%%%%%%%%%%%
    \multirow{6}{*}{\textbf{\makecell[l]{Reasoning \\\& Planning}}} & \textbf{Sequential Plan Retrieval} & Please generate an image where a girl is reading a book, and her pose is the same as the boy in "example.jpg" & \cite{shen2024hugginggpt, shen2023taskbench, song2023restgpt, wu2024can, zhao2024large} \\
    & \textbf{Asynchronous Planning} & To make calzones, here are the steps and times required; please calculate the optimal plan for completion & \cite{lin2024graph} \\
    & \textbf{Commonsense Reasoning} & Infer the stance and generate/retrieve the corresponding commonsense explanation graph & \cite{saha2021explagraphs} \\
    % that explains the inferred stance
    & \textbf{Defeasible Inference} & Given a premise, a hypothesis may be weakened or overturned in light of new evidence & \cite{madaan2021could} \\
    & \textbf{Tool Usage} & Take Shower -> Walkto (bathroom) -> Walkto(shower) -> Find(shower) -> TurnTo(shower) & \cite{zhuang2023toolchain, hao2023reasoning} \\
    & \textbf{Embodied Planning} & Cook a potato and put it into the recycle bin & \cite{song2023llm, xu2024p} \\
    %%%%%%%%%%%%%%%%%%%%%%%%%%%%
    \midrule
    %%%%%%%%%%%%%%%%%%%%%%%%%%%%
    \multirow{8}{*}{\textbf{\makecell[l]{Tabular}}} & \textbf{Cell Type Prediction} &  What is the type of Cell B17? & \cite{jin2023tabprompt} \\
    & \textbf{Fraud Detection} & Is the following transaction legit or fraudulent? & \cite{rao2020xfraud,singh2023graphfc} \\
    & \textbf{Outlier Detection} & Is the data normal or anomalous? &\cite{goodge2022lunar,cheng2023anti, zhou2023detecting} \\
    & \textbf{CTR Prediction} & Given the history and advertisement features, predict the probability that a user will click on a specific ad. & \cite{li2019fi, du2022learning, guo2021dual} \\
    & \textbf{Data Imputation} & What is the missing value in cell B17 & \cite{zhong2023data, you2020handling, wu2021towards, cappuzzo2024relational}. \\
    & \textbf{Table Type Classification} & What is the category of the table? & \cite{wang2021tuta, jia2023getpt, jin2023tabprompt}. \\
    & \textbf{Table QA} & What is the average amount purchased and value purchased for the supplier who supplies the most products? & \cite{zhang2020cfgnn, zhang2020graph} \\
    & \textbf{Table Retrieval} & Retrieve tables containing information about the currencies of Asian countries & \cite{wang2021retrieving, cheng2021hitab} \\
    %%%%%%%%%%%%%%%%%%%%%%%%%%%%
    \midrule
    %%%%%%%%%%%%%%%%%%%%%%%%%%%%
    \multirow{1}{*}{\textbf{\makecell[l]{Infrastructure}}} & \textbf{Users Behavior Forecasting}  & pedestrians’ crossing actions;  lane change maneuvers & \cite{hussien2024rag} \\
    % Driver behavior forecasting &  & \cite{hussien2024rag} \\
    % & Edge Level Prediction &  & \cite{} \\
    % & Graph Level Prediction &  & \cite{} \\
    % & Utility Forecasting &  & \cite{} \\
    % & Vulnerability Analysis &  & \cite{} \\
    % & Reliability Analysis &  & \cite{} \\
    % & Resilience Analysis &  & \cite{} \\
    %%%%%%%%%%%%%%%%%%%%%%%%%%%%
    \midrule
    %%%%%%%%%%%%%%%%%%%%%%%%%%%%
    \multirow{2}{*}{\textbf{\makecell[l]{Scene}}} & \textbf{Visual QA} & Write a 500-word advertisement for this place in the scene graph that would make people want to visit it & \cite{he2024g} \\
    & \textbf{Embodied Planning} & Cook a potato and put it into the recycle bin & \cite{xu2024p} \\
    %%%%%%%%%%%%%%%%%%%%%%%%%%%%
    \midrule
    \multirow{5}{*}{\textbf{\makecell[l]{Biological}}} & \textbf{Gene Imputation} & What's the imputed gene value of the unsequenced gene KRAS for all cells?   & \cite{rao2021imputing, huang2020scgnn, wang2021scgnn, xu2021efficient, wen2022bi} \\
    %%%%%%%%%%%%%%%%%%%%%%%%%%%%
    & \textbf{Clustering} & What's the cluster for cell B17? & \cite{ciortan2022gnn, yu2022zinb, gan2022deep, cheng2022scgac} \\
    & \textbf{Multi-omics Prediction} & What's the protein expression provided by single-cell RNA-seq expression for cell B17? & \cite{wen2022graph} \\
    & \textbf{Cell Type Deconvolution} & What's the cell type proportion for spot A in the spatial transcriptomics data? & \cite{ding2024spatialctd, song2021dstg} \\
    & \textbf{Spatial Domain Identification} & What's the spatial domain for cell B17 in the spatial transcriptomics data? & \cite{hu2021spagcn, dong2022deciphering} \\
    %%%%%%%%%%%%%%%%%%%%%%%%%%%%
    \midrule
    %%%%%%%%%%%%%%%%%%%%%%%%%%%%
    \multirow{1}{*}{\textbf{\makecell[l]{Random}}} & \textbf{Graph Query} & What is the clustering coefficient of node P357 & \cite{fatemi2023talk, chai2023graphllm, bachmann2024pitfalls, dai2024revisiting, wang2024can, guo2023gpt4graph, luo2024graphinstruct} \\
    \bottomrule
    \end{tabular}
    \end{adjustbox}
    }

    \label{tab-QA-summary}
    
\end{table*}
% \end{sidewaystable}

\begin{figure*}[t!]
    \centering
    \includegraphics[width=1\linewidth]{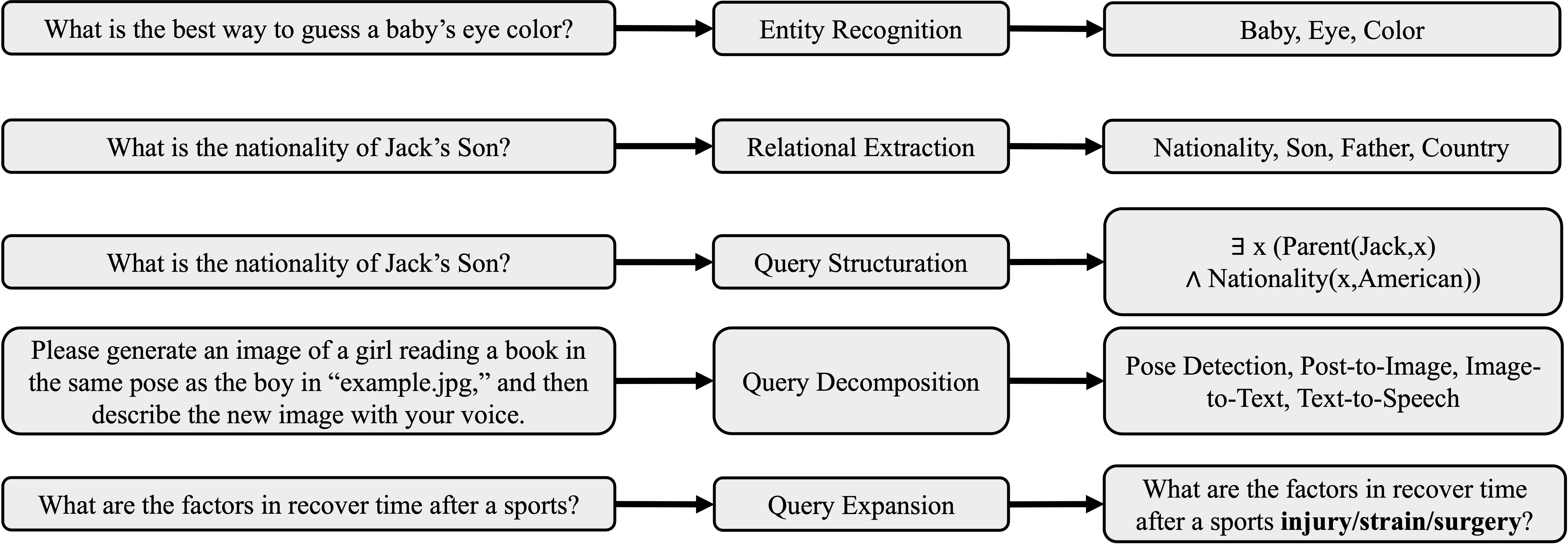}
    \caption{Existing techniques of query processor $\boldsymbol{\Omega}^{\text{Processor}}$ in GraphRAG.}
    \label{fig-query-process}
\end{figure*}

\begin{table}[t!]
\caption{Difference of query processor $\boldsymbol{\Omega}^{\text{Processor}}$ between RAG and GraphRAG.}
\resizebox{\linewidth}{!}{
\begin{tabular}{lcc}
\hline
\textbf{Technique} & \textbf{RAG} & \textbf{GraphRAG} \\
\hline
Entity Recognition & Extracting mentions in knowledge bases & Extracting mentioned nodes in graphs. \\
Relational Extraction & Extracting textual relations  & Extracting graph edge relations\\
Query Structuration & Structuring text query to SQL, SPARQL & Structuring text query to GQL  \\
Query Decomposition & Decomposed queries are separate  & Decomposed queries are logically related \\
Query Expansion & Expansion based on semantic knowledge & Expansion based on relational knowledge\\
\hline
\end{tabular}
}
\label{tab-query-process}
\end{table}

\newpage
\subsection{Query Processor \texorpdfstring{$\boldsymbol{\Omega}^{\text{Processor}}$}{Omega Processor}}

Unlike RAG, where both queries and data sources are purely text-formatted, data sources used in GraphRAG are graph-structured, which raises challenges in bridging text-formatted queries and graph-structured data sources. For example, the information that connects the knowledge graph and the query "Who is Justin Bieber's brother?" is not a specific passage but instead the entity "Justin Bieber" and the relation "brother of". Many techniques are proposed to correctly extract this information from the query, including entity recognition, relational extraction, query structuration, query decomposition, and query expansion. In the following, we first review each of these five query processing techniques within the broader NLP domain, followed by a focused examination of their unique adaptations for GraphRAG.

\subsubsection{Name Entity Recognition} \label{sec-ner}
Named Entity Recognition (NER) aims to identify mentions of entities from the text that belong to predefined categories, such as persons, locations, or organizations, and it serves as a fundamental component for numerous natural language applications~\cite{huang2021few, keraghel2024survey, li2020survey, meng2021distantly}. NER techniques can be broadly categorized into four main approaches: (1) rule-based methods, which rely entirely on handcrafted rules and require no annotated data; (2) unsupervised learning methods, which use unsupervised algorithms without labeled training examples; (3) feature-based supervised learning methods, which depend on supervised algorithms and careful feature engineering; and (4) deep learning approaches, which automatically discover representations needed after (un)-supervised training the deep learning models. Recent LLMs fall into the category of deep learning approaches and have demonstrated unprecedented success for NER. More details about these techniques and their resources can be found in~\citet{li2020survey}. 

Specifically, in the GraphRAG context, entity recognition primarily uses deep learning techniques (e.g., EntityLinker~\cite{tian2024graph, yasunaga2022deep} and LLM-based extraction~\cite{jin2024graph}) to identify entities in queries grounded by nodes in the given graph data sources. This step is vital for applications such as knowledge graph-based question answering~\cite{tian2024graph, yasunaga2021qa, yasunaga2022deep}. For example, given the question, "What is the best way to guess the color of the eye of the baby?", NER extracts entities such as "baby", "eye", and "color", which correspond to nodes in the knowledge graph and are treated as the seed nodes to initialize the retrieval process thereafter~\cite{wen2023mindmap, sanmartin2024kg}. For more recent GraphRAG research, NER has evolved beyond identifying the entity names but instead their structures. For example, \citet{jin2024graph} leverages LLMs to recognize node types in the graph, which further guides the retriever to identify nodes that match the recognized types for next-round exploration. For example, given the question "Who are the authors of `Language Models are Unsupervised Multi-task Learners'?" the initially recognized entity should not only be based on the semantic name "Language Models are Unsupervised Multi-task Learners" but also be based on the type of that entity, which is the paper node in this case. Accurately recognizing the names and structures of entities in GraphRAG reduces cascading errors and provides a solid foundation for subsequent retrieval and generation steps.

\subsubsection{Relational Extraction}
Similar to NER, relational extraction (RE) is a long-standing technique in NLP to identify relations among entities and is widely applied to structured search, sentiment analysis, question answering, summarization, and knowledge graph construction~\cite{chen2018knowedu, li2020real, nasar2021named}. Recent advances in RE have been largely driven by deep learning techniques, and they can be summarized into three perspectives: text representation, context encoding, and triplet prediction, more details of which can be found in \citet{pawar2017relation, nasar2021named, han2020more}. 

For GraphRAG, RE serves two key purposes: constructing graph-structured data sources (e.g., knowledge graphs) by extracting triplets and matching the relations mentioned in the query and the graph data source to guide the graph search. For instance, given a query like "What is the capital of China?", relational extraction identifies the relation "capital of" and searches for corresponding edges via vector similarity in the knowledge graph, which guides the neighborhood selection and graph traversal direction~\cite{gao2024two, kim2023kg, luo2023reasoning, luo2024graph}.

\subsubsection{Query Structuration}

Query structuration transforms queries into formats tailored to specific data sources and tasks. It often converts natural language queries into structured formats like SQL or SPARQL~\cite{jiang2023structgpt, li2023chain} to interact with relational databases. Recent advancements leverage pre-trained and fine-tuned LLMs to generate structured queries from natural language input to query databases. For graph-structured data, Graph Query Language (GQL) has emerged, such as Cypher, GraphQL, and SPARQL, which enables complex interactions with property graph databases. Additionally, \citet{jin2024graph} introduced a technique that decomposes complex queries into multiple structured operations, including node retrieval, feature fetching, neighbor checks, and degree assessment, enhancing precision and adaptability in querying.

\subsubsection{Query Decomposition}
Query decomposition~\cite{wong1976decomposition} aims to split the input query into multiple distinct subqueries, which are used to first retrieve sub-results and aggregate these sub-results together for the final results. In most existing RAG and GraphRAG, decomposed queries usually possess explicit logic connections that can handle complex tasks that require multistep reasoning and planning~\cite{lin2024graph, park2023graph, shen2023taskbench, song2023restgpt, xu2024p}. For example, a query like "Please generate an image where a girl is reading a book, and her pose is the same as the boy in `example.jpg' then describe the new image with your voice" involves multiple subtasks~\cite{xu2024p}, each of which would be completed by a specific sub-query. In addition, \citet{park2023graph} enhance the decomposition of the query by building a question graph where each sub-query is represented as a triplet within the graph. These graph-structured sub-queries effectively guide the retriever/generator through multi-step promptings.

\subsubsection{Query Expansion}
Query Expansion enriches a query by adding meaningful terms with similar significance~\cite{azad2019query}, which primarily addresses three challenges: (1) user-submitted queries are ambiguous and relate to multiple topics; (2) queries may be too brief to fully capture user intent; and (3) users are often uncertain about what they are seeking. Generally, it can be categorized into manual query expansion, automatic query expansion, and interactive query expansion. More recently, LLM-based query expansion has been a prominent area due to the creativity of the generated content\cite{chen2024analyze, jagerman2023query, lei2024corpus}

Unlike existing methods that mostly focus on textual similarities and overlook relations, QE in GraphRAG augments LLM expansion with structured relations. For example \citet{xia2024knowledge} expands the query by leveraging neighboring nodes of the mentioned entities in the query. Alternatively, \citet{keqing} convert the query into several sub-queries using pre-defined templates.

\subsection{Retriever \texorpdfstring{$\boldsymbol{\Omega}^{\textbf{Retriever}}$}{Omega Retriever}}
\label{sec:retriver}

After obtaining the processed query $\hat{Q}$, the retriever $\boldsymbol{\Omega}^{\textbf{Retriever}}$ identifies and retrieves relevant content $C$ from external graph sources $G$ to augment the downstream task execution:
\begin{equation}
    C = \boldsymbol{\Omega}^{\textbf{Retriever}}(\hat{Q}, G)
\end{equation}

Recently, retrievers have been increasingly integrated with LLMs to mitigate hallucination issues~\cite{tonmoy2024comprehensive}, address privacy concerns~\cite{zeng2024good}, and enhance explainability and dynamic adaptability~\cite{shi2024retrieval, wang2023knowledge}. While effective, they are predominantly designed for texts and images and not readily transferable to graph-structured data for GraphRAG for two reasons. First, the input/output format of GraphRAG differs significantly from that of traditional RAG. While most retrievers in RAG use NLP tokenizers for encoders and adhere to the "Text-in, Text-out" workflow, the workflow of GraphRAG is more diverse, including "Text-in, Text-out"~\cite{tian2024graph, wang2024knowledge}, "Text-in, Graph-out"~\cite{wu2024can, zhuangtoolchain}, "Graph-in, Text-out" and "Graph-in, Graph-out" processes~\cite{wangretrieval}. Secondly, retrievers in traditional RAGs do not capture graph structure signals. Methods like BM25 and TF-IDF~\cite{robertson2004simple, ramos2003using} primarily focus on lexical signals, and deep-learning-based retrievers~\cite  {karpukhin2020dense} usually capture semantic signals, both of which overlook the graph structure signals. This motivates us to review existing GraphRAG retrievers, i.e., heuristic-based, learning-based, and domain-specific retrievers, with a particular emphasis on their unique technical design adapted to graph-structured data.

\begin{figure*}[t!]
    \centering
    \includegraphics[width=1\linewidth]{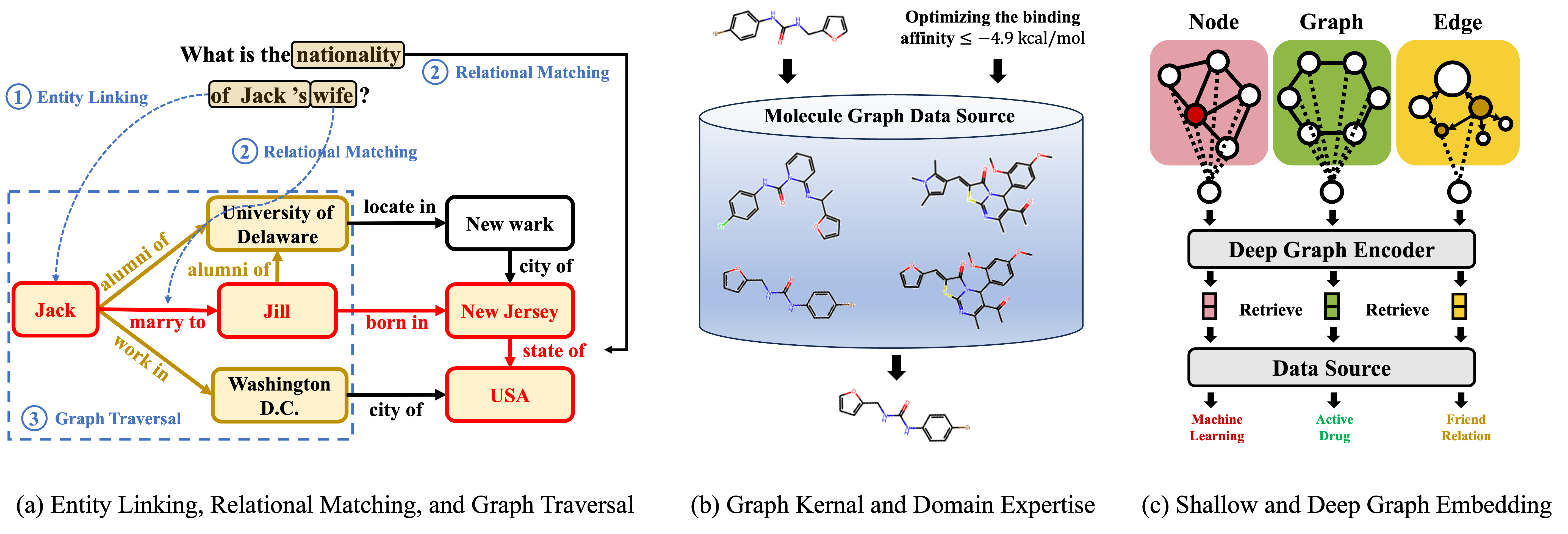}
    \caption{Visualizing representative retrievers used in GraphRAG.}
    \label{fig-retriever-tech}
\end{figure*}

\begin{table}[t!]
\caption{Categorizing representative retrievers used in GraphRAG.}
\resizebox{\linewidth}{!}{
\begin{tabular}{lcccc}
\hline
\textbf{Method/Strategy} & \textbf{Input} & \textbf{Output} & \textbf{Description} \\
\hline
Entity Linking & Entity Mention & Node & Match query entity and graph node\\
Relational Matching & Relation Mention & Edge & Match query relation and graph edge \\
Graph Traversal & Node/Edge & Graph & Expand seed nodes/edges into subgraphs \\
Graph Kernel & (Sub)Graph & (Sub)Graph & Match query graph and candidate graph\\
\hline
Shallow Embedding & Any & Any & Embedding similarity match query and candidate\\
Deep Embedding & Any & Any & Embedding similarity match query and candidate\\
\hline
Domain Expertise & Expertise Rule & Any & Match Domain Expertise with nodes/edges/graphs\\
\hline
\end{tabular}
}
\label{tab-retriever}
\end{table}

\subsubsection{Heuristic-based Retriever}
Heuristic-based retrievers primarily use predefined rules, domain-specific insights, and hard-coded algorithms to extract relevant information from graph data sources. Their reliance on explicit rules often makes them more time/resource-efficient compared to deep learning models. For instance, simple graph traversal methods like BFS or DFS can be executed in linear time without needing training data. However, this same reliance on fixed heuristics also limits their adaptability to generalize to unseen scenarios. In the following, we review the heuristic-based retrievers commonly used in GraphRAG.

\textbf{Entity Linking}: In heuristic-based retrievers, entity linking involves mapping entities identified in the query to corresponding nodes in graph data sources. This mapping forms an initial bridge between the query and the graph, serving as either the retriever by itself or as a foundation for further graph traversal to broaden the scope of the retrieval. The effectiveness of this approach relies on accurate entity recognition conducted by the query processor and the quality of labeled entities on graph nodes. This technique is commonly applied in knowledge graphs, where Top-K nodes are selected as starting points based on their textual similarity to the query. The similarity metric can be computed using vector embeddings~\cite{wen2023mindmap, sanmartin2024kg} and lexical features~\cite{wang2024knowledge}. More recently, LLMs have been used as knowledgeable context augmenters to generate mention-centered descriptions as additional input to augment the long-tail entities where their limited training data usually cause the entity linking model to struggle to disambiguate~\cite{xin2024llmael}.

\textbf{Relational Matching}:
Relational matching, similar to entity linking, is a heuristic-based retrieval approach designed to identify edges within graph data sources that align with the relations specified in a query. This method is crucial for tasks that focus on identifying relationships among entities in a graph. The matched edges guide the traversal process by indicating which edges to explore next based on the entities and relations encountered in the graph data sources. Similar to entity linking, Top-K edges are selected based on their similarity to each edge in the graph~\cite{kim2023kg, gao2024two}.

In addition to the efficiency and simplicity of the above two types of heuristic-based retrievers, another key advantage is their ability to overcome ambiguity. For example, although machine/deep learning-based retrievers are difficult to differentiate semantically/lexically similar entities/relations (e.g., Byte vs. Bit, and President of vs. Resident of), these heuristic methods can easily distinguish them based on pre-defined rules, even in cases where semantic/lexical differences are subtle.

\textbf{Graph Traversal}:
After performing entity linking and relational matching to identify initial nodes and relations in graph data sources, graph traversal algorithms (e.g., BFS, DFS) can expand this set to uncover additional query-relevant information. However, a core challenge for traversal-based retrieval is the risk of information overload, as the exponentially expanding neighborhood often includes substantial irrelevant content. To address this, current traversal techniques integrate adaptive retrieval and filtering processes, selectively exploring the most relevant neighboring nodes and incrementally refining the retrieved content to minimize noise. This graph traversal is mainly used in GraphRAG for knowledge and document graphs. When traversing on these two types of graphs, many methods extract all paths less than length $l$ {\it between} the nodes identified by entity linking~\cite{yasunaga2021qa, yasunaga2022deep, zhanggreaselm, jiang2024hykge, feng2020scalable}, while others consider the $l$-hop subgraph around the initial entities~\cite{niu2024mitigating, tian2024graph, jiang2023structgpt, kim2024causal}. To more efficiently traverse the KG, other methods prune irrelevant paths via the use of a LLM~\cite{luoreasoning, sanmartin2024kg, wang2024knowledge, knowledgenavigator} and others use pre-defined rules or templates to traverse the graph~\cite{keqing, liao2024gentkg, dai2024counter}.

\textbf{Graph Kernel}:
Compared with the above heuristic-based retrievers for retrieving nodes, edges, and their combined subgraphs, some earlier works (e.g., graph extraction and image retrieval)~\cite{wu2016gksh, lebrun2008image, glavavs2013event} treat the text and image as the entire graph and use graph-level heuristics such as graph kernels to measure similarity and retrieve. Graph kernels measure pairwise similarities by calculating inner products between graphs, aligning both structural and semantic aspects of the query and the retrieved graphs. Notable examples include the random-walk kernel and the Weisfeiler Leman kernel~\cite{shervashidze2011weisfeiler, vishwanathan2010graph}. The random walk kernel computes similarity by performing simultaneous random walks on two graphs and counting the number of matching paths. The 
Weisfeiler Leman kernel iteratively applies the Weisfeiler Leman algorithm to produce color distributions of node labels at each iteration and then calculates similarity based on the inner products of these histogram vectors. For example, \citet{wu2016gksh} constructs event graphs of both documents and queries and uses a product graph kernel that counts walks between two graphs to measure the query-document similarity and rank the documents. \citet{lebrun2008image} conducts event graph matching by introducing a fast and efficient graph-matching kernel for image retrieval. Similarly, \citet{glavavs2013event} translates images into representative attribute structural graphs that capture spatial relations among regions and perform graph kernel based on random walks to derive hash codes for image retrieval.

\textbf{Domain Expertise}:
The domain-agnostic nature of traditional heuristic-based methods restricts their effectiveness in areas that require specialized expertise. For instance, in drug discovery, chemists typically design drugs by referencing existing molecules with desirable properties rather than constructing molecular structures from scratch. These molecules are selected based on domain knowledge that guides the retrieval of structures with similar characteristics. Following this intuition, many GraphRAG systems incorporate domain expertise to enhance retriever design. \citet{wang2022retrieval} develop a hybrid retrieval system that integrates heuristic-based and learning-based retrieval to retrieve exemplar molecules that partially meet the target design criteria.

\subsubsection{Learning-based Retriever}
One significant limitation of heuristic-based retrievers is their over-reliance on pre-defined rules, which limits their generalizability to data that does not strictly adhere to these rules. For example, when confronted with entities that have slight semantic or structural variations, such as "doctor" and "physician", heuristic-based retrievers like entity linking may treat them differently due to their distinct lexical representations, despite their shared underlying meaning. To overcome this limitation, learning-based retrievers have been proposed to capture deeper, more abstract, and task-relevant relations between the query and objects in data sources, which avoid relying solely on hard-coded rules. These retrievers often work by uniformly compressing information of various formats (e.g., texts and images) into embeddings based on machine learning encoders and then fetching relevant information by conducting an embedding-based similarity search. Notably, some entity linking and relational matching methods that employ machine learning encoders to generate embeddings for matching should also be considered as learning-based retrievers. 

In conventional RAG, assuming the query $q$ and data sources that contain $n$ instances $S$ are embedded by corresponding encoders as $\mathbf{q} = \mathcal{F}_{q}(q) \in \mathbb{R}^{d}$ and $\mathbf{S} = \mathcal{F}_{S}(S) \in \mathbb{R}^{n\times d}$, we retrieve top-$k$ instances by similarity search according to the pre-defined similarity function $\phi$ in the embedding space. 
\begin{equation}
    \mathcal{S}^{*} = \argmax_{k}\phi(\mathbf{q}, \mathbf{S}),
\end{equation}

Unlike RAGs that use language and vision encoders to embed texts and images, encoders used in GraphRAG retrieval extend beyond independently and identically distributed (i.i.d.) data by embedding nodes, edges, and (sub)graphs. Depending on the input format, the encoder could be a text encoder for query, a graph-based encoder for graph structure, and an integrated text-and-graph encoder for the textual attributed graph~\cite{chen2023label, chen2024exploring}. We specifically focus on graph-based encoders. Existing graph-based encoders can be broadly categorized into shallow embedding methods -- such as Node2Vec and DeepWalk -- and deep embedding methods like Graph Neural Networks (GNNs). Below, we review these two encoders and their unique roles in GraphRAG.

\textbf{Shallow Embedding Methods}:  
Shallow embedding methods~\cite{fey2019fast}, like Node2Vec~\cite{grover2016node2vec} and Role2Vec~\cite{ahmed2018learning}, learn node, edge, and graph embeddings that retain the essential structural information of the original graph. Based on the type of structural information that can be extracted, these methods generally fall into two categories: proximity/role-based embeddings. Proximity-based methods, such as DeepWalk and Node2Vec~\cite{grover2016node2vec, perozzi2014deepwalk}, focus on preserving the proximity of connected nodes, ensuring that nodes close in the graph also remain close in the embedding space. Role-based methods, like Role2Vec and GraphWave~\cite{ahmed2018learning, donnat2018learning}, generate node embeddings based on their structural roles rather than their proximity relations. In general, these methods initialize each node with a latent embedding vector and conduct unsupervised training to squeeze structural signals derived from graph structure into the embedding. In GraphRAG, proximity-based shallow embeddings can effectively retrieve entities that are proximally close, while role-based embeddings can capture entities that share similar roles. For instance, proximity-based embeddings could be used to retrieve academic papers by fetching papers sharing similar research topics or retrieve reviews from products that are co-purchased with the current product~\cite{salemi2023lamp, wang2024augmenting}. Meanwhile, role-based embeddings could support tasks like generating company emails by retrieving similar emails based on shared roles or tones~\cite{wang2024augmenting}.

\textbf{Deep Embedding Methods}:
Although shallow embedding methods incorporate structural signals into learned embeddings for nodes, edges, or entire graphs, they struggle to leverage semantic features—like bag-of-words representations for academic paper retrieval or atomic numbers for molecular retrieval~\cite{fey2019fast}. Additionally, these methods lack inductivity, requiring re-initialization and retraining whenever new nodes, edges, or graphs are added. This limitation significantly reduces their applicability in GraphRAG retrieval tasks as real-world knowledge evolves dynamically where new information continually replaces outdated content, such as in citation networks, social graphs, and knowledge graphs~\cite{cheng2024multi, wang2023knowledge, wu2024updating}. To address these limitations, deep embedding methods have been proposed, which not only jointly fuse features and graph structures to obtain embeddings for retrieval but also inherent inductive property as the newly coming nodes/edges/graphs share common feature space with the ones during the training phase. One of the most representative and powerful approaches in this category is GNN, which combines the power of message-passing to encode structural signals and feature transformation to extract task-relevant information. Mathematically, $l^{\text{th}}$-layer graph convolution can be formulated as:

\begin{align}
    \mathbf{x}_i^l &= \gamma_{\boldsymbol{\Theta}_{\gamma}} \left( \mathbf{x}_i^{l-1} \oplus \sum_{j \in \mathcal{N}_{i}} \phi_{\boldsymbol{\Theta}_{\phi}} \left( \mathbf{x}_i^{l-1}, \mathbf{x}_j^{l-1}, \mathbf{e}_{ij} \right) \right),~~~~~~~~~~~~~~~~~~\text{Node-level} \label{eq-node_level}\\
    \mathbf{e}_{ij}^l &= \gamma_{\boldsymbol{\Theta}_{\gamma}} \left( \mathbf{e}_{ij}^{l-1} \oplus \sum_{e_{mn} \in \mathcal{N}_{ij}^{e}} \phi_{\boldsymbol{\Theta}_{\phi}} \left( \mathbf{e}_{ij}^{l-1}, \mathbf{e}_{mn}^{l-1}, \mathbf{x}_{e_{ij}\cap e_{mn}} \right) \right),~~~~\text{Edge-level} \label{eq-edge_level}\\
    \mathbf{G}^l &= \rho_{\boldsymbol{\Theta}_{\rho_G}}(\{\mathbf{x}_i^l, \mathbf{e}_{ij}^l \mid v_i \in \mathcal{V}_G, e_{ij} \in \mathcal{E}_{G}\}),~~~~~~~~~~~~~~~~~~~~~~~~~~~~~~~\text{Graph-level} \label{eq-graph_level}
\end{align}

In node-level graph convolution, each node $v_i$ adaptively aggregates the embeddings of its neighboring nodes $\mathcal{N}_i$, with weights based on edge features via the weighting function $\phi_{\boldsymbol{\Theta}_{\phi}}$. The aggregated neighborhood embeddings are then combined with the node's own embedding from the previous layer $\mathbf{x}_i^{l-1}$, using a combination function $\gamma_{\boldsymbol{\Theta}_{\gamma}}$, as shown by Eq~\eqref{eq-node_level}. Optimizing loss from training downstream tasks would enable the weighting function $\phi_{\boldsymbol{\Theta}_{\phi}}$ to prioritize the most important neighbors and enable the combination function $\gamma_{\boldsymbol{\Theta}_{\gamma}}$ to balance contributions from the node’s neighborhood and its own embedding. Similarly, in edge-level graph convolution, the same aggregation principle applies, but the neighbors of an edge are edges incident to the same ending points of that edge $\mathcal{N}^e_{ij}$, as shown by Eq~\eqref{eq-edge_level}. Graph-level embeddings could be obtained by further applying pooling operation $\rho_{\boldsymbol{\Theta}_{\rho}}$ over node and edge embeddings, as shown by Eq~\eqref{eq-graph_level}. Following this GNN-based embedding paradigm, various forms of graph knowledge from diverse sources—such as nodes, edges, and (sub)graphs—can be uniformly embedded into vector representations, as shown in Figure~\ref{fig-retriever-tech}(c) where we derive embeddings for nodes ($\mathbf{X}$), edges ($\mathbf{E}$), and graphs ($\mathbf{G}$).

Having obtained these node/edge/graph-level embeddings further enables us to create embeddings for different types of structures ($\mathbf{S}$) by combining these sub-structure embeddings according to specific configurations for each structure. For instance, if the retrieved subgraph is a path within a knowledge graph, we can aggregate the embeddings of the nodes and relations along that path to form a cohesive path embedding.\footnote{Incorporating structural signals may be necessary, a consideration to be addressed in future work.} Eventually, the resulting embeddings for different structures can be utilized either during the training phase to optimize query alignment or during the testing phase to enable similarity-based neural search. For example, GNN-RAG~\cite{sanmartin2024kg} uses a GNN to perform retrieval, where a separate round of message passing is performed for each query. The query $\hat{Q}$ is incorporated into the message passing by, including its embedding in the message computation. A set of ``candidate'' nodes is chosen which have a probability of being relevant greater than some threshold. The shortest path from the query nodes to each candidate node is retrieved as context. \citet{gnn_llm} consider the use of a conditional GNN~\cite{conditional_mpnn} where only the linked entities from the query are initialized to a non-zero representation. The candidate nodes are chosen in a similar manner to~\cite{sanmartin2024kg}. A single path is then retrieved for each candidate node and is extracted by backtracking until we reach a query node. REANO~\cite{fang2024reano} encodes the query information into an edge-specific attention weight, conditional on the query. After aggregation, the top k triples most similar to the query are chosen as context.

\subsubsection{Advanced Retrieval Strategies}
Real-world queries are often complex and encode multi-aspect intentions, possess structure patterns, and desire multi-hop reasoning that the aforementioned basic retrievers struggle to address. For example, answering "What is the name of the fight song of the university whose main campus is in Lawrence, Kansas, and whose branch campuses are in the Kansas City metropolitan area?" demands multi-hop reasoning to identify the university based on location and retrieve information about its fight song~\cite{he2024make, trivedi2022interleaving}. Similarly, a query like "What are the main themes in the dataset?" requires understanding the product community structure, retrieving themes for each community, and aggregating the identified themes together to summarize the main theme~\cite{edge2024local}. Furthermore, when asking "Who is the most impactful research scholar in deep learning?" the answer could vary depending on multiple aspects~\cite{zhang2023pre}, such as the number of citations, the volume of published papers, or the number of co-authors. Accurately addressing such queries requires a deeper understanding of the underlying data distribution to discern which aspect the query prioritizes. To address these highly complex queries, advanced retrieval strategies have been proposed, and we review them as follows:

\textbf{Integrated Retrieval:} Integrated retrieval combines various types of retrievers to capture relevant information by balancing their strengths and weaknesses. Typically, integrated retrieval approaches are categorized according to which individual retrievers are used in combination, with notable examples including neural-symbolic retrieval~\cite{dietz2023neuro, lee2024diakop, wang2024knowledge} and multimodal retrieval~\cite{cui2024more, long2024generative}. 

Since the knowledge stored in graph-structured data exists mostly in symbolic format, neural-symbolic retrieval is a natural choice for the integrated retrieval strategy in GraphRAG. This strategy interleaves rule-based patterns for retrieving symbolic knowledge with neural-based signals for retrieving more abstract and deep knowledge. For example, \citet{luo2023reasoning, wen2023mindmap} first expands the neighbors based on the knowledge of the symbolic knowledge graph and then performs path retrieval using neural matching. In contrast, \citet{mavromatis2024gnn} first utilizes GNNs to retrieve seed entities (neural retrieval) and then extract the shortest paths from seed entities (symbolic retrieval). Similarly, \citet{tian2024graph, yasunaga2021qa, yasunaga2022deep, wang2024knowledge2, luo2023reasoning} fetch the k-hop neighborhood of the entities mentioned in the current question-answering pair and the session of user-generated items as the answer candidates (symbolic retrieval) and compute attention between the query and the extracted subgraph to differentiate candidate relevance (neural retrieval).

\textbf{Iterative Retrieval:} Iterative retrieval is a multistep process where consecutive retrieval operations share common dependencies such as causal, resource, and temporal dependency. These dependencies can be implicitly characterized by the retrieval order in RAG~\cite{trivedi2022interleaving, yang2024rag} or explicitly modeled as a graph structure in GraphRAG~\cite{he2024make, wu2024can}. Consequently, iterative retrieval is primarily utilized in GraphRAG to capture these dependencies. For example, KGP~\cite{wang2023knowledge} alternates between generating the next piece of evidence for the question and selecting the most promising neighbor. ToG~\cite{sunthink} starts by identifying initial entities and then iteratively expands reasoning paths until enough information is gathered to answer the question. StructGPT~\cite{jiang2023structgpt} pre-defines graph interfaces and prompts LLMs to iteratively invoke these interfaces until sufficient information is collected.

\textbf{Adaptive Retrieval:} While retrieved external knowledge offers benefits, it also introduces risks. If the generator already possesses sufficient internal knowledge for a task, the retrieved external information may be unnecessary or even conflicting~\cite{chen2022rich, xu2024knowledge}. Specifically, when internal knowledge fully covers the necessary information, retrieval becomes redundant and may introduce contradictions. To mitigate this, knowledge checking has been proposed in RAG systems~\cite{jeong2024adaptive, jin2024tug, wang2024astute, yao2024seakr}. This approach allows the system to adaptively assess when and how much external information is needed. By equipping the retriever with this adaptability, RAG can provide more intelligent, flexible, and context-aware responses, fostering better harmony between internal and external knowledge sources.

One of the adaptive retrievals in GraphRAG is designed by considering different reasoning depths for different queries, i.e., too few hops of graph traversal might overlook critical reasoning relations, while too many can introduce unnecessary noise. \citet{knowledgenavigator, wu2023retrieve} address this by training models to predict the required number of hops for a given query and retrieving the relevant graph content accordingly. No existing works focus on resolving knowledge conflicts in GraphRAG, and therefore, we leave this discussion to future work.

\subsection{Organizer}\label{sec:organizer}
After retrieving the relevant content $C$ from external graph data sources, which may be in the format of entities, relations, triplets, paths or subgraphs, the organizer $\boldsymbol{\Omega}^{\textbf{Organizer}}$ processes this content in conjunction with the processed query $\hat{Q}$. The aim is to post-process and refine the retrieved content to better adapt it for generator consumption, thereby further improving the quality of the downstream content generation. Formally, the organizer is represented as follows:
\begin{align}
    \hat{C} = \boldsymbol{\Omega}^{\textbf{Organizer}}(\hat{Q}, C)
\end{align}

In GraphRAG, the need for fine-grained organization and refinement of retrieved content is driven by several key reasons. Firstly, when the retrieved contents are subgraphs, their heterogeneous format of knowledge in terms of node/edge features and graph structures becomes more likely to include irrelevant and noisy information, which poses significant difficulty for LLM to digest and thus compromises the generation quality. This raises the desire for graph pruning techniques to polish the retrieved subgraph and remove task-irrelevant knowledge. Secondly, LLMs have been widely demonstrated to possess attention biases toward certain positions of relevant information within the retrieved context~\cite{chen2023understanding}. Therefore, the exponentially growing neighbors as the receptive field enlarges (i.e., the number of hops increases) in the retrieved subgraphs would also exponentially increase the amount of context length in the prompt and dilute the focus of LLMs on the task-relevant knowledge~\cite{shi2023large}. This poses a new requirement for the graph-based reranking mechanism to prioritize the most important content within the retrieved graph. Thirdly, the retrieved content might be incomplete in terms of both the semantic content and the structural content, which necessitates graph augmentation for the enhancement. Finally, the retrieved content is often a graph, which not only possesses semantic content information but also owns its unique structure. This complex structural content is not easily consumed by LLMs that are trained by next-token prediction coupled with linearized prompting, which requires structure-aware verbalization techniques to reorganize. We will formally review each of the above-motivated organizer techniques in the following sections.

\subsubsection{Graph Pruning}
In GraphRAG, the retrieved graph can be large and potentially contain a significant amount of noisy and redundant information. For example, when graph traversal methods are applied in retrieval, the size of the retrieved subgraph exponentially increases with the number of hops. Large subgraph sizes not only increase computational costs but can also reduce generation quality due to the inclusion of noisy information. In contrast, if the number of hops is too small, the retrieved subgraph may be too small to include crucial knowledge required by tasks. To achieve a better trade-off between the size of the retrieved subgraph and the amount of its encoded task-relevant information, various graph pruning methods have been proposed to reduce the size of subgraphs by removing irrelevant nodes and edges while preserving the essential information.

\begin{itemize}[leftmargin=*]
    \item \textbf{Semantic-based pruning:}  Semantic-based pruning focuses on reducing the graph size by removing nodes and edge relations that are semantically irrelevant to the query. For example, QA-GNN~\cite{yasunaga2021qa} prunes irrelevant nodes with low relevance scores by encoding the query context and node labels using LLMs, followed by a linear projection. GraphQA~\cite{taunk2023grapeqa} further removes clusters of nodes with the lowest relevance to the query. KnowledgeNavigator~\cite{guo2024knowledgenavigator} scores the relations in the retrieved graph based on the query and prunes irrelevant relations to reduce graph size. Additionally, \citet{gao2022graph} partition the retrieved subgraph into smaller subgraphs and then ranks them with only the top-k smaller subgraphs retained for generations. G-Retriever~\cite{he2024g} defines a semantic score for each retrieved node and edge, then refines the graph by solving the prize-collecting Steiner tree problem to construct a more compact and relevant subgraph.
    \item \textbf{Syntactic-based pruning:} Syntactic-based pruning removes irrelevant nodes from a syntactic perspective. For instance, \citet{su2024pipenet} leverages dependency analysis to generate a parsing tree of the context and then filters the retrieved nodes based on their span distance from the parsing tree.
    \item \textbf{Structure-based pruning:} Structure-based pruning methods focus on pruning the retrieved graph based on its structural properties. For example, RoK~\cite{wang2024reasoning} filters out reasoning paths in the subgraph by calculating the average PageRank score for each path. Other works, such as \citet{jiang2023structgpt} and \citet{he2021improving}, also leverage PageRank to extract the most relevant entities.
    \item \textbf{Dynamic pruning:} Unlike the aforementioned methods, which typically prune the graph once, dynamic pruning removes noisy nodes dynamically during training.  For example, JointLK~\cite{sun2021jointlk} uses attention weights to recursively remove irrelevant nodes at each layer, keeping only a fixed ratio of nodes. Similarly, DHLK~\cite{wang2023dynamic} filters out nodes with attention scores below a certain threshold dynamically during the learning process.
\end{itemize}

\subsubsection{Reranker}

The performance of LLMs can be influenced by the position of relevant information within the context, whether it appears at the beginning, middle, or end~\cite{chen2023understanding}. Additionally, LLMs' generation is impacted by the order in which in-context knowledge is provided, with later documents contributing less than earlier ones~\cite{ivgi2023efficient, liu2024lost}. While retrieved information is typically ordered by relevance scores during the retrieval process, these scores are often based on coarse-grained rankings across a large set of candidates. Enhancing reordering solely among the retrieved information at a fine-grained level, a process known as re-ranking, is essential to achieve optimal downstream performance. For example, \citet{li2023graph} rerank retrieved triples using a pre-trained cross-encoder. \citet{jiang2024hykge} and \citet{liu2024knowledge} employ pre-trained reranker models to rerank retrieved paths. \citet{yu2021kg} train a GNN to rerank the retrieved passages.  \citet{liao2024gentkg} order the paths by the time they occurred, giving more emphasis to recent paths. 

\subsubsection{Graph Augmentation}
Graph augmentation aims to enrich the retrieved graph to either enhance the content or improve the robustness of the generator. This process can involve adding supplementary information to the retrieved graph, sourced from external data or knowledge embedded within LLMs.  There are two main categories of methods:
\begin{itemize}[leftmargin=*]
    \item \textbf{Graph Structure Augmentation:} Graph structure augmentation methods involve adding new nodes and edges to the retrieved graph. For instance, GraphQA~\cite{taunk2023grapeqa} augments the retrieved subgraph by incorporating noun phrase chunk nodes extracted from the context. Moreover, \citet{yasunaga2021qa} and \citet{taunk2023grapeqa} treat the query as a node, integrating it into the retrieved graph to create direct connections between the query and relevant information. \citet{tang2024cross} augment the graph structure based on pretrained diffusion models. 
    \item \textbf{Graph Feature Augmentation:} Graph feature augmentation methods focus on enriching the features of the nodes and edges in the graph. Since the original features might be lengthy or sparse, data augmenters can be employed to summarize or provide additional details for these features. For example, Once~\cite{liu2024once} uses LLMs as Content Summarizers, User Profilers, and Personalized Content Generators in recommendation systems. Similarly, LLM-Rec~\cite{lyu2023llm} and KAR~\cite{xi2024towards} apply various prompting techniques to enrich node features, making them more informative for downstream tasks.
\end{itemize}

Additionally, some graph augmentation techniques focus solely on the retrieved graph itself, such as randomly dropping nodes, edges, or features to improve model robustness. \citet{ding2022data} provide a systematic review of these data augmentation methods.

\subsubsection{Verbalizing}
\label{sec:verbalizing}
Verbalizing refers to converting retrieved triples, paths or graphs into natural language that can be consumed by LLMs. There are two main approaches to verbalization: linear verbalization and model-based verbalization.

Linear verbalization methods typically convert graphs into text using predefined rules. The primary techniques for linear verbalization include:

\begin{itemize}[leftmargin=*]
    \item \textbf{Tuple-based:} These methods place the different pieces of retrieved information and order them in a tuple~\cite{baek2023knowledge, oguz2020unik}. For example, when performing retrieval on a KG, many methods retrieve a set of facts. A single fact is verbalized in the generation prompt as the tuple $(\text{entity} \: 1, \text{relation} \: 1, \text{entity} \: 2)$~\cite{niu2024mitigating,  tian2024graph}. For a set of facts, we first sort them in a specific order, and then verbalize them one at a time as an individual tuple. Each piece of information is typically separated by line in the prompt. Note that the same logic can be applied to paths, nodes, and so on.

    \item \textbf{Template-based:} These methods verbalize paths or graphs using predefined templates to generate more natural text. For example, LLaGA~\cite{chen2024llaga} proposes some templates such as Hop-Field Overview Template to convert graph into sequence. For KGs, several methods~\cite{knowledgenavigator, unioqa} convert individual facts into natural text. For example, \citet{knowledgenavigator} convert a fact (entity 1, relation, entity 2) to text using the template {\it ``The \{relation\} of \{entity 1\} is/are: \{entity 2\}''}.
\end{itemize}

Model-based verbalization methods typically use fine-tuned models or LLMs to convert input facts into coherent and natural language. These methods generally fall into two categories:

\begin{itemize}[leftmargin=*]
    \item \textbf{Graph-to-text verbalization}: These methods focus on converting retrieved graphs into natural language while preserving all the information. For instance, \citet{koncel2019text} and \citet{wang2020amr} leverage graph transformers to generate text from knowledge graph. \citet{ribeiro2020investigating} evaluate several pretrained language models for graph-to-text generation, while \citet{wu2023retrieve}, and \citet{agarwal2020knowledge} fine-tune LLMs to transform graphs into sentences, ensuring a faithful representation of the graph content in textual form. 
    
    \item \textbf{Graph Summarization:}  In contrast to Graph-to-Text Verbalization, which retains all details, Graph Summarization methods aim to generate concise summaries based on the retrieved graph and the query. EFSum~\cite{ko2024evidence} proposes two approaches: one directly prompts LLMs to summarize the retrieved facts and query, while the other fine-tunes LLMs specifically for summarization tasks. CoTKR~\cite{wu2024cotkr}, on the other hand, alternates between two operations: Reasoning, where it decomposes the question, generates a reasoning trace, and identifies the specific knowledge needed for the current step; and Summarization, where it summarizes the relevant knowledge from the subgraph that retrieved based on the current reasoning trace.
\end{itemize}

\subsection{Generator}
\label{sec:generator}

The generator aims to produce the desired output for specific tasks based on the query and the retrieved information. These tasks can range from discrimination tasks (e.g., node/edge/graph classification) to generation tasks (e.g., KG-based question answering) and graph generation (e.g., molecular generation). Due to the uniqueness of different tasks, different generators are often desired. We categorize generators into three main types:  Discriminative-based Generators, which leverage models like GNNs and Graph Transformers for tasks like classification; LLM-based Generators, which utilize the capabilities of LLMs to generate answers for text-based tasks; and Graph-based Generators, which generate new graphs using generative models such as diffusion models. Next, we provide a detailed illustration of these generators.

\subsubsection{Discrimination-based Generator}
Discrimination-based generators focus on discriminative and regression tasks, which can typically be modeled as graph tasks, such as node, edge, or graph classification and regression. Models designed for graph data, such as GNNs and Graph Transformers, are widely used as discrimination-based generators. The choice of GNN depends on the graph type and task.  For instance, GCN~\cite{kipf2016semi}, GraphSAGE~\cite{hamilton2017inductive}, and GAT~\cite{velivckovic2017graph} are typically applied to homogeneous graphs, whereas models like RGCN~\cite{schlichtkrull2018modeling} and HAN~\cite{wang2019heterogeneous} are used for heterogeneous graphs, and HGNN~\cite{feng2019hypergraph} and Hyper-Attention~\cite{bai2021hypergraph} are suitable for hypergraphs. Additionally, graph transformers~\cite{min2022transformer, shehzad2024graph} have gained popularity for their ability to capture global dependencies.
Additionally, different training strategies, such as (semi-)supervised learning~\cite{ma2021deep} and graph contrastive learning~\cite{ju2024towards, liu2022graph}, are employed depending on the specific requirements of the task.

\subsubsection{LLM-based Generator}
LLMs have demonstrated remarkable capabilities in understanding and generating natural language across a wide range of tasks. However, LLMs are inherently designed to process sequential data, while the retrieved information in GraphRAG is typically structured as graphs. Although various GraphRAG organizers, such as verbalization methods, convert retrieved graph information into text, these transformations may result in the loss of important graph structure information, which could be crucial for certain tasks. To take advantage of the ability of LLMs, many research efforts have been proposed to feed the graph information into LLMs, and we summarize them into the following categories:

\begin{itemize}[leftmargin=*]
    \item \textbf{Verbalizing:} Verbalizing aims to convert the retrieved information in GraphRAG into sequences that can be processed by LLMs. These methods are detailed in Section~\ref{sec:verbalizing}. 

    \item \textbf{Embedding-fusion:} Embedding-fusion integrates graph embeddings and text embeddings within LLMs. The graph embeddings can be obtained using GNNs or Graph Transformers\cite{chai2023graphllm}. To align graph embeddings with text embeddings, a domain projector is typically learned to map graph embeddings to the text embedding space. Embedding fusion can occur at different layers of LLMs. For example, \citet{he2024g} feed the projected graph embeddings through the self-attention layers of LLMs, while \citet{tian2024graph} prepend the projected graph embeddings with the text tokens. \cite{argatu2024joint} fuse the text and projected graph embeddings before the prediction layers of LLMs.
    Additionally, LLMs can either be fine-tuned along with the domain projector using methods such as LoRA, or the LLM can remain fixed, training only the graph embedding model and domain projector.

    \item \textbf{Positional embedding-fusion:} Directly converting the graph into sequences by Verbalization may lose graph structure information, which can be crucial in some tasks. Positional embedding-fusion aims to add the position of nodes in the retrieved graph to the LLMs. GIMLET\cite{zhao2023gimlet}, as a unified graph-text model, employs a generalized position embedding to encode both graph structures and textual instructions as unified tokens. LINKGPT~\cite{he2024linkgpt} leverages the pairwise encoding in LPFormer~\cite{shomer2024lpformer} to encode the pairwise information between two nodes.
\end{itemize}

\subsubsection{Graph-based Generator}

In the scientific graph domain, GraphRAG generators often go beyond LLM-based methods due to the need for accurate structure generation. RetMol~\cite{wangretrieval} is particularly versatile because it can work with various encoder and decoder architectures, supporting multiple generative models and molecule representations. For example, generators can be transformer-based or utilize Graph VAE architectures. \citet{huanginteraction} highlight the use of a diffusion model, specifically the 3D molecular diffusion model IRDIF. In the generation process, SE(3)-equivariance is achieved through architectures like Equivariant Graph Neural Networks (EGNNs)~\cite{satorras2021n}, which ensure that the geometric properties of molecular structures remain invariant to spatial transformations such as rotation, translation, and reflection at each step. Incorporating SE(3)-equivariance into the diffusion model guarantees that the generated molecular structures maintain geometric consistency under these transformations. For KGs, multiple works~\cite{feng2020scalable, yasunaga2021qa, taunk2023grapeqa} use a GNN to generate the answer. The GNNs used in these works are conditional on the query, thereby making the final predictions relevant to it.

\subsection{Graph Datasources}
We have conducted a comprehensive review of the primary techniques applied in the initial four model-centric components of GraphRAG—namely, the query processor, retriever, organizer, and generator. However, even with the best configurations of these components, a GraphRAG system may still fall short of optimal performance if the underlying graph data sources, from which external knowledge is retrieved, are not meticulously curated. This also underscores the recent significant shift in AI research from a model-centric to a data-centric perspective, where enhancing data quality and relevance becomes equally, if not more, crucial for achieving superior results. Adopting this data-centric perspective, the following section provides an overview of existing GraphRAG research on constructing graph data sources from a high-level perspective, with a detailed discussion of domain-specific graph construction methods reserved for the subsequent domain-specific section.

\begin{itemize}[leftmargin=*]
    \item \textbf{Explicit Construction:} Explicit construction refers to building graphs based on explicit and predefined relationships in the data. This method is widely adopted across various domains. For example, molecule graphs are constructed from the connections between atoms; knowledge graphs are formed based on explicit relationships between entities; citation graphs are built by linking papers through citation relationships; and recommendation graphs model interactions between users and items.

    \item \textbf{Implicit Construction:} Implicit construction is used when there are no explicit relationships between nodes, but instead, implicit connections can be derived. For instance, word co-occurrence in a document can suggest shared semantic information, and feature interaction in Tabular data can indicate the correlation between features. Graphs can explicitly model these connections, which might be beneficial to the downstream tasks.
\end{itemize}

After the graph is constructed, there are also several ways to formally represent graphs. 
\begin{itemize}[leftmargin=*]
    \item \textbf{Adjacency matrix:} The adjacency matrix is one of the most popular ways to denote a graph. Specifically, the adjacency matrix $\mathbf{A}\in\mathbb{R}^{||\mathcal{V}\times |\mathcal{V}|}$ denotes the graph connections among nodes in $\mathcal{V}$, where $|\mathcal{V}|$ is the number of nodes.

    \item \textbf{Edge list:} The edge list represents each edge in the graph, typically in the form of tuples or triples, such as $(i, j)$ or $(i, r, j)$, where $i$ and $j$ are nodes, and $r$ is the relation between nodes $i$ and $j$.

    \item \textbf{Adjacency list:} The adjacency list is a node-centric representation where each node is associated with a list of its neighbors. It is typically represented as a dictionary $\{i: \mathcal{N}_i\}$, where $\mathcal{N}_i$ is the neighbor list of node $i$.
    
    \item \textbf{Node Sequence:} A node sequence transforms a graph into a sequence of nodes in either an irreversible or reversible manner. Most serialization methods are irreversible and do not allow for complete recovery of the original graph structure. For example, there are also some serialization methods that are reversible which can recover the whole graph structure. For example, \citet{zhao2023graphgpt} propose serializing graphs using Eulerian paths by first applying Eulerization to the graph. Besides, if the graph establishes a tree structure, the BFS/DFS can also serialize the graph in a reversible manner. 
    
    \item \textbf{Nature language:}  With the growing popularity of LLMs for processing text-based information, various methods have been developed to describe graphs using natural language.
\end{itemize}

Note that the above-mentioned data structures can only represent basic graphs without support for complex scenarios such as multi-relational edges or edge attributes. For instance, using an adjacency matrix to represent a multi-relational attributed graph requires an expanded structure: $\mathbf{A} \in \mathbb{R}^{|\mathcal{V}| \times |\mathcal{V}| \times |\mathcal{R}|}$, where $\mathcal{R}$ denotes the set of possible relationships. Here, $\mathbf{A}_{i,j,r}$ represents the weight of the edge connecting node $i$ and node $j$ under relation $r$. 

Selecting an appropriate graph representation is essential for task-specific requirements. For example, \citet{ge2024sequential} finds that the order of graph descriptions significantly impacts LLMs' comprehension of graph structures and their performance across different tasks.

\section{Knowledge Graph \texorpdfstring{\emojikgtitle}{Knowledge Graph}}
\label{sec-kg}
A knowledge graph is a structured database that connects entities through well-defined relationships. It can either encompass a broad spectrum of general knowledge, such as the widely recognized Google Knowledge Graph~\cite{chen2020review, liu2020k}, or delve deeply into specialized domains, like the BioASQ dataset~\cite{tsatsaronis2015overview} for biomedical reasoning. The diverse information contained in a knowledge graph -- represented as entities, relationships, paths, and subgraphs -- serves as a valuable resource for enhancing various downstream tasks across different sectors, including question-answering~\cite{tian2024graph, wang2024knowledge, yasunaga2022deep}, commonsense reasoning~\cite{ilievski2021cskg}, fact-checking~\cite{kim2023factkg}, recommender systems~\cite{guo2020survey}, drug discovery~\cite{bonner2022review}, healthcare~\cite{chandak2023building}, and fraud detection~\cite{mao2022financial}.

\subsection{Application Tasks}
This section reviews representative applications that GraphRAG on KGs is used for.

\begin{itemize}[leftmargin=*]
    \item \textbf{Question-answering}: Question-answering (QA) can focus on a single domain or span across global knowledge. Typically, a query in text format is given, such as "What is the best way to predict a baby's eye color?" or "Were there fossil fuels in the ground when humans evolved?"~\cite{tian2024graph} -- the answer can be a sentence generated by a large language model (LLM), a selected text span from relevant documents, or even a specific choice in a multiple-choice QA scenario. In all these contexts, GraphRAG leverages knowledge graphs to retrieve relevant information, providing the necessary context or supporting facts to generate accurate answers.
    
    \item \textbf{Fact-Checking}: Fact-checking is to verify the truthfulness of statements by cross-referencing them with reliable sources of information. GraphRAG enhances this task by querying a knowledge graph to retrieve relevant facts and relational structures that either support or refute the given claim. GraphRAG identifies discrepancies or confirmations within the data by mapping the statement onto the knowledge graph, providing a thorough and evidence-based validation process.

    \item \textbf{Knowledge Graph Completion}: Knowledge graph completion is the task of predicting new facts to enhance the comprehensiveness of the graph and infer missing facts~\cite{zhang2023making}. GraphRAG addresses this task by retrieving structural knowledge around the triplets for inference, supplying essential structural knowledge, and enhancing the LLM inference.

    \item \textbf{Cybersecurity Analysis and Defense}: Cybersecurity Analysis and Defense aims to analyze and respond to vulnerabilities, weaknesses, attack patterns, and threat tactics. With the increasing complexity and volume of cybersecurity data, GraphRAG has been proposed to provide cybersecurity analysis with more comprehensive insights into potential attack vectors and mitigation strategies~\cite{rahman2024retrieval}.

\end{itemize}

\subsection{Knowledge Graph Construction}

We discuss how KGs are typically constructed. For each type of construction technique, we give examples of common KG databases. How a KG is constructed is important, as it can affect both its usefulness and function in different downstream tasks. We describe the main techniques below:
\begin{itemize}[leftmargin=*]
    \item {\bf Manual construction}: Some KGs are constructed manually via human annotation. WikiData~\cite{vrandevcic2014wikidata} is a KG that uses crowd-sourced efforts to gather a variety of knowledge. Each entity corresponds to a page in the Wikipedia encyclopedia. Another KG, the Unified Medical Language System (UMLS)~\cite{umls}, contains biomedical facts collected from numerous sources.

    \item {\bf Rule-based construction}: Many traditional approaches use rule-based techniques for constructing the graph. This takes the form of custom parsers and manually defined rules used to extract facts from raw text. Note that these parsers can differ depending on the source of the text. Prominent examples include ConceptNet~\cite{speer2017conceptnet}, which links different words together via assertions, and Freebase~\cite{bollacker2008freebase} which contains a wide variety of general facts. REANO~\cite{fang2024reano} extracts the entities and relations from a set of passages using traditional entity recognition (ER) and relation extraction (RE) methods, respectively. This includes SpaCy~\cite{honnibal2017spacy} for ER and TAGME~\cite{ferragina2010tagme} for RE. To extract the facts that connect two entities via a relation, they use DocuNet~\cite{zhang2021document}.

    \item {\bf LLM-based construction}: Recently, work has explored how LLMs can be used to construct KGs from a set of documents. In such a way, the LLM can automatically extract the entities and relations and link those together to form facts in the given text. Of note is that no ground-truth KG exists for these methods. Rather, they simply use a KG as a way to organize and represent a set of documents. For example, CuriousLLM~\cite{yang2024curiousllm} considers passages in the text as entities and determines whether two entities should be connected based on their encoded textual similarity. On the other hand, \citet{cheng2024structure} uses a manually-defined prompt to convert a piece of text into a KG. Graph-RAG~\cite{edge2024local} first divides each document into chunks and then uses an LLM to detect all the entities in each chunk, including their name, type, and description. To identify the relation between any two entities, both entities and a description of their relationship are passed to an LLM. An LLM is then used again to summarize the content of each entity and relation to arrive at their final title. Lastly, AutoKG~\cite{chen2023autokg} uses a combination of LLM embeddings and clustering techniques to construct a KG from a set of texts.
\end{itemize}

\subsection{Retriever}

Real-world facts in KGs can provide grounded information for generative models, enhancing the reliability of the model output. Given the structured nature of KGs, they are naturally well-suited for retrieval. The goal is for a given question or query to retrieve either relevant facts~\footnote{Throughout this paper, we will refer to facts as triples or edges, interchangeably} or entities that can help answer that question. Multiple considerations need to be considered during retrieval, including the type of facts we want to retrieve, the efficiency, and the amount of facts retrieved. In general, retrieval of KGs has two stages: identifying seed entities and retrieving facts or entities. We describe both below.

{\bf Identifying seed entities}: The first step in retrieving the relevant facts for a given query is to identify a set of ``seed entities'', which we'll refer to as $V_{\text{seed}}$. Seed entities are the initial entities that are chosen to be highly relevant to the original query. Given such, we expect that triples that contain any of these entities or are nearby in the graph should provide helpful context. Multiple techniques exist for identifying the seed entities. Some works~\cite{jiang2023structgpt, kim2023kg, gnn_llm, pullnet, zhang2022subgraph} assume that we are given a set of initial entities for each query. However, most works~\cite{feng2020scalable, sunthink, wen2023mindmap, zhanggreaselm, niu2024mitigating, yasunaga2022deep} attempt to extract the entities from the query. One approach is through entity extraction~\cite{al2020named}, which uses methods specifically designed for extracting entities from a given text. Most works only extract entities from the original query. Another common approach is to extract a set of entities that are semantically similar to the original query~\cite{wen2023mindmap, sanmartin2024kg}. HyKGE~\cite{jiang2024hykge} first generates a hypothesis and extracts entities from the original query and the hypothesis. Similarly, in order to reduce the possibility of hallucination, \citet{knowledgenavigator} uses an LLM to generate two similar questions and retrieves all entities found in the original and generated questions. In a similar vein, RoK~\cite{wang2024reasoning} first uses chain of thoughts reasoning to expand the original query, extracting the seed entities from the expanded query.

{\bf Retrieval Methods}: The outcome of the previous step provides us with a set of entities that are related in some capacity to the query. These entities are then leveraged to retrieve a set of facts or entities that can aid us in answering the query. We summarize the core retrieval methods below.
\begin{itemize}[leftmargin=*]
    \item {\bf Traversal-based retriever}: These methods traverse the graph and extract paths to aid in answering a specific question. Given the set of seed entities, $V_{\text{seed}}$, \citet{yasunaga2021qa, yasunaga2022deep, zhanggreaselm} extract all paths up to length two between the entities in $V_{\text{seed}}$, resulting in a final entity set $V$. They further augment $V$ by including all triples that connect any two entities in $V$. Given $V$, both \cite{yasunaga2021qa, zhanggreaselm} only keep the top $k$ entities by relevance scores. This is calculated by training a separate model that takes the text embedding of the query and entity as input and outputs how relevant the entity is to the query. For \citet{yasunaga2022deep}, if $\lvert V \rvert > 200$, they randomly sample 200 entities. \citet{sunthink} use a version of beam-search to explore the KG. \citet{jiang2024hykge, feng2020scalable} extract all paths of length $k \leq 2$ between seed entities. Alternatively, LARK~\cite{choudhary2023complex} retrieves all facts that lie on paths $\leq k$ in length starting from the seed entities.     \citet{delile2024graph} first extract the shortest paths connecting all seed entities. They further prioritize some entities over others by considering the recency, importance, and relevance to the query of their associated text. OREOLM~\cite{hu2022empowering} traverse $k$ hops from the seed entities, contextualizing the importance of each relation and entity to a path via a learnable $d$-dimensional embedding and it's LM-encoded representation. \citet{zhang2022subgraph} introduce a trainable retriever that traverses the graph starting from each seed entity. They also train a model to score each newly visited edge, only keeping a portion of them. KG-RAG~\cite{sanmartin2024kg} works in a similar manner, scoring each edge by its relevance and similarity to the query via a dense retriever. They then use an LLM to decide which paths to explore in the next step.
    RoG~\cite{luoreasoning} uses instruction tuning to fine-tune an LLM to generate useful relation paths, which can be retrieved from the KG. 
    KnowledgeNavigator~\cite{knowledgenavigator} first uses the query to predict the expected number of hops, $h_Q$ needed to traverse in the retrieval stage. It then traverses $h_Q$ hops starting from the seed entities, using an LLM to score and prune irrelevant nodes. \citet{retrieve_rewrite_answer} operate in a similar manner; however, they choose which paths to traverse based solely on the relations. Furthermore, when scoring a path, all relations that lie on that path are considered when computing the score. RoK~\cite{wang2024reasoning} considers a different approach, using the Personalized PageRank (PPR) score to identify useful paths. They further augment these paths by including the 1-hop neighbors of the seed entities. PullNet~\cite{pullnet} assumes that each entity has an associated set of documents. Given a single seed entity, PullNet, traverses k hops, where in each iteration it extracts the facts for the newly observed entities. It also extracts any entities that are contained in documents associated with an entity found in the traversal. Furthermore, for each entity, only the top N facts are used, which are ranked via similarity to the query. KG-R3~\cite{pahuja2023retrieve} uses MINERVA~\cite{minerva}, a reinforcement learning approach to mining paths between entities, to retrieve a set of important paths between both entities in the fact. 
    \citet{wang2024knowledge} use an LLM to traverse the graph starting from the seed entities. At each iteration in the traversal, we choose the next node to visit by prompting an LLM. Specifically, given the information already collected in the traversal, the LLM is prompted to generate the remaining information needed to correctly answer the question. The neighboring node that best matches the required information is chosen as the next node to visit. They further instruction-tune the LLM.

    \item {\bf Subgraph-based retriever}: These methods extract a subgraph of size $k$ around each of the seed entities. Facts that contain one of the seed entities or are nearby, should be highly relevant to answering the question. Furthermore, they may actually contain the answer itself. Each of \cite{niu2024mitigating, tian2024graph, jiang2023structgpt, kim2024causal} extract either the one or two hop subgraph around each seed entity. The final set of facts is the union of each individual subgraph. \citet{gao2022graph} propose to first extract the subgraph containing the seed entity and potential answers using the method in \citet{sun2018open}. This is then partitioned into a set of smaller subgraphs. Then, they design a framework to rank the subgraphs, keeping the top $k$ subgraphs for generation. For a question-choice pair, MVP-Tuning~\cite{mvp_tuning} considers the triple that contains the highest number of seed and choice entities. They further augment this by extracting the top $k$ most similar questions in the dataset using BM25~\cite{bm25}, and extract the triples for each of them.

    \item {\bf Rule-based retriever}: These methods use pre-defined rules or templates to extract paths from the graph. GenTKG~\cite{liao2024gentkg} considers a temporal KG, where they first extract logical rules from the KG, and use the top k rules to extract paths in a given time interval for the seed entities. Both \cite{dai2024counter, luo2023chatkbqa} generate queries using SPARQL, which are then used to retrieve import paths. KEQING~\cite{keqing} decomposes the original query into $k$ sub-queries using using an LLM fine-tuned via LoRA~\cite{hulora}. For each sub-query, they find the most similar question templates, which are predefined. For each template, they further pre-define a set of logical chains, which are then used to extract matching paths for the seed entities in the sub-query from the KG. 
    
    \item {\bf GNN-based retriever}: GNN-RAG~\cite{mavromatis2024gnn} trains a GNN for the retrieval task. A separate round of message passing is done for each query $q$, which is incorporated in the message computation along with the relation and entity representations. The GNN is then trained as in the node classification task, where the correct answer entity for $q$ has a label of 1 and 0 otherwise. During inference, the entities with probability above some threshold are treated as candidate answers, and the shortest path from the seed entity is extracted. \citet{gnn_llm} use a conditional GNN~\cite{conditional_mpnn} for retrieval, where for each query, only the seed entity (they assume there is only one) is initialized to a non-zero representation based on the LLM-encoded query. They then run $L$ rounds of message passing where after each layer $l$ only the top-K new edges are kept, resulting in a set of entities $C_q^l$. This is determined by a learnable attention weight, which prunes the other edges from the graph. The final set of candidate entities is the union of candidate entities at each layer $l$, $C_q = \cup_{l=1}^L C_q^l$. It is optimized in a similar manner to~\cite{mavromatis2024gnn}. For each candidate entity, they retrieve an evidence chain by backtracking from the entity until it reaches the seed entity, choosing those edges with the highest attention weight.
    REANO~\cite{fang2024reano} initializes the entity and relation representations via the mean-pooled representations of all mentions of that entity/relation in the texts, encoded by T5. They then run a GNN, which includes an attention weight that considers the relevance of a given triple to the original question (also encoded by T5). After running the GNN, they retrieve the top K triples in the KG that are most relevant to the question, where relevance is defined via the dot product between the triple encoded by the GNN and the question.  

    \item {\bf Similarity-based retriever}: STaRK~\cite{wu2024stark} considers the vector similarity of the query to each entity. Each entity embeds both the textual and relational information together. They further consider multi-vector similarity, where the entities are encoded using multiple vectors. This is done by chunking the textual and relational information of each entity, with each chunk being embedded into its own vector. Both REALM~\cite{zhu2024realm} and EMERGE~\cite{zhu2024emerge} extract the entities most similar to the query. While REALM only retrieves the entities themselves, EMERGE further retrieves the 1-hop subgraph around each entity.

    \item {\bf Relation-based retriever}: \citet{kim2023kg} propose a general framework for reasoning on KGs using LLMs. They first use an LLM to segment the original query into a set of $i \in I$ sub-sentences, where each sub-sentence $S_i$ has an associated set of entities $\mathcal{E}_i$. For each sub-sentence, they further use an LLM to retrieve the top-$k$ most relevant relations $\mathcal{R}_{i, k}$. Given the set of $k$ relations, for each sub-sentence, they retrieve all triples that contain a relation in $\mathcal{R}_{i, k}$ and whose entities are in $\bigcup_{i \in I} S_i$. GenTKGQA~\cite{gao2024two} focuses on temporal KG QA. Like~\cite{kim2023kg}, they retrieve the top-k relations for the query. They then retrieve all facts that contain one of the top k relations and satisfy the temporal constraints. 

    \item {\bf Fusion-based retriever}: These techniques consider a combination of different retrieval techniques. Mindmap~\cite{wen2023mindmap} considers extracting some paths $\leq k$ hops from the seed entity and the 1-hop subgraph of each seed. These two extracted components are combined into one subgraph. DALK~\cite{dalk}
    uses a procedure similar to Mindmap, where they extract both paths and the 1-hop subgraph around each seed entity. However, they argue that this procedure often results in the retrieval of redundant or unnecessary information. To remove these facts, they use an LLM to rank the retrieved facts given both the original question and the subgraph. Only the Top-k most relevant facts are kept.
    UniOQA~\cite{unioqa} considers two branches for retrieving. The first is a translator, which is a fine-tuned LLM that generates the answer in a CQL format. the second is a searcher that retrieves the 1-hop subgraph around the seed entities. When determining the answer, answers from the translator are prioritized over those from the searcher. KG-Rank~\cite{kgrank} considers ranking all triples in the 1-hop neighborhood of the seed entities via the similarity of the relation to the query, the similarity of each triple to the encoded output of $a = \text{LLM}(q)$, and an MMR ranking~\cite{carbonell1998use} that uses the similarity score. Only the top-ranked triples are kept. GrapeQA~\cite{taunk2023grapeqa} extends~\cite{yasunaga2021qa} by further including a set of ``extra nodes'', which are the common neighbors of the entities retrieved via a path-based retriever. They further introduce a clustering-based method for pruning entities that may be irrelevant to the query. SubgraphRAG~\cite{li2024simple} considers both GNN and textual information. For the GNN, they consider initializing the node representations using a one-hot encoding to differentiate between seed entities and others. A GNN is then run for L layers, resulting in the final representation $s_v$ for a node $v$. To retrieve the relevant triples, they consider first concatenating the final node representations for each triple $(h, r, t)$ such that $z_{\tau} = [s_h, s_t]$. The probability of choosing this triple is then given by  $p(h, r, t) = \text{MLP}([z_q, z_h, z_r, z_t, z_{\tau}])$, where $z_q$, $z_h$, $z_r$, $z_t$ are the encoded textual representations of the query and the triple $(h, r, t)$, respectively. Only the top K triples are chosen.
    
    \item {\bf Agent-based retriever}: These techniques use LLM agents to retrieve facts from the KG. KnowledGPT~\cite{wang2023knowledgpt} defines a set of tools for searching over a KG. Given a query, they generate a piece of code to search over the KG that considers the seed entities. The code is then executed over the KG to find the correct answer. KG-Agent~\cite{kgagent} focuses on fine-tuning an LLM to generate the SQL code for retrieving the correct answer. Using a set of tools, they extract a set of paths that contain the seed entities. KnowAgent~\cite{zhu2024knowagent} first identifies the relevant actions for the query via a planning module. Using these actions, they then generate a set of paths that are used for generation. 

    {\bf Other retrievers}: KICGPT~\cite{kicgpt} is concerned with the task of knowledge graph completion, where given a partial fact $(h, r, *)$, we want to predict the correct entity $\hat{e}$. KICGPT retrieves the entities by first scoring all possible entities using a traditional KG embedding score function. That is, for a score function $f(\cdot)$ and a partial fact $(h, r, *)$, they compute the set of scores $\{f(h, r, e) \; \forall e \in \mathcal{V} \}$. They use RotatE~\cite{sunrotate} for the function  $f(\cdot)$, a popular approach. Only the top $k$ entities by score are retrieved. To supplement their knowledge, they also retrieve all triples with (a) the same relation as the query and (b) all triples that contain the entity $h$ in the query. These are referred to as the analogous and supplement triple pools, respectively.

\end{itemize}

\subsection{Organizer}

In this subsection, we describe how the retrieved knowledge is organized for generation. More concretely, this is how the information is formatted when given to the generator. Note that not every method necessarily has an explicit organizer. We summarize the common methods below:
\begin{itemize}[leftmargin=*]
    \item {\bf Tuple-based organizer}: These methods consider each piece of retrieved information as an ordered triple. For example, it would include a triple in the generation prompt as ``$(\text{entity} \: 1, \text{relation} \: 1, \text{entity} \: 2)$''. Similarly, a path of length $m$ is given by ``$(\text{entity} \: 1, \text{relation} \: 1, \text{entity} \: 2, \text{relation} \: 2, \cdots, \text{entity} \: m)$''. The entities and relations are usually represented either as their names or IDs. Each triple or path is usually listed on a separate line. Many works append the retrieved paths to the original query as additional context \cite{sunthink, jiang2024hykge, mavromatis2024gnn, sanmartin2024kg, luoreasoning, choudhary2023complex, gnn_llm, wang2024reasoning, zhu2024knowagent}. Other works that retrieve facts instead of paths operate in a similar manner, where instead they append the triples~\cite{niu2024mitigating, zhu2024realm, dai2024counter, liao2024gentkg, kgrank, jiang2023structgpt, kim2023kg, gao2024two}. Some methods~\cite{zhu2024emerge, tian2024graph} consider only including the retrieved entities as the context. Given a set of facts, KG-R3~\cite{pahuja2023retrieve} first lists all entities and then relations, i.e., ``$(\text{entity} \: 1, \text{entity} \: 2, \cdots , \text{entity} \: m, \cdots, \text{relation} \: 1, \cdots, \text{relation} \: m-1)$''. 
    \citet{delile2024graph} consider a KG where each entity has an associated chunk of text. Each text chunk for an entity is considered as a different piece of information to be included in the context. Both \cite{tian2024graph, gao2024two} represent each entity and relation as an embedding, which is the combination of the LLM and GNN embedding. \citet{gnn_llm} further include the probability of each path containing the correct answer given by the GNN model. MVP-Tuning~\cite{mvp_tuning} considers combining multiple facts that share the same subject and relation to remove redundant information. That is, for a subject-relation pair $(\text{subject}, \text{relation})$, they denote the facts for $k$ possible objects as ``$\text{subject} \: \text{relation} \: \{\text{object} \: 1,  \cdots, \text{object} \: k\}$''. KG-Agent~\cite{kgagent} stores the current KG information and the historical reasoning programs in lists.

    \item {\bf Text organizer}: \citet{retrieve_rewrite_answer} verbalize the retrieved subgraph by passing each triple to the LLM and prompting it to convert it to a text representation. MindMap~\citet{wen2023mindmap} uses a similar procedure for subgraphs, where each is organized as a path before being passed to the LLM. Some methods use a set of pre-defined templates to verbalize the triples or paths \cite{knowledgenavigator, unioqa, kim2024causal}. \citet{keqing} experiment with verbalizing either via an LLM or pre-defined question templates, finding that LLM-based verbalizing works better for ChatGPT while template-based works better for LLaMA~\cite{touvron2023llama}. KICGPT~\cite{kicgpt} uses a combination of data preprocessing and LLM prompting to convert the triples to text. StaRK~\cite{wu2024stark} uses an LLM to synthesize each entity with its relational and textual information. Note that they use some pre-defined templates that depend on the specific task. CoTKR~\cite{wu2024cotkr} uses an LLM to summarize and then re-write a subgraph of facts for a question through a ``knowledge rewriter''. To train the rewriter, preference alignment is used, which optimizes the rewriter's output to match our preferred output. First, $k$ representations of the retrieved subgraph are produced, with ChatGPT choosing the best and work representations as the most and least preferred solutions.  
    
    \item {\bf Other organizer}: There are some exceptions to the previous classification. KnowledGPT~\cite{wang2023knowledgpt} represents the information in the form of a python class format. They also experiment with including additional information like the entity description and entity-aspect information.  

    \item {\bf Re-Ranking}: Some methods also {\it re-rank} the information in a specific order. This is done as the order of information can have a subtle impact on LLM performance. \citet{delile2024graph} order the text chunks of each entity based on the impact (measured by \# of citations of the parent paper) and the recency. \citet{dai2024counter} sort the triples by the relevancy score to the triple. \citet{choudhary2023complex} attempt to order the paths in a logical matter, such that for a given path, the subsequent paths build upon it. \citet{kgrank} re-rank the retrieved triples using a task-specific Cross-Encoder~\cite{jin2023medcpt}. STaRK~\cite{wu2024stark} considers re-ranking the retrieved entities using an LLM. The LLM is given the relational and textual information of each, and is asked to give it a score from 0 to 1, which is then used for re-ranking. GenTKG~\cite{liao2024gentkg} orders the paths by the time they occurred, further including the time with each. KICGPT~\cite{kicgpt} ranks all entities using the score of the KG embedding score function, keeping only the top $k$ entities. KICGPT re-rank the entities using in-context learning, where they prompt the LLM with examples from the analogy and supplement pool, as prior knowledge to aid the LLM in how to re-rank the entities. 
    
\end{itemize}

\subsection{Generator}

In this section we describe how the retrieved and organized data is used to generate a final response to the query. We categorize these generators according to the type of methods used to create these responses.

\begin{itemize}[leftmargin=*]
\item {\bf LLM-based generator}: The vast majority of works use a LLM to generate the response. The input to the LLM is the original query and retrieved and organized context, formatted using a specific template. The most commonly used LLMs include ChatGPT~\cite{chatgpt_paper}, Gemini~\cite{team2023gemini}, Mistral~\cite{jiang2023mistral}, Gemma~\cite{team2024gemma}, among others. For open-source models where the weights are publicly available, fine-tuning is sometimes used to modify the weights for a specific task \cite{unioqa, zhu2024knowagent}. This is often done through LoRA~\cite{hulora}, which allows for efficient fine-tuning.

\item {\bf GNN-based generator}: Some methods use graph neural networks (GNNs)~\cite{kipfgcn} to conduct the generation. \citet{yasunaga2021qa, taunk2023grapeqa, feng2020scalable} extract both the language and GNN embeddings for each potential answer (i.e., entity) conditional on the query. The probability of a single entity being the answer is then learnt based on the fusion of the two types of embeddings.

\item {\bf Other generators}: ~\citet{zhanggreaselm, yasunaga2022deep, hu2022empowering} formulate the prediction as a masked language modeling (MLM) problem. The goal is to predict the correct value (i.e., entity) for the masked token which answers the query. To do so, they fine-tune RoBERTa~\cite{liu2019roberta} language model. KG-R3~\cite{pahuja2023retrieve} scores the potential answer entities by performing cross-attention the representations of the query and each individual entity.
PullNet~\cite{pullnet} uses GraftNet~\cite{sun2018open} to score the different entities. \citet{gao2022graph} first selects the correct subgraph by computing the cosine similarity between the query and subgraph representations. For the subgraph with the highest similarity, it's fed to GraftNet~\cite{sun2018open} to select the most probable entity. REANO~\cite{fang2024reano} passes the encoded triples and their associated text passages to the T5 decoder. The task is framed as a classification problem, where the goal is to assign the highest probability to the triple with the correct answer.
\end{itemize}

\subsection{Resources and Tools} 

In this section, we list common tools and KGs that are used in graph RAG systems. For each, we give a brief description and a link to the project.

\subsubsection{Data Resources}
\begin{itemize}[leftmargin=*]
    \item {\bf Freebase}~\footnote{\href{https://developers.google.com/freebase}{https://developers.google.com/freebase}}~\cite{bollacker2008freebase} is an encyclopedic KG that contains a large variety of general and basic facts.
    \item {\bf ConceptNet}~\footnote{\href{https://conceptnet.io/}{https://conceptnet.io/}}~\cite{speer2017conceptnet} is a semantic graph, where the links in the graph are used to describe the meaning of different words or ideas.
    \item {\bf WikiData}~\footnote{\href{https://www.wikidata.org/wiki/Wikidata:Main\_Page}{https://www.wikidata.org/wiki/Wikidata:Main\_Page}}~\cite{vrandevcic2014wikidata} is a crowdsourced knowledge base that functions as a structured analog to the Wikipedia encyclopedia.
\end{itemize}

\subsubsection{ Tools}
\begin{itemize}[leftmargin=*]
    \item {\bf Graph RAG}~\footnote{\href{https://github.com/microsoft/graphrag}{https://github.com/microsoft/graphrag}}~\cite{edge2024local} is an official open-source implementation of the Graph RAG~\cite{edge2024local} framework. It can further be installed via the {\it graphrag} python package.
    \item {\bf LangChain}~\footnote{\href{https://python.langchain.com/v0.1/docs/use\_cases/graph/constructing/}{https://python.langchain.com/v0.1/docs/use\_cases/graph/constructing/}} is an open-source framework for using LLMs with various components and applications, including RAG, where using RAG on KGs is supportive. 
\end{itemize}

\section{Document Graph \texorpdfstring{\emojidoctitle}{Document Graph}}
\label{sec-doc}

A document graph typically models the connections between different documents or various granularity of documents. It is widely observed in real-world scenarios~\citep{schuhmacher2014knowledge, sonawane2014graph, xu2020document}, such as hyperlinks connecting different websites and citations linking one paper to another. Additionally, a document graph's relationships between sentences and entities can explicitly capture semantic and syntactic contextual information. The structural information in documents can serve as a valuable resource for GraphRAG, aiding LLMs in various tasks. For example, in the retrieval process in RAG, the document containing the answer may have less apparent connections with the question, such as in multi-hop question answering~\citep{mavi2024multi}. However, we can identify these relevant documents through their connections to other documents whose context is strongly aligned with the question's context~\citep{dong2024don}. In this section, we will systematically review the document graph.

\subsection{Application Tasks}
Document graphs are beneficial for a wide range of tasks. In this subsection, we will review the tasks where document graphs can play a significant role. Although document graphs have not yet been fully integrated with LLMs, they still offer great potential for enhancing the capabilities of LLMs in various applications.

\begin{itemize}[leftmargin=*]
\item \textbf{Multi-document Summarization(MDS)}: Multi-document summarization aims to condense the contents of multiple documents into a cohesive summary. Summarizing an entire corpus can involve large volumes of text, often exceeding the context window limitations of LLMs. Document graphs can help compress the corpus by extracting key components and their relationships, proving highly beneficial for MDS~\citep{yasunaga2017graph, wang2020heterogeneous, li2020leveraging, zhang2023contrastive, xie2022gretel, zhang2023enhancing, yang2018integrated, li2023compressed, chen2021sgsum}. These graphs also provide different levels of granularity for summarization through hierarchical clustering~\citep{edge2024local}.

\item \textbf{Text generation}: Text generation focuses on producing coherent and meaningful text. While text-based RAG models have been widely used to generate more reliable text, document graphs can further enhance this process by retrieving similar documents using graph topology. For instance, when writing the abstract of a paper, access to its cited papers in the related work section can significantly improve writing efficiency, as these referenced papers often contain relevant knowledge and context~\citep{wang2024augmenting, pan2020semantic}.
 
\item \textbf{Document Retrieval}: Document retrieval aims to find a list of documents relevant to a given query, which is a key task in Information Retrieval (IR). The exact query terms may not always appear together in the candidate document; however, by leveraging the connections between documents, we can retrieve related documents through those documents that are strongly linked to the query. Thus, it becomes essential to consider document-level relationships in the retrieval process. Several works~\citep{dong2024don, yu2021graph, zhang2018graph, liu2018matching} have leveraged document graphs with various granularities to improve document retrieval and ranking. Rather than retrieving the entire document, graphs can also be used to retrieve specific segments, such as chunks.

\item \textbf{Document Classification:} Document classification is a fundamental task in natural language processing. Traditional methods often focus on the locality of words, limiting their ability to capture long-distance and non-consecutive word interactions. Moreover, these methods typically focus on individual documents, overlooking the relationships between them, where connected documents often exhibit homophily, meaning they are more likely to share similar labels. Building a graph can help enhance document classification by leveraging both local and global relationships between words within a single document~\citep{zhang2020every, liu2022hierarchical} and by utilizing document-level relationships~\citep{zhang2020text, xiao2022graph}.

\item \textbf{Question Answering}: Question answering aims to provide answers to questions based on information from documents and is a fundamental task for RAG. However, documents used for question answering can be long, and traditional methods often focus on local structures within these documents, neglecting their global structure, which is crucial for long-range understanding. Additionally, some multi-hop questions require reasoning across multiple documents, necessitating the use of document-level relationships.  Humans often consolidate scattered information into structured knowledge to streamline the reasoning process and make more accurate judgments, in line with cognitive load theory~\citep{sweller1988cognitive, li2024structrag}. Graph-based methods are well-suited for this task by constructing word-level document graphs~\citep{nie2022capturing, wang2023docgraphlm}, utilizing document-level relationships~\citep{he2024g, wang2024knowledge} and leveraging hierarchical interactions~\citep{fang2019hierarchical}.

\item \textbf{Relation Extraction:} Relation extraction aims to extract semantic relationships between entities in text, which often requires local, global, syntactic, and semantic dependencies, especially in the case of document-level relation extraction. Graphs have been proven to be helpful for document-level relation extraction~\citep{wang2020global, nan2020reasoning, sahu2019inter, christopoulou2019connecting, zhou2020global} by capturing these dependencies more effectively.

In addition to the aforementioned tasks, graphs have also been shown to be helpful in other areas, such as fake news detection~\citep{hu2021compare}, coherence assessment~\citep{liu2023modeling}, and machine translation~\citep{xu2020document}.
\end{itemize}

\subsection{Document Graph Construction}

Different tasks may require different types of document graphs, such as document-level, sentence-level, or word-level graphs. Therefore, the method of obtaining the document graph plays a crucial role. There are primarily two approaches to constructing a document graph: explicit construction and implicit construction.

\begin{itemize}[leftmargin=*]
\item \textbf{Explicit Construction.}  In real-world scenarios, many documents have explicit connections. Examples include web pages linked through hyperlinks, academic papers citing other works, and social media posts connected by reposts, comments, and interactions.  In these cases, it is natural to connect the documents to build a document graph, as the connected documents often share semantic relationships~\citep{peng2024connecting}. For instance, \citet{asai2019learning} constructed a Wikipedia graph based on hyperlinks between Wikipedia articles. \citet{li2022dynamic, li2021graph} follow the same way to leverage hyperlinks to construct graphs. \citet{yu2021kg} built a graph using an external knowledge graph (KG), where each node represents a retrieved passage mapped to entities within the KG, and two passage nodes are connected if their mapped entities are linked in the KG. 
These connections have been leveraged to pretrain large language models (LLMs), enhancing their performance across various tasks~\citep{yasunaga2022linkbert, zou2023pretraining, hu2021relation}.

\item \textbf{Implicit Construction.} Despite the presence of explicit connections between documents, different components within those documents also exhibit important semantic and syntactic relationships. The structure of these components can be highly beneficial for various tasks in natural language processing~\citep{wu2023graph, mihalcea2011graph}. For example, semantic parsing graphs, such as Abstract Meaning Representation (AMR) graphs, can be leveraged to enhance information extraction~\citep{zhang2021abstract, huang2017zero} and text summarization~\citep{dohare2017text}. Additionally, syntactic parsing trees are exploited to understand grammatical structure and resolve ambiguities~\citep{zhang2023survey}.

The implicit construction of document graphs is highly diverse, adapting to the specific requirements of different tasks. For instance, nodes in document graphs can represent various granularity, such as words, entities, sentences, text segments, paragraphs, documents, or topics. In addition, document graphs often exhibit heterogeneity, where edges connect different types of nodes. Next, we will detail the construction of different types of edges:

\begin{itemize}
    \item {\it Word-word edge:} The word-word edges connect words with semantic or syntactic relations or dependencies. There are various methods to establish these connections, including word co-occurrence within a sliding window~\citep{wang2020heterogeneous, yu2021graph, zhang2020text, liu2018matching, de2018question}, dependency parsing graphs~\citep{xu2018graph2seq, tang2020integration, xu2020document, zhang2019aspect, qiu2019dynamically, tu2020select} generated by NLP parsing tools~\citep{lee2011stanford, kumawat2015pos}, Abstract Meaning Representation~\citep{xu2021dynamic, zhang2021abstract, huang2017zero, dohare2017text, xu2021exploiting} and semantic graphs~\citep{dong2024don, song2018graph, pan2020semantic} based on different representations~\citep{banarescu2013abstract, shi2019simple}. Additional techniques include coreference resolution~\citep{xu2020document, sahu2019inter, leskovec2005extracting, dhingra2018neural, song2018exploring}, word embedding similarity~\citep{wang2023docgraphlm, wang2022me} and using large language models (LLMs) to extract relationships between words~\citep{edge2024local}. For entities, the construction process is similar to that for words, though entity extraction methods are often required to first identify entities~\citep{nasar2021named, al2020named, pawar2017relation}.

    \item {\it Word-Sentence edge:} Word-sentence edges connect words to sentences based primarily on the relationship of belonging. These edges are constructed by linking words to the sentences they appear in~\citep{wang2020heterogeneous, hu2021compare, christopoulou2019connecting}, with edge weights often measured using term frequency-inverse document frequency (TF-IDF). Besides, \citet{ramesh2023single} connects the entities with passage titles via hyperlinks with existing out-of-box entity linkers.

    \item {\it Sentence-Sentence edge:} Sentence-sentence edges connect sentences based on their semantic similarity or relationships. For example, these edges can be constructed through sentence interactions~\citep{yasunaga2017graph, maheshwari2024presentations, huang2021breadth}, similarities between TF-IDF representations~\citep{li2020leveraging, chen2021sgsum}, BM25~\citep{min2019knowledge}, sentence embeddings~\citep{liu2023modeling, wang2023docgraphlm, zhang2024coarse}, Part of speech (Pos) feature and N-Gram feature~\citep{yang2018integrated}. Additionally, \citet{zheng2020srlgrn} construct a Semantic Role Labeling(SRL) graph using  AllenNLP-SRL model~\citep{shi2019simple}. This allows for connecting sentences within long documents or across different documents, which is particularly useful for LLMs with limited context length.

    \item {\it Sentence-document edge:} Sentence-document edges are constructed by linking sentences to the specific documents they belong to~\citep{thayaparan2019identifying}.

    \item {\it Document-document edge:} Document-document edges connect documents based on their similarities. These edges can be constructed when entities in one document are referenced or shared by another~\citep{thayaparan2019identifying, dong2024don}, through document clustering~\citep{wang2023docgraphlm}, embedding similarity~\citep{li2020leveraging}, topic similarity~\citep{zhang2024coarse} or structural similarity~\citep{liu2023modeling}.
\end{itemize}
\end{itemize}

Document graphs typically exhibit heterogeneity, consisting of multiple types of edges. In addition, several hierarchical graphs have been proposed~\citep{weninger2012document, zhang2024teleclass}, where different levels of abstraction are captured. The graph structure can also be dynamic or updated during the learning process~\citep{wang2020heterogeneous, nan2020reasoning, wang2023docgraphlm, liu2022hierarchical}. Moreover, the node in the document graph can also be the response of LLMs. For example, GoR~\citep{zhang2024graph} connects the history responses of LLMs with the responding chunks of documents for long-context summarization.

\subsection{Retriever}

The retrievers for document graphs typically follow the general retriever design, as described in Section~\ref{sec:retriver}. However, there are some special retrieval methods as follows:

\begin{itemize}[leftmargin=*]
    \item \textbf{Pre-Retrieval:} As introduced earlier, there are many scenarios where building a document graph is necessary. However, constructing a fine-grained graph for a large volume of documents can be inefficient and unnecessary. The pre-retrieval aims to first retrieve relevant documents based on the query and then construct a graph. For example, \citet{thayaparan2019identifying} use pre-trained GloVe vectors to first extract relevant sentences and then construct a graph based on the retrieved sentences. \citet{zheng2020srlgrn, yu2021kg} also construct graphs based on the retrieved information.
    
    \item \textbf{Graph similarity-based retriever:}  Graph similarity aims to measure the similarity between two graphs. If both the query and the retrieval data source are graphs, calculating the graph similarity is essential to retrieve relevant information. For instance, \cite{zhang2018graph} leverages the General Maximum Common Subgraph (GMCS) method to retrieve related graphs.
    
    \item \textbf{Iterative Retriever:} In certain tasks, such as multi-step question answering, the node containing the answer might not be directly similar to the query. Iterative retrievers address this by first retrieving nodes related to the query and then using the retrieved information to iteratively retrieve subsequent nodes. For example, \citet{wang2024knowledge, zhang2021answering} utilize iterative retrieval methods for multi-step question-answering tasks, and \citet{ma2024think} iteratively retrieve the documents based on the existing knowledge graph. \citet{asai2019learning} train a Recurrent Neural Network (RNN) to recurrently retrieve relevant information with the Bayesian Personalized Ranking (BRR) loss.
    
    \item \textbf{Topology-based Retriever:} Various topological relationships in a graph can be used to measure different types of similarity. For example, proximity-based topological similarity measures the structural distance between two nodes, while role-based topological similarity assesses the similarity of the roles nodes play within the graph. These similarities can also be leveraged in the retrieval process~\citep{wang2024augmenting}.
\end{itemize}

\subsection{Organizer}

In this section, we describe the organizers for document graphs. GraphRAG on document graphs is still in its early stage, and many works do not incorporate explicit organizers, as shown in Section~\ref{sec:organizer}. We summarize the existing methods below:

\begin{itemize}[leftmargin=*]
    \item \textbf{Graph Pruning}: Graph Pruning refines the retrieved subgraph to reduce irrelevant information and improve computational efficiency. For example, \citet{hemmati2023multi} prune graphs based on the local clustering coefficient, while \citet{zhang2018graph1} employs path-centric pruning to incorporate off-path information. \citet{li2022graph} dynamically drop irrelevant nodes during decoding, and \citet{angelova2006graph} prune edges based on a similarity threshold. Additionally, \citet{edge2024local} uses community detection algorithms to create distinct communities that are then fed into the generators.
    \item \textbf{Reranking}:  Reranking methods aim to reorder retrieved information to facilitate the generation. For example, \citet{yu2021kg}, \citet{zhang2021answering}, and \citet{dong2024don} use GNNs to rerank retrieved passages. \citet{li2021graph} perform listwise reranking to reorder passages expanded via the graph structure.
\end{itemize}

\subsection{Generator}
In this section, we summarize commonly used generators for document graphs. Various methods are applied within document graphs, depending on the input format. Some approaches use the entire graph as input, in which case GNNs and Graph Transformers are commonly employed. Other methods take individual sentences as input, where RNNs or (Large) Language Models are suitable. Additionally, some works leverage both graph and text as inputs, requiring integrated methods that can process multimodal data effectively. We detail each category in the following:

\begin{itemize}[leftmargin=*]
    \item \textbf{GNN-based generator}: Numerous works model tasks as graph-related problems, leveraging GNNs as generators. Conventional GNNs, such as GCN~\citep{kipf2016semi}, GraphSAGE~\citep{hamilton2017inductive}, and GAT~\citep{velivckovic2017graph}, are widely adopted in various studies~\citep{peng2024connecting, thayaparan2019identifying, yasunaga2017graph, munikoti2023atlantic, pan2020semantic, dong2024don}. When graphs contain edge relations, Relational Graph Convolutional Networks (R-GCNs) are often used to capture these relationships effectively~\citep{de2018question, min2019knowledge}. In addition, some works incorporate graph contrastive learning techniques to enhance performance~\citep{xie2022gretel, zhang2022ke, hemmati2023multi}.
    \item \textbf{Graph Transformer-based generator}: Graph transformers~\citep{yun2019graph}, which capture global information across the graph, are used to encode graph structures for various tasks~\citep{zhang2020text, mei2021graph}. These models leverage the transformer’s self-attention mechanism to capture dependencies beyond local neighborhoods, making them well-suited for tasks requiring global context.
    \item  \textbf{RNN-based generator}: Recurrent Neural Networks (RNNs), such as LSTMs, are popular for processing sequences. Therefore, RNNs are employed when the input is in text form~\citep{song2018exploring}.
    \item \textbf{LLM-based generator}: LLM-based generators typically transform the retrieved subgraph into text before using large language models (LLMs) for generation~\citep{wang2024knowledge, edge2024local}. Various models are used in this approach. For instance, BERT~\citep{devlin2018bert} has been applied by \cite{tu2020select, li2022mutually, zhang2021answering} to support generation tasks, while \citet{yu2021kg} and \citet{ju2022grape} leverage the T5 model~\citep{raffel2020exploring}, and \citet{chen2020improving} use RoBERTa~\citep{liu2019roberta}
    \item \textbf{Integrated Generator}: Some works leverage both graph and text data simultaneously for generation, using integrated generators that combine graph models with text generation models. For example, \citet{xu2021dynamic} and \citet{wang2023docgraphlm} employ a combination of RoBERTa and GCN, while \citet{ramesh2023single} fuse GNN with T5 to harness the strengths of both graph structure and language model capabilities.
\end{itemize}

In addition to the aforementioned generators, some approaches embed graphs directly into text generation models. For example, \citet{li2020leveraging} proposes replacing traditional self-attention layers in transformers with graph-informed self-attention, enabling the model to integrate graph structure directly into the generation process.

\subsection{Resources and Tools}
The data resources for document GraphRAG can include any type of document, and thus, we do not provide a comprehensive list here. Instead, in the following, we introduce several tools specifically designed or commonly used for GraphRAG on document graphs:

\begin{itemize}[leftmargin=*]
    \item \textbf{CoreNLP: }\footnote{\url{https://github.com/stanfordnlp/CoreNLP}} The Stanford CoreNLP natural language processing toolkit offers a comprehensive set of natural language processing tools, including token and sentence boundaries, parts of speech, named entities, numeric and time values, dependency and constituency parses, coreference, sentiment, quote attributions, and relations. These tools are valuable for constructing document graphs. 
    \item \textbf{spaCy: }\footnote{\url{https://github.com/explosion/spaCy}} The spaCy is an advanced natural language processing library known for its speed and neural network models, which are optimized for tasks such as tagging, parsing, named entity recognition, text classification, part-of-speech tagging, dependency parsing, sentence segmentation, lemmatization, morphological analysis, and entity linking. 
    \item \textbf{BLINK}: \footnote{\url{https://github.com/facebookresearch/BLINK}} BLINK is an entity linking python library that uses Wikipedia as the target knowledge base.
    \item \textbf{OpenIE}\footnote{\url{https://github.com/dair-iitd/OpenIE-standalone}}: Open Information Extraction (OpenIE) is a tool for extracting structured information from text. It is valuable for generating triples from unstructured text, which can be used as nodes and edges in document graphs.
    \item \textbf{CogComp NLP}\footnote{\url{https://github.com/CogComp/cogcomp-nlp}}: Developed by the Cognitive Computation Group, this suite includes tools for natural language processing tasks such as named entity recognition, sentiment analysis, and coreference resolution.
    \item \textbf{GraphRAG}~\citep{edge2024local}\footnote{\url{https://github.com/microsoft/graphrag}}: GraphRAG is a data pipeline and transformation suite that is designed to extract meaningful, structured data from unstructured text using the power of LLMs. The process involves extracting a knowledge graph out of raw text, building a community hierarchy, generating summaries for these communities, and then leveraging these structures when performing RAG-based tasks.
    \item \textbf{LangChain}~\footnote{\url{https://www.langchain.com/}}: LangChain is a framework for developing applications powered by LLMs, with use cases in document analysis and summarization, RAG, chatbots, and code analysis. Notably, LangChain supports Graph Transformers, which convert documents into graph-structured formats, making it highly suitable for processing document graphs.
    \item \textbf{Neo4j}\footnote{\url{https://neo4j.com/}}: Neo4j is a graph database platform offering comprehensive tools for storing, visualizing, managing, and querying graph data. It includes an LLM Graph Builder, which can extract graphs with LLMs. It also provides GraphRAG demos to demonstrate how to implement a LLMs and RAG system with Neo4j.
    \item \textbf{LlamaIndex}~\citep{Liu_LlamaIndex_2022}\footnote{\url{https://www.llamaindex.ai/}}: LlamaIndex is a data framework designed to support the development of applications based on LLMs. It allows developers to seamlessly integrate data sources, ranging from various file formats to applications and databases, with LLMs. LlamaIndex features an efficient data retrieval and query interface, enabling developers to input any LLM prompt and receive context-rich, knowledge-enhanced outputs. Notably, it includes a Property Graph Index that facilitates the building, modeling, storage, and querying of graphs.
    \item \textbf{Haystack}\footnote{\url{https://haystack.deepset.ai/}}: Haystack is an end-to-end framework for building applications powered by LLMs, Transformer models, vector search, and more. It supports a variety of use cases, including RAG, document search, question answering, and answer generation. Haystack enables the orchestration of state-of-the-art embedding models and LLMs into custom pipelines for creating comprehensive NLP applications. Additionally, it supports Neo4j as a DocumentStore, making it suitable for document graph storage and query operations.
\end{itemize}

\section{Scientific Graph \texorpdfstring{\emojiscititle}{Scientific Graph}}
\label{sec-sci}
Scientific graphs refer to graph-structured data used in domains such as drug discovery~\cite{wangretrieval,huanginteraction,liu2024moleculargpt} and biomedicine~\cite{yang2024kg, liu2024drak,delile2024graph,jiang2024hykge,byambasuren2019preliminary,wen2023mindmap}, both of which are common application areas for GraphRAG. Therefore, in this section, scientific graphs specifically refer to molecular graphs and medical graphs.

In recent years, significant progress has been made in the development of artificial intelligence for science~\cite{abramson2024accurate,jumper2021highly,thirunavukarasu2023large,madani2023large}. Machine learning (ML) and deep neural network technologies are increasingly driving scientific discovery from experimental data. Notably, generative models such as Large Language Models (LLMs) have achieved remarkable success in working with scientific graph data, including molecular graphs and biomedical graphs. In molecular graphs, for example, atoms serve as nodes, while chemical bonds represent edges, capturing the structure of molecules. AI technologies can handle both prediction and generation tasks on these graphs, driving advancements in fields like drug discovery.

Despite the impressive capabilities of generative models like LLMs, several challenges still hinder their applications to scientific domains. One of the most prominent challenges is the lack of domain-specific expertise. In fields like drug discovery, molecule generation is crucial, yet traditional generative models often struggle with producing incorrect or scientifically invalid structures. In medical question-answering (QA) tasks, incorrect answers, hallucinations, and limited interpretability are frequently encountered. To address these challenges, recent works have proposed leveraging external knowledge databases to enhance generation via GraphRAG. GraphRAG improves accuracy by retrieving relevant scientific graphs from extensive databases to guide the generation or answering process. This approach ensures scientific validity by incorporating known valid graph structures, leverages existing knowledge for practical applications, and accelerates the generation process by narrowing the search space.

\subsection{Application Tasks}

Scientific graphs are beneficial for a wide range of tasks. In this subsection, we will review some representative tasks where scientific graphs can play a significant role. 

\begin{itemize}[leftmargin=*]
\item \textbf{Molecule generation:} Molecule generation refers to the process of creating or designing new molecular structures, often using generative models~\cite{wangretrieval,huanginteraction}. It plays a crucial role in fields like drug discovery. The use of scientific graphs, especially molecular graphs, can enhance the rationality and accuracy of generated molecular structures in molecular generation. Typically, an inquiry such as a molecule is given to retrieve the most relevant molecular structures to guide molecular generation. 

\item \textbf{Molecule property prediction:} Molecule property prediction refers to the use of computational methods to estimate the physical, chemical, or biological properties of molecules based on their structure~\cite{liu2024moleculargpt}. 
It has proven highly effective in accelerating the drug discovery process while significantly reducing associated costs. The use of scientific graphs, especially molecular graphs, can enhance the accuracy of prediction. To achieve this, a query molecule is provided, and similar molecules are identified as demonstrations to improve the prediction.

\item \textbf{Question answering:} Question answering (QA) in the scientific domain refers to the use of computational methods to provide accurate and context-specific answers to scientific questions~\cite{yang2024kg, jiang2024hykge, wen2023mindmap, li2024dalk, jeong2024improving,wu2024medical,soman2024biomedical}. This involves retrieving or generating information from scientific literature, databases, or other resources to address complex, domain-specific queries, such as "After meals, I feel a bit of stomach reflux. What medication should I take for it?" In graghRAG, the scientific literature is usually converted into a knowledge graph to provide a basis for answering questions or to enrich the query.
\end{itemize}

\subsection{ Scientific Graph Construction}
In GraphRAG, the choice and construction of data sources are crucial and typically include both public and private datasets. Public datasets are often derived from widely recognized resources, such as molecular databases like PubChem~\cite{kim2019pubchem}, ChEMBL~\cite{gaulton2012chembl}, and ZINC~\cite{irwin2005zinc}, as well as biomedical literature and data sources like PubMed~\cite{lu2011pubmed} and ClinicalTrials~\cite{zarin2011clinicaltrials}. These datasets provide a broad foundation of information across various domains, offering reliable and authoritative references for the model. 
On the other hand, private datasets contain user-specific information, such as medical records from hospitals or confidential clinical trial data~\cite{wu2024medical}. These datasets are highly confidential and unique, enabling GraphRAG models to offer personalized and proprietary knowledge support.

For chemistry, molecules can be represented in four main forms: 1D SMILES (Simplified Molecular Input Line Entry System)~\cite{wangretrieval,liu2024moleculargpt}, 2D molecular graphs~\cite{luo2021predicting}, 3D molecular graphs with coordinates~\cite{huanginteraction}, and text captions describing molecular structures~\cite{delile2024graph}.
1D SMILES is a linear string generated through depth-first search (DFS) on the molecular graph, following specific rules.
2D molecular graphs represent atoms as nodes and bonds as edges, visually showing the connectivity of atoms.
3D molecular graphs incorporate spatial coordinates for each atom, reflecting the molecule’s structure in three-dimensional space, which is crucial for tasks such as molecular docking and reaction prediction.
Text captions, on the other hand, provide a natural language description of the molecular structure, which can be used in tasks that involve textual data interpreting chemical structures from text. Scientific graphs can typically be constructed as follows:

\begin{itemize}[leftmargin=*]
\item \textbf{Text-based construction}: Text-based graph construction is the most commonly used method, which can transform textual scientific knowledge into knowledge graphs. \citet{delile2024graph} believe that text captions suffer from issues of information redundancy and imbalance. By constructing text as a knowledge graph, it is possible to rebalance the retrievable information and reduce redundancy.
Building a knowledge graph typically involves two key steps:

\begin{itemize}[leftmargin=*]
    \item {\it Entity Extraction:} This step involves identifying and extracting key entities from the text, such as domain-specific terms (e.g., chemical compounds, genes).

    \item {\it Relationship Extraction:} After identifying the entities, the next step is to extract the relationships between these entities, determining how they are linked in the given context (e.g., "A is related to B"). This step forms the structural backbone of the knowledge graph.
\end{itemize}

\item \textbf{SMILES-based construction}:
Many chemical databases store data in analytical descriptor formats such as SMILES  ~\cite{wang2022retrieval}. To model graphs, SMILES representations need to be transformed into graph structures.
SMILES-based construction utilizes a library such as RDKit~\cite{landrum2013rdkit} to read the SMILES notation, create a molecular object, and extract atomic and bonding information to construct a 2D graph. Each atom in the molecule is represented as a node, and each bond is represented as an edge between nodes. The atoms carry features such as atom type, degree, charge, and aromaticity, while the bonds can include bond type (single, double, etc.) and whether they are part of a ring.

\item \textbf{3D graph construction}:
\citet{huanginteraction} introduce IRDIFF, an interaction-based retrieval-augmented 3D molecular diffusion model designed for target-specific molecule generation. For pretraining, PMINet is utilized to capture interactive structural context information with binding affinity signals, leveraging the PDBbind v2016 dataset. This dataset offers 3D protein structures and a substantial set of experimentally validated protein-ligand complexes. These protein structures are primarily obtained via techniques like X-ray crystallography, nuclear magnetic resonance (NMR), or cryo-electron microscopy, containing detailed atomic coordinates, bond angles, and secondary structure information.
\end{itemize}

\subsection{Retriever}
The retriever is responsible for locating relevant information based on the input query. In GraphRAG, this information typically consists of graph-structured data.  The retriever can be roughly classified as heuristic-based retriever and deep learning-based retriever.

\begin{itemize}[leftmargin=*]
    \item \textbf{Heuristic-based retriever:} A heuristic-based retriever employs predefined rules, algorithms, and heuristics to identify and retrieve relevant knowledge from graph-structured data sources. Heuristic-based retrieval can be categorized into several types as below:
    
    \begin{itemize}[leftmargin=*]
        \item \textit{Similarity-based retriever:} \citet{wangretrieval} retrieve exemplar molecules based on their similarity to the input molecule, using cosine similarity as the metric. MindMap~\cite{wen2023mindmap} encodes the entities from the query and the external knowledge graph into dense embeddings using BERT, and then retrieves the entity set with the highest similarity scores.
    
        \item \textit{Matching-based retriever:} A matching-based method provides an alternative approach, enabling LLMs to generate evidence-based responses by leveraging comprehensive private data. Specifically, \citet{wu2024medical} identify the most relevant graph through a top-down matching process. Once the relevant content is identified, the LLM generates an intermediate response with its assistance, enhancing both the transparency and interpretability of the results.

        \item \textit{Knowledge graph-based retriever:} A knowledge graph-based method can effectively identify and retrieve the necessary information to respond to a query. \citet{pelletier2024explainable,li2024dalk} use named entity recognition and relation extraction to connect user queries with relevant entities in the knowledge graph, thereby uncovering interpretable and actionable insights from existing biomedical knowledge. This approach significantly enhances the transparency and utility of predictive models. \citet{delile2024graph} map the text chunks to the knowledge graph, then utilize graph distances to find the chunks most relevant to the user's question. In addition,  this work introduces a scoring metric that balances the data by giving each concept mapped along the shortest path an equal opportunity. This metric prioritizes text chunks based on both their recency and their impact. HyKGE~\cite{jiang2024hykge} first queries the LLM to generate a hypothetical output and extracts entities from both the output and the query. Then, HyKGE retrieves reasoning chains between any two anchor entities in an existing knowledge graph, such as CMeKG~\cite{byambasuren2019preliminary} and CPubMed-KG, and feeds the reasoning chains along with the query into the LLM. KG-Rank~\cite{yang2024kg} identifies entities within the query and retrieves the related triples or sub-graphs from the KG to gather factual information.

        \item \textit{Fusion-based retriever:} \citet{soman2024biomedical} begins by retrieving the relevant node in the knowledge graph based on vector similarity to the query's entity. Then, it retrieves the context triples (Subject, Predicate, Object) linked to this node within the knowledge graph.
    \end{itemize}

\item \textbf{Deep Learning-based Retriever:}
Deep learning-based retrievers can extract relevant knowledge to guide the generation process, such as the generation of molecules or proteins. Specifically, \citet{huanginteraction} uses a pre-trained protein-molecule interaction network
named PMINet to extract interactive structural context information between the target protein and ligands in the reference pool to guide the generation of target-aware ligands. \citet{jeong2024improving}
utilize the off-the-shelf MedCPT retriever, a tool specifically tailored for retrieving documents in response to biomedical queries, capable of retrieving up to ten relevant pieces of evidence for each input. DALK~\cite{li2024dalk} filters out noise and retrieves the most relevant knowledge by utilizing the ranking capabilities of LLMs.
\end{itemize}

\subsection{Organizer}
Basically, there are two types
of the organizer: embedding-based and query-based organizers according to how the retrieved knowledge is utilized.

\begin{itemize}[leftmargin=*]
    \item \textbf{Query-based organizer.} The simplest and most direct organizer is a query-based organizer, which integrates the retrieved information with input queries to generate responses. Specifically, HyKGE~\cite{jiang2024hykge} enhances the reasoning process by incorporating external knowledge into the large language model. In this framework, the retrieved reasoning chains are fed into the LLM alongside the original query. This enables the model to ground its responses in structured, contextually relevant information, improving both the accuracy and depth of the generated output. \citet{wu2024medical} prompt the LLM to answer the question by providing the retrieved entity names and their relationships in a concatenated form. This approach involves retrieving relevant entities and the relationships between them from an external knowledge source and structuring this information as a cohesive input for the LLM. DALK~\cite{li2024dalk} integrates the query and retrieves knowledge and then feeds it into LLMs for reasoning and getting the predicted answer. Similarly, KG-Rank~\cite{yang2024kg} combines the re-ranked triplets with the task prompt and inputs them into LLMs for answer generation. In contrast to the methods mentioned above, \citet{soman2024biomedical} begin with context pruning. Specifically, this approach refines the retrieved context by selecting only the most semantically relevant elements needed to answer the query promptly. The input prompt is then combined with this pruning-aware information, producing an enriched prompt that is fed into the LLM for text generation. \citet{delile2024graph} develop a data rebalancing mechanism to ensure that each entity relevant to a question has an equal opportunity of being represented while also highlighting recent significant discoveries. This rebalanced knowledge is then combined with the query prompt and provided as input to the LLM.

    \item \textbf{Embeding-based organizer.} Embedding-based organizer integrates the retrieved information with input embedding to generate responses. Specifically, \citet{wangretrieval} uses a lightweight, trainable standard cross-attention mechanism to fuse the embedding of the input molecule and the retrieved example molecules. Similarly, \citet{huanginteraction} uses a trainable cross-attention mechanism to fuse the enhanced embeddings of the retrieved example ligands and the generated molecules.  

\end{itemize}
\subsection{Generator}

The generator is a core component of the GraphRAG model and is responsible for producing the final output by integrating retrieved evidence with the input data.
Generators can be broadly classified into three categories according to the type of generated models they use: transformer-based generators, diffusion model-based generators, and large language models-based generators. 

\begin{itemize}[leftmargin=*]
    \item \textbf{Transformer-based generator:} RetMol~\cite{wangretrieval} utilizes the Megatron version of the molecule generative model Chemformer for drug discovery. Specifically,  the Chemformer model is a Transformer-based model that can be efficiently applied to tasks in chemistry.

    \item \textbf{Diffusion-based generator:} \citet{huanginteraction} introduce a novel interaction-based retrieval-augmented diffusion model (IRDIFF) for structure-based drug design. Specifically, IRDIFF is able to generate molecules that bind strongly to the target pocket by utilizing protein-molecule interaction data between reference proteins and the target protein to guide the diffusion model.

    \item \textbf{LLM-based generator:} Some approaches leverage large language models, such as LLaMA2, LLaMA3, GPT-4, and Gemini, etc~\cite{wu2024medical,jeong2024improving,pelletier2024explainable,liu2024moleculargpt,li2024dalk,soman2024biomedical}. For instance, MedGraphRAG~\cite{wu2024medical} utilizes GraphRAG to improve the answering capabilities of models such as LLaMA2, LLaMA3, Gemini, and GPT-4 for medical question answering. MolecularGPT~\cite{liu2024moleculargpt} utilizes GraphRAG to improve GPT-3's predictive capabilities for molecular property prediction. DALK~\cite{li2024dalk} employs GraphRAG to boost GPT-3.5-turbo's ability to answer questions related to Alzheimer’s Disease. HyKGE~\cite{jiang2024hykge} utilizes GraphRAG to enhance the medical question-answering capabilities of GPT-3.5 and Baichuan13B.
\end{itemize}

\subsection{Resources and Tools}
In this section, we provide an overview of common data sources and tools utilized in graph RAG systems within the scientific domain, along with a brief description of each project.
\subsubsection{Data Resources}
The public datasets are commonly from well-known resources, such as molecular databases: 

\begin{itemize}[leftmargin=*]
\item \textbf{PubChem}~\cite{kim2019pubchem}: PubChem encompasses a wide range of chemical data, including 2D and 3D structures, chemical and physical properties, bioactivity, pharmacology, toxicology, drug targets, metabolism, safety guidelines, associated patents, and scientific literature. Most of its entries pertain to small molecules, with a primary emphasis on those containing fewer than 100 atoms and 1,000 bonds.

\item \textbf{ChEMBL}~\cite{gaulton2012chembl}: It is an openly accessible database containing detailed information on drugs, drug-like small molecules, and their bioactivity. This curated resource stands out for its comprehensive coverage of the drug discovery process, encompassing data on more than 2.2 million compounds and over 18 million records documenting their effects on biological systems.
ChEMBL provides insights into the interactions between small molecules and their protein targets, along with data on how these compounds influence cellular and organismal functions. It also includes information on ADMET (absorption, distribution, metabolism, excretion, and toxicity) profiles. The database stores two-dimensional structures, calculated molecular properties (such as logP, molecular weight, and Lipinski’s Rule of Five parameters), and bioactivity data like binding affinities and pharmacological effects.

\item \textbf{ZINC}~\citep{irwin2005zinc}: The ZINC dataset is a curated collection of commercially available chemical compounds designed specifically for virtual screening purposes. It offers over 230 million purchasable compounds in 3D formats that are ready for docking, as well as more than 750 million compounds available for analog searches. Each molecule is adjusted to biologically relevant protonation states and is annotated with properties such as molecular weight, calculated LogP, and rotatable bonds. The library includes vendor and purchasing details, making it compatible with several widely used docking software programs. Compounds are provided in multiple protonation states and tautomeric forms within certain constraints, and some formats even offer multiple conformations per molecule. The ZINC database is available for free download~\footnote{\href{http://zinc.docking.org}{http://zinc.docking.org}} in multiple common file formats, including SMILES, mol2, 3D SDF, and DOCK flexibase.
\end{itemize}

There are also biomedical data sources as below:
\begin{itemize}[leftmargin=*]
    \item \textbf{PubMed}~\cite{lu2011pubmed}: PubMed is a freely accessible database that primarily houses the MEDLINE collection, containing references and abstracts in life sciences and biomedicine. Managed by the United States National Library of Medicine (NLM) within the National Institutes of Health, PubMed is part of the Entrez retrieval system. As of May 23, 2023, PubMed contains over 35 million citations and abstracts, with records dating back to 1966 and selectively to 1865, and a few even to 1809. On this date, approximately 24.6 million records include abstracts, and 26.8 million provide links to full-text articles, with around 10.9 million available freely. In the last decade (up to December 31, 2019), PubMed added nearly one million new records annually on average.
    
    \item \textbf{ClinicalTrials}~\cite{zarin2011clinicaltrials}: ClinicalTrials is a global registry and database that provides information on clinical studies funded by both private and public sources. Managed by the U.S. National Library of Medicine, this resource includes a summary of each study's protocol, and for some studies, results are available in a tabular format. Users can search studies by criteria such as study status, condition or disease, country, and other keywords. Continuously updated, the database now includes over 300,000 research studies conducted across all U.S. states and in more than 200 countries worldwide.

    \item  \textbf{Open-source medical KGs}: CMeKG (Clinical Medicine Knowledge Graph)~\footnote{\href{https://cmekg.pcl.ac.cn/}{https://cmekg.pcl.ac.cn/}}, CPubMed-KG (Large-scale Chinese Open Medical  Knowledge Graph)~\footnote{\href{https://cpubmed.openi.org.cn/graph/wiki}{https://cpubmed.openi.org.cn/graph/wiki}} and Disease-KG (Chinese disease Knowledge Graph)~\footnote{\href{https://github.com/nuolade/disease-kb}{https://github.com/nuolade/disease-kb}} are open-source medical knowledge graphs that consolidate a vast amount of medical text data, covering areas such as diseases, medications, symptoms, and diagnostic treatments. The combined knowledge graph includes 1,288,721 entities and 3,569,427 relations.
\end{itemize}

\subsubsection{Tools}
\begin{itemize}[leftmargin=*]
    \item \textbf{RDKit}~\cite{landrum2013rdkit}: RDKit is open-source toolkit for cheminformatics. RDKit has several key features: it can process chemical structures by reading and writing various file formats, such as SMILES, InChI, and Mol files. It generates multiple types of molecular fingerprints, enabling chemical structure comparison and similarity searches. RDKit also provides algorithms for calculating molecular similarity and supports the representation and processing of chemical reactions, including the identification of reactants and products. Although it does not offer molecular docking capabilities on its own, RDKit can be integrated with other docking tools. Additionally, RDKit supports machine learning algorithms, allowing for pattern recognition in chemical data and the construction of predictive models.
    \item \textbf{CADRO}: The Common Alzheimer’s and Related Dementias Research Ontology (CADRO)~\footnote{\href{https://iadrp.nia.nih.gov/about/cadro}{https://iadrp.nia.nih.gov/about/cadro}} is employed to extract a subset of Alzheimer’s disease (AD)-related samples from the medical QA datasets for evaluation. CADRO organizes terms into a three-tiered classification system with eight main categories and multiple subcategories focused on AD and related dementias, containing frequently used terminologies or keywords in the field. From CADRO, users obtain a list of AD-related keywords most relevant to the medical QA datasets.

\end{itemize}

\section{Social Graph \texorpdfstring{\emojisoctitle}{Social Graph}}
\label{sec-soc}

The social graph typically consists of entities connected by their social relations and is ubiquitous across real-world applications. A primary example is social networks like Twitter and Facebook, where entities represent individuals linked by social interactions (e.g., friendships, followers/followees, likes, and mentions). These social graphs extend beyond human interactions and are not confined to living entities, such as tortoises co-using the same burrow in animal social networks~\cite{rossi2015network}, complementary products co-purchased by the same customer in E-commerce recommender systems~\cite{wang2024knowledge2, wu2024stark}, or even the LLM-simulated social agents~\cite{zhang2023exploring, li2023metaagents}. The wealth of social-relational knowledge in these social graphs is the golden resource for GraphRAG\cite{zeng2024large, jiang2023social, xie2023factual, wang2024augmenting, zeng2024federated, wang2024knowledge2, deldjoo2024review, du2024large, wu2024stark, huang2021graph, zhang2021learning, qiu2020exploiting, wei2024llmrec, kim2020retrieval, tian2021joint}
, which is reviewed in this section.

\subsection{Application Tasks}

\begin{itemize}[leftmargin=*]
    \item \textbf{Entity Property Prediction}: 
    Entity property prediction focuses on predicting properties and classifying categories for social entities in social networks, examples of which include the prediction of partnership compatibility, assessment of morality, detection of account suspensions, identification of toxic behaviors~\cite{jiang2023social} and product property prediction~\cite{wang2024knowledge2}.

    \item \textbf{Text Generation}: 
    Text generation aims to produce text that aligns with social contexts and norms. Typically, the interplay between structural and textual information, such as proximity-based network homophily and role-based similarity~\cite{ahmed2018learning}, serves as the foundation for text generation. For instance, \citet{wang2024augmenting} retrieves the texts of proximity/role-similar nodes to enhance the text generation of the target node. \citet{kim2020retrieval, xie2023factual} generate personalized recommendation explanations by retrieving customers'/products' historical reviews. In addition, some other works also leverage semantic similarity to retrieve reference/attributes/opinions and augment the downstream review generation~\cite{sharma2018cyclegen, dong2017learning, park2015retrieval}.

    \item \textbf{Recommendation}:
    The recommendation task aims to find the most relevant items to satisfy user demands. Due to the missing and sparse customer/item interactions (e.g., cold-start issues), GraphRAG can be naturally applied to boost customer/item sparse interaction by retrieving additional meta-knowledge. Depending on the concrete recommendation scenario, GraphRAG has been used for graph-based recommendation~\cite{wei2024llmrec, du2024large}, next-item recommendation~\cite{wang2024knowledge2, huang2021graph, zhang2021learning, qiu2020exploiting, zeng2024federated, wang2023zero}, and conversational recommendation~\cite{friedman2023leveraging}

    \item \textbf{Question-answering}: Question-answering tasks are encountered not only in knowledge and document graph domains but also in social graphs. For example, a user might ask, "What are the best parks for family gatherings around Los Gatos?" and expect a personalized query answer~\cite{zeng2024large, kemper2024retrieval}. Furthermore, such questions might explicitly require graph-structured reasoning. For instance, \citet{wu2024stark} build a semi-structured knowledge base, and some queries might ask, "Can you list the products made by Nike?" Answering this question requires not only a deep understanding of the query but also familiarity with the structural information of the data.

    \item \textbf{Fake News Detection}: Detecting fake news necessitates considering both semantic content (e.g., the content of the news) and structural interactions (e.g., interactions among news). For instance, \citet{ram2024credirag} evaluates the credibility of a Reddit post based on the credibility of other Reddit posts retrieved by their common interactions with the same common commenters.
    
\end{itemize}

\subsection{Social Graph Construction}
The relations that GraphRAG leverages to derive additional information from social graphs can be mainly summarized into three rationales: proximity-based, role-based, and personalization-based rationale. The proximity-based rationale, rooted in the adage "birds of a feather flock together," suggests that nodes close to each other in a social network often share similar properties~\cite{mcpherson2001birds}. For example, close friends tend to possess similar hobbies. The role-based rationale focuses on nodes with similar local subgraph structures sharing similar features or label distributions~\cite{donnat2018learning}. For example, managers at the same hierarchical level within a company generally have comparable job titles and responsibilities, and hub airports exhibit similar operational characteristics and strategic importance. Lastly, the personalization-based rationale refers to individuals' uniqueness regarding their characteristics and interactions. For example, in recommender systems, users interact with items in various ways, such as clicking, viewing, adding to a cart, purchasing, and reviewing, each interaction providing valuable knowledge that GraphRAG can leverage to personalize generated content. Based on the above three relational rationales in the social graphs, the social graph construction methods can be summarized as follows:

\begin{itemize}[leftmargin=*]
    \item \textbf{User-User-Interaction}~\cite{jiang2023social}: This type of social graph represents user-to-user interactions commonly seen on social networks such as Twitter, Reddit, and Facebook. Examples include follower-followee relationships on Twitter, friendship relationships on Facebook, and user comments on other users' threads on Reddit. Note that this user-user interaction naturally exists in the real world without human curation or modification compared to documents or knowledge graphs.

    \item \textbf{User-Item-Interaction}~\cite{wang2024knowledge2, wang2024augmenting, xie2023factual, kim2020retrieval}: This type of social graph represents user-to-item interactions commonly found on e-commerce platforms like Amazon and eBay. These interactions, including purchasing, adding-to-the-cart, and viewing, can be modeled as a bipartite graph, where each type of interaction reflects a unique user intention.

    \item \textbf{Item-Item-Interaction}~\cite{wang2024knowledge2, zhang2021learning, wu2024stark}: This social graph captures interactions between items, typically identified by shared interactions from the same customer or user. For example, a "co-view" interaction between two products indicates that they are viewed by the same customer, while a "view-add-to-cart" interaction indicates that one product is first viewed and the other is added to the cart next by the same customer. In e-commerce networks, these co-interactions between two items can be broadly categorized into complementary and substitute relationships~\cite{zhao2024recommender}.

    \item \textbf{Metadata-Interaction}~\cite{wu2024stark, yang2024common}: Items and users often possess metadata. For example, products on platforms like Amazon may include brand, manufacturer, and color attributes. This metadata can be represented as additional node types and the corresponding edges indicating relations between products and attributes, such as ownership or association.

    \item \textbf{Agent-Agent Interaction}~\cite{chen2023agentverse, zhang2023exploring}: With the increasing intelligence of LLMs, recent literature has explored the potential of using LLM-powered agents to simulate social behaviors, such as collaboration, debate, and reflection. These agent-agent interactions could also be used for GraphRAG.
    
\end{itemize}

Note that among the five types of interaction mentioned above, the user-user, user-item, and metadata relations are naturally formulated without human curation. In contrast, item-item interactions are generated by manual extraction, and agent-agent interactions require simulations.

\subsection{Retriever}
The relations of the social graphs constructed according to the above methods can be leveraged in GraphRAG to enhance downstream tasks by retrieving additional information. For example, retrieving historical metadata and customer interactions can improve recommendations. Next, we review representative retrieval methods in GraphRAG for social graphs.

\begin{itemize}[leftmargin=*]
    \item \textbf{ID-based Retriever}: Similar to entity linking, the ID-based retriever works by retrieving content specifically generated by a user/item, examples of which include retrieving historical item interactions of a specific customer~\cite{zeng2024large, jiang2023social, zeng2024federated}, reviews posted on a specific product~\cite{xie2023factual}, and meta-data information about a specific user~\cite{wu2024stark, salemi2023lamp}.

    \item \textbf{Filtering-based Retriever}: Based on the ID-based retriever, the Filtering-based Retriever retrieves additional content based on collaborative filtering~\cite{wang2023zero}. For user filtering, Top K users are identified by comparing the similarity of their historical item interactions with those of the target user demanding recommendation. To enhance the context of the target user, it retrieves the most popular items from those Top K users. In contrast, for item filtering, it identifies the Top K items that are most similar to the current item by comparing their user-interaction history. Among them, the most popular ones would be fetched to augment the current item recommendation.

    \item \textbf{Social Relational Retriever}: Like the ID-based Retriever, the Social Relational Retriever focuses on retrieving knowledge from entities sharing certain relations with the target entities on hand. For example, \citet{du2024large} hierarchically retrieve neighboring texts several hops from the central node to augment the semantic information of the target user/item. Meanwhile, \citet{wang2024augmenting} retrieve texts corresponding to nodes that share high proximity-based and role-based similarity, respectively.
    
    \item \textbf{Integrated Neural-Symbolic Retriever}: This approach leverages both symbolic and neural retrievers to improve retrieval effectiveness~\cite{wang2024knowledge2, zeng2024federated, wang2020global}. The symbolic retriever retrieves information by following explicitly defined rules, such as retrieving based on identifiers, structured relationships, or interaction patterns, ensuring that the retrieved data strictly aligns with specific criteria. Meanwhile, the neural retriever complements this by using embedding-based similarity, capturing nuanced patterns and contextual relationships that may not be directly encoded in rules. Integrating them together provides a better trade-off between rule-based precision and neural-based adaptability and generalizability. For instance, \citet{wang2020global, wang2024knowledge2, huang2021graph, qiu2020exploiting} first retrieve K-hop neighboring products from the product knowledge graph (symbolic retriever) for products in the user session and further utilize neural-based adaptive filtering to aggregate items that are most relevant to the current sequence (i.e., neural retriever). Similarly, \citet{zeng2024federated} address data sparsity and heterogeneity by combining ID-based retrieval that retrieves movies based on the user ID of one party with a text-based retriever that enables movie retrieval between parties.

\end{itemize}

\subsection{Organizer}
To further enhance the retrieved content, the organizer for the social graphs employs specialized techniques beyond the typical re-ranking and filtering used in other graph domains~\cite{zeng2024large, hou2024large, deldjoo2024review}. For social graphs, the organizer of GraphRAG often uses Keyword Extraction, Profile Summarization, and Hierarchical Graph Aggregation to curate the retrieved content:

\begin{itemize}[leftmargin=*]
    \item \textbf{Keyword Extraction}: Keyword extraction identifies the most relevant and informative keywords from the retrieved content. These extracted keywords guide the downstream generator in prioritizing attention and reduce the risk of overwhelming LLMs with excessive context. For example, \citet{xie2023factual} use an embedding estimator to pinpoint keywords aligned with a personalized latent query, which are then used to generate explanations.

    \item \textbf{Profile Summarization}: Profile summarization creates rich and detailed user profiles that capture attributes such as age, gender, preferred and disliked genres, favorite directors, country, and language derived from users' past interactions and item metadata~\cite{wei2024llmrec}. This enriched user data increases the informativeness of the retrieved content while safeguarding privacy by using rewritten profiles. Additionally, after retrieving related movies via user/item filtering, \citet{wang2023zero} apply a three-step prompting strategy to extract features tailored to the user and select representative movies by directly instructing LLMs. These extracted features and representative movies are further used as the user profile to guide the final recommendation generation. \citet{guo2024lightrag} leverage LLMs to generate more details about the entity profile based on its keywords/short phrases from external data to aid text generation.

    \item \textbf{Hierarchical Graph Aggregation and Summarization}: 
    When retrieving neighborhood textual information, the exponentially expanding receptive field at higher neighborhood layers often causes the aggregated content to exceed manageable limits, potentially diminishing the LLM's effectiveness~\cite{li2024long}. This challenge highlights the need for hierarchical graph aggregation and summarization. The primary idea is to first summarize neighborhood information retrieved at each layer before propagating it to the next layer. This approach keeps the volume of text each node receives within consistent bounds. For instance, \citet{du2024large} recursively aggregate information from neighboring nodes at higher layers and leverage LLMs to rephrase and compress the content before sharing it with relevant nodes in lower layers. Consequently, this strategy expands the receptive field while optimizing the computational budget for downstream generation tasks.
\end{itemize}
 
\subsection{Generator}
Once the relevant content is retrieved and organized appropriately, it is further processed by a downstream generator to produce the final content. Depending on the desired output, existing generators used for social graphs can be categorized into LLM-based text generators and Prediction-based generators. The choice between these two depends on the format of the desired output and the requirements of the social graph applications.

\begin{itemize}[leftmargin=*]
    \item \textbf{LLM-based Text Generation}: This generator is typically used when the downstream application requires text outputs~\cite{zeng2024federated, xie2023factual, wang2023zero, wang2024augmenting}, such as item recommendations based on names, recommendation explanations, and review generation. Due to the probabilistic nature of the next token prediction, the generated text may not always precisely align with the desired output. To address this hallucination issue, the ground-truth text is often used for grounding the generated texts, such as matching generated items with the ground-truth items on the platform~\cite{hou2024large}.

    \item \textbf{Prediction-based Generation}: The generator directly predicts outputs and is primarily used for non-textual tasks~\cite{du2024large}, such as item recommendation and social prediction. For instance, Graph Neural Networks have become popular choices for graph-based recommendation~\cite{du2024large, wei2024llmrec} and transformers are used for item recommendation tasks~\cite{wang2024knowledge2}. Furthermore, in social property prediction~\cite{jiang2023social}, the generator could simply be a multilayer perception used for classification (e.g., partisanship classification) or regression (e.g., morality regression).
    
\end{itemize}

\subsection{Resources and Tools}

This section lists common resources and tools used for GraphRAG on social graphs. The summary and the link of  are itemized as follows:

\subsubsection{Data Resources}
\begin{itemize}[leftmargin=*]
    \item \textbf{STARK-Amazon}\footnote{\href{https://stark.stanford.edu/dataset_amazon.html}{https://stark.stanford.edu/dataset\_amazon.html}}~\cite{wu2024stark}: STARK-Amazon is a large-scale, semi-structured retrieval benchmark dataset for product search on the Amazon platform, integrating textual and relational knowledge bases. The nodes in the dataset represent products, colors, brands, and categories, while edges capture relationships like also\_bought, also\_viewed, has\_brand, has\_category, and has\_color. Rich textual information, including product descriptions, customer reviews, and entity names, provides valuable context for retrieval tasks. Queries are generated by sampling relational templates and grounding them with specific entities, then leveraging LLMs to synthesize relevant textual and relational information. This process results in queries that capture customer interests, interpret specialized descriptions, and deduce relationships involving multiple entities within the query.
    
    \item \textbf{Amazon-Review}\footnote{\href{https://jmcauley.ucsd.edu/data/amazon/}{https://jmcauley.ucsd.edu/data/amazon/}}~\cite{ni2019justifying, he2016ups, mcauley2015image, xie2023factual, zeng2024federated}: The Amazon-Review dataset includes reviews with ratings, text, and helpfulness votes, along with product metadata such as descriptions, categories, price, brand, and image features. Additionally, it provides link information through “also viewed” and “also bought” graphs. This dataset has two versions: the initial release logging the review data between 1996 and 2014~\cite{he2016ups, mcauley2015image}, and a more recent version that logs ongoing data in 2014~\cite{ni2019justifying}. The latest version also adds new metadata, including detailed product information, bullet points, and an ethical details table. Extensive GraphRAG work~\cite{} in social graphs has leveraged this dataset to build RAG frameworks for recommendation research.
    
    \item \textbf{MovieLens}\footnote{\href{https://grouplens.org/datasets/movielens/}{https://grouplens.org/datasets/movielens/}}~\cite{wang2023zero, wei2024llmrec, zeng2024federated}: The MovieLens datasets describe people’s expressed preferences for movies~\cite{harper2015movielens}. These preferences take the form of tuples, each resulting in a person expressing a preference (a 0-5 star rating) for a movie at a particular time. These preferences were entered through the MovieLens website — a recommender system that asks its users to give movie ratings to receive personalized movie recommendations. Side information includes movie title, year, and genre in textual format. \cite{wei2024llmrec} further crawl the visual content of movie posters accessible from \href{https://github.com/HKUDS/LLMRec}{here}. Note that Movielen-100k is frequently used~\cite{wei2024llmrec, zeng2024federated} and provides users with demographic information such as user ID, age, gender, occupation, and zip code.

    \item \textbf{Netflix}\footnote{\href{https://www.kaggle.com/datasets/netflix-inc/netflix-prize-data}{https://www.kaggle.com/datasets/netflix-inc/netflix-prize-data}}~\cite{wei2024llmrec}: This dataset was constructed to support participants in the Netflix Prize. The movie rating files contain over 100 million ratings from 480 thousand randomly chosen, anonymous Netflix customers over 17 thousand movie titles.  The data were collected between October 1998 and December 2005 and reflect the distribution of all ratings received during this period. The date of each rating, the title, and the year of release for each movie ID are also provided. The multi-model side information is collected through web crawling~\cite{wei2024llmrec}, which is further stored \href{https://github.com/HKUDS/LLMRec}{here}.

    \item \textbf{Yelp}\footnote{\href{https://www.yelp.com/dataset}{https://www.yelp.com/dataset}}~\cite{xie2023factual}: The Yelp dataset is a rich resource for academic research, particularly in fields like data science, natural language processing (NLP), and machine learning. This comprehensive dataset includes over 6.9 million reviews, 150,346 businesses, and 200,100 photos, covering 11 metropolitan areas. Researchers and students can explore real-world business and user data, with extensive details about businesses such as location, categories, attributes (like hours, parking, and ambiance), and aggregated check-ins. Reviews include full-text content, user ratings, and metadata, while user profiles offer insights into social interactions and behaviors, including friends, compliments, and review history. Additionally, the dataset contains 908,915 tips, providing quick recommendations and insights, and 1.2 million business attributes, offering granular information about service offerings.

    \item \textbf{Weibo}\footnote{\href{https://www.aminer.cn/influencelocality}{https://www.aminer.cn/influencelocality}}~\cite{zhang2013social}: The Weibo dataset offers a rich snapshot of user interactions, behaviors and social connections within the Sina Weibo platform, capturing both static and dynamic facets of the social network. Starting with 100 randomly selected seed users, the dataset scales to encompass 1.7 million users and around 0.4 billion following relationships, averaging 200 followers per user. Each user is characterized by social profile details, including name, gender, verification status, and follower/followee counts. Tweet content is provided in both original Chinese and indexed formats, supporting robust research in social network analysis, content diffusion, and user interaction dynamics, making it an invaluable resource for RAG studies.

    \item \textbf{Brexit}\footnote{\href{https://github.com/somethingx01/TopicalAttentionBrexit?tab=readme-ov-file}{https://github.com/somethingx01/TopicalAttentionBrexit?tab=readme-ov-file}}~\cite{zhu2020neural}: This dataset includes a portion of the X (Twitter) network, specifically the remain-leave discourse before the 2016 UK Referendum on exiting the EU. It comprises a network with 7,589 users, 532,459 directed follow relationships, and 19,963 tweets, each associated with a binary stance. The dataset is preprocessed according to \cite{minici2022cascade} to assign each user a scalar value between 0 and 1, referred to as opinion, representing the average stance of the tweets retweeted by the user. The stance of each tweet is either 0 (“Remain”) or 1 (“Leave”).

    \item \textbf{Diginetica}\footnote{\href{https://competitions.codalab.org/competitions/11161\#learn_the_details-data2}{https://competitions.codalab.org/competitions/11161\#learn\_the\_details-data2}}~\cite{wang2024knowledge2}: This dataset comprises user session logs from an e-commerce search engine. The data spans six months and captures user interactions, including clicks, product views, and purchases. Each user session, defined by a one-hour inactivity period, contains anonymized user IDs, hashed queries, product descriptions, metadata (price, hashed product names, image identifiers, and product categories), and log-scaled prices.

    \item \textbf{Yoochoose}\footnote{\href{https://www.kaggle.com/datasets/chadgostopp/recsys-challenge-2015}{https://www.kaggle.com/datasets/chadgostopp/recsys-challenge-2015}}~\cite{wang2024knowledge2}: The Yoochoose dataset contains session data from an online European retailer. Each session records the user's click events, with some sessions also including purchase events. Collected over several months in 2014, the data captures user interactions with the retailer's website. The product meta-data includes categories.
\end{itemize}

\subsubsection{Tools} 
\begin{itemize}[leftmargin=*]
    \item \textbf{X-Developer Platform}\footnote{\href{https://developer.x.com/en/docs/x-api/getting-started/about-x-api}{https://developer.x.com/en/docs/x-api/getting-started/about-x-api}}: The X Developer Platform provides powerful tools and resources for developers to integrate X's real-time, historical, and global data into their own applications. With three main products—X API, X Ads API, and X for Websites—the platform supports a wide range of use cases, from retrieving and analyzing Tweets to managing ad campaigns and embedding X content directly into websites. The X API offers endpoints for managing conversations, exploring trends, and engaging with users, while the X Ads API enables businesses to manage ads with custom targeting and analytics. X for Websites allows seamless embedding of live content to enhance website engagement. Through comprehensive documentation, libraries, and community support, the X Developer Platform enables developers to create innovative solutions using X’s data and engagement tools.
    
    \item \textbf{Reddit-API}\footnote{\href{https://support.reddithelp.com/hc/en-us/articles/16160319875092-Reddit-Data-API-Wiki}{https://support.reddithelp.com/hc/en-us/articles/16160319875092-Reddit-Data-API-Wiki}}: Reddit is a news aggregation and discussion platform where posts are organized into "subreddits," user-created boards moderated by the community. The Reddit API provides developers access to the site’s extensive collection of posts and comments. This free API has enabled the development of moderation tools, third-party applications, and training datasets for LLMs such as ChatGPT, Google, and Gemini. By using this API, we can query tremendous user interactions (i.e., comments and posts) and their corresponding textual contents (i.e., subreddit threads), which serve as golden resources for GraphRAG.
    
    \item \textbf{Rec-Bole}\footnote{\href{https://recbole.io/}{https://recbole.io/}}~\cite{zhao2021recbole}: RecBole is an open-source, unified, and comprehensive library designed for developing and benchmarking recommendation algorithms. Built with Python and PyTorch, RecBole offers researchers a streamlined, efficient framework to experiment with over 100 recommendation algorithms across four main types: General, Sequential, Context-Aware, and Knowledge-Based Recommendations. The platform simplifies data handling by providing pre-processed copies of 43 benchmark datasets, making it easy for users to dive into model testing and development. The provided user-product interactions, as well as text-formatted user/product meta-data, could also be used for GraphRAG.
\end{itemize}

\section{Planning and Reasoning Graph \texorpdfstring{\emojiplantitle}{Planning and Reasoning Graph}}
\label{sec-reasonplan}

A planning or reasoning graph characterizes the inherent logical flow among different entities, where entities typically represent concrete planning or reasoning substeps, and edges denote their logical relations. For the planning graph~\cite{shen2024hugginggpt, zhuangtoolchain}, a common example is a set of API tools used to achieve certain goals, where nodes represent actions, and edges denote their relational dependencies. For reasoning graphs, a notable example is the recent proposed chain/tree/graph of thoughts techniques~\cite{besta2024graph, wei2022chain, yao2024tree} where each node represents a decision-making thinking step connected by the reasoning flow. The dependency constraint and reasoning flow in the planning/reasoning graphs can be naturally represented as relational knowledge, which forms the foundation for GraphRAG in fulfilling planning/reasoning tasks. This section reviews GraphRAG for the planning and reasoning graph~\cite{besta2024graph, lin2024graph, madaan2021could, saha2021explagraphs, sakib2024consolidating, shen2024hugginggpt, shen2023taskbench, song2023llm, song2023restgpt, wu2024can, wei2022chain, xu2024p, yao2024tree, zhao2024large, zhuangtoolchain, hao2023reasoning}.

\subsection{Application Tasks}
The representative tasks conducted on planning and reasoning graphs are summarized as follows:
\begin{itemize}[leftmargin=*]
    \item \textbf{Sequential Plan Retrieval}~\cite{shen2024hugginggpt, shen2023taskbench, song2023restgpt, wu2024can, zhao2024large}:
    As one of the most frequently encountered tasks, plan retrieval aims to retrieve the plan of steps or tools in the format of subgraphs to complete user queries. For example, given the user query "Please generate an image where a girl is reading a book, and her pose is the same as the boy in "example.jpg," then please describe the new image with your voice.", the retrieved final plan from the global plan graph would be "Post Detection" $\rightarrow$ "Pose-to-Image" $\rightarrow$ "Image-to-Text" $\rightarrow$ "Text-to-Speech."

    \item \textbf{Naturalistic Asynchronous Planning}~\cite{lin2024graph}: In contrast to plan retrieval, which considers only dependency constraints among plans, incorporating time constraints introduces a greater challenge. Naturalistic Asynchronous Planning aims to produce a plan that meets dependency requirements and optimizes task completion efficiency, using time summation, time comparison, and constrained reasoning. For example, a user might request, "To make calzones, here are the steps and times required; please calculate the optimal plan for completion." An efficient plan would execute "roll dough," "add filling," and "fill dough" sequentially, with "preheat oven" in parallel, and then conclude with "bake."

    \item \textbf{Structured Commonsense Reasoning}~\cite{saha2021explagraphs}: Given a belief and an argument, structured common sense reasoning aims to infer the stance and generate/retrieve the corresponding commonsense explanation graph that explains the inferred stance. 

    \item \textbf{Defeasible Inference}~\cite{madaan2021could}: Defeasible Inference is a mode of reasoning in which, given a premise, a hypothesis may be weakened or overturned in light of new evidence. A prominent approach is to support defeasible inference through argumentation by constructing inference graphs.

    \item \textbf{Tool Usage}~\cite{zhuang2023toolchain, hao2023reasoning}: Instructing LLMs to use external tools for complex real-world problems has gained increasing importance. Recent research has explored advanced planning strategies to enhance LLMs' tool-use intelligence. Notably, two approaches employ A* search~\cite{zhuang2023toolchain} and Monte Carlo Tree Search~\cite{hao2023reasoning}, both utilizing graph-structured reasoning to adaptively retrieve the next tool based on the LLM's internal evaluations and environmental feedback. These methods enable dynamic tool retrieval, refining the model's problem-solving precision and flexibility.

    \item \textbf{Embodied Planning}~\cite{song2023llm, xu2024p}: Embodied planning tasks in Embodied AI involve guiding agents to perform sequences of actions based on natural language instructions and visual cues in simulated or real-world environments. These tasks, such as organizing or cleaning, challenge agents due to ambiguous instructions, limited task-specific knowledge, sparse feedback, and complex, variable action spaces.
\end{itemize}

\subsection{Reasoning and Planning Graph Construction}
Most existing methods for constructing reasoning and planning graphs begin by analyzing relational dependencies and subsequently adding edges based on these hard-coded rules. Therefore, rather than limiting our focus to reviewing this only rule-based construction method, we review various dependency categories used for edge addition.

\begin{itemize}[leftmargin=*]
    \item \textbf{Resource Dependency}~\cite{shen2023taskbench, song2023restgpt, wu2024can}: This dependency is defined as the shared resources among different actions/decisions. For example, two tools are connected if the output of one tool matches the input of the other, enabling a seamless transition from one process to another. The decision to add edges in existing graph construction methods is made by checking whether one node's input matches another node's output (e.g., plans, tools, or some other abstract processes).

    \item \textbf{Temporal Dependency}~\cite{shen2023taskbench}: This relation ensures that the sequence of events follows certain orders within the planning and reasoning process. For example, connections in some collected datasets denote the successive order between two APIs for daily life.

    \item \textbf{Inclusive Dependency}~\cite{ma2024harec}: The dependency described indicates that two connected nodes belong to the same category or environment. For example, cobblestones and birdhouses are both part of the category of garden decorations~\cite{wang2024knowledge2}. Hypergraphs can effectively capture such belonging relationships where one entity belongs to multiple environments~\cite{feng2019hypergraph}. Furthermore, these dependencies often form hierarchical structures, where "grandparent" entities encompass "parent" entities, which in turn encompass their "children." As the depth of the hierarchy increases, the space of possible dependencies grows exponentially, presenting a significant computational challenge. To address this, many previous works have proposed encoding such hierarchies in hyperbolic space~\cite{ma2024harec, liu2019hyperbolic, zhang2024hierarchical}. To the best of our knowledge, no prior research has explored inclusive dependencies in RAG systems.

    \item \textbf{Causual Dependency}~\cite{lin2024graph}: This dependency indicates the cause-and-effect logic within the graph, where one action/decision causes the trigger of another action/decision. A long-standing example is the casual graph to encode assumptions about the data-generating process.

    \item \textbf{Analogy Dependency}~\cite{yu2023thought, yuan2023analogykb}: This dependency underscores analogical reasoning, where relationships take the form "A is to B as C is to D." By recognizing and leveraging such dependency, humans build on existing knowledge to forge new insights across domains. A powerful historical example is the discovery of Coulomb's Law, inspired by the analogy between gravitational forces affecting celestial bodies and electrical forces between charged particles~\cite{priestley1775history}.
\end{itemize}

While resource-dependent and causal relations involve a sequentially structured relation (former step/tool/decision leads to the later step/tool/decision), they are inherently different. For instance, if Tool A generates a report in PDF format and Tool B is designed to extract data from PDF files, Tool A and B share resource dependency because the output of Tool A (the PDF) matches the input of Tool B. However, this connection does not imply a direct cause-and-effect relationship between the two tools, i.e., using Tool A would not necessarily cause us to use Tool B.

\subsection{Retriever}
Retrievers for handling tasks on reasoning and planning graphs are often modeled as graph traversers. The query or task instruction locates the initial seeding nodes to initialize the graph traversal; then a traversal expands the graph's scope either until a preset budget is exhausted or specific criteria are met. A core step throughout is selecting the most relevant neighbors from all potential candidates. Based on the criterion for neighborhood selection, retrievers fall into two main categories: embedding-based methods, which prioritize neighbors based on embedding similarity, and heuristic-based methods, which use local and global reward functions to determine neighbor importance.

\begin{itemize}[leftmargin=*]
\item \textbf{Embedding-based}: \citet{wu2024can} decompose the query, subsequently perform the embedding similarity match between the concatenated subquery with the current retrieved task API and each of the existing APIs, and then select the top one from the existing neighboring APIs. It explores the strategy of training and the one without training.

\item \textbf{Heuristic-based}: Compared to embedding-based methods, which rely on dedicated training data for mapping, heuristic-based methods define rules to guide the graph retriever effectively\cite{zhao2024large, zhuangtoolchain, hao2023reasoning}. \citet{zhuangtoolchain} model tool planning as a tree search algorithm, incorporating A* search to adaptively retrieve the most promising tool for subsequent use based on accumulated and anticipated costs. Both cost functions are heuristically designed, drawing on prior literature and practical insights.

\item \textbf{Thought Propagation Retrieval}~\cite{yu2023thought}: Given an input problem, thought propagation retrieval prompts LLMs to propose a set of analogous problems, and then applies established prompting techniques, like Chain-of-Thought (CoT), to derive solutions. The aggregation module subsequently consolidates solutions from these analogous problems, enhancing the problem-solving process for the original input.
\end{itemize}

\subsection{Organizer}
The current GraphRAG literature on planning and reasoning graphs generally omits organizer mechanisms, as the retrieval process alone achieves sufficient precision, eliminating the need for reranking. Unlike document or knowledge graphs, which typically apply one-shot embedding-based similarity retrieval to select the top-K relevant content~\cite{wu2024stark}, planning and reasoning graphs use a multi-round embedding similarity process integrated with reasoning steps, enhancing plan fidelity. Moreover, reward-based retrieval involves a sophisticated search that further boosts accuracy. Together, these high-quality strategies reduce the need for fine-grained reranking or filtering.

\subsection{Generator}
Most existing GraphRAG approaches for reasoning and planning tasks either output the retrieved plan directly or integrate it into the LLM for downstream solution generation. For instance, \citet{wu2024can} outputs the retrieved graph-structured plan as the final result, while \citet{shen2024hugginggpt} compiles executed results from expert tools to generate the response. Similarly, after constructing a tool invocation graph, \citet{shen2023taskbench} directly prompts the LLM to generate the parameters, and \citet{lin2024graph} leverages the LLM to produce asynchronous plans based on task dependencies, time, and graph constraints. Notably, most of these works focus on fusing textual information, primarily using different graph structure formats (e.g., adjacency matrix, adjacency list, edge list, CSR) presented in textual form.

\subsection{ Resources and Tools}

We summarize the useful resources and tools for GraphRAG on planning and reasoning graphs. 

\subsection{Data Resources}
\begin{itemize}[leftmargin=*]
    \item \textbf{Hugging Face}\footnote{\href{https://github.com/microsoft/JARVIS}{https://github.com/microsoft/JARVIS}}~\cite{shen2023taskbench}: Hugging Face offers a wide array of AI models covering multi-modality tasks in language, vision, audio, video, and more. Each task corresponds to a tool node that handles specific input and output. If tools A and B are connected, the output type of A must match the input type of B. Thus, edges in the Hugging Face plan graph represent the resource-dependent relation. \cite{shen2023taskbench} firstly collects the tool repository and builds a tool graph with a collection of tools and their dependencies. Then, to generate each question, they sample a subgraph from the tool graph in the three basic formats: node, chain, and directed acyclic graph (DAG), each of which embodies a specific pattern for tool invocation. After that, the sampled subgraph is sent to LLMs to synthesize user instructions and populate the parameters for the tool subgraphs. At last, LLM-based and rule-based self-critic mechanisms are used to check and filer out the generated instruction to guarantee quality.

    \item \textbf{Multimedia}\footnote{\href{https://github.com/microsoft/JARVIS}{https://github.com/microsoft/JARVIS}}~\cite{shen2023taskbench}: Unlike the AI-focused tools of Hugging Face, multimedia tools serve a broader range of user-centric tasks such as file downloading and video editing. The edges remain consistent with the Hugging Face domain and the tool connections denote the resource-dependent relation similar to Hugging Face. The construction of Multimedia is similar to Hugging Face.
    
    \item \textbf{Daily Life APIs}\footnote{\href{https://github.com/microsoft/JARVIS}{https://github.com/microsoft/JARVIS}}~\cite{shen2023taskbench}: Daily life services, including web search and shopping, can also be viewed as tools for specific tasks. The dependencies among these APIs are primarily temporal, meaning that two daily life APIs are connected if one follows the other in sequence. The construction of Daily Life APIs is mostly similar to Hugging Face except that the edges are constructed manually because of the scarcity of publically available APIs.

    \item \textbf{RestBench}\footnote{\href{https://restgpt.github.io/}{https://restgpt.github.io/}}~\cite{song2023restgpt, wu2024can}: A dataset consisting of multiple APIs to address complex real-world user instructions in two scenarios: TMDB movie database and Spotify music player. The TMDB movie database offers RESTful APIs encompassing information about movies, TVs, actors, and images. Spotify Music Player provides API endpoints to retrieve content metadata, receive recommendations, create and manage playlists, and control playback. For RestBench, \cite{song2023restgpt} employed NLP experts to brainstorm instructions for different combinations of APIs and correspondingly annotate the gold API solution path for each instruction. Two additional experts are further employed to thoroughly verify the solvability of each instruction and the correctness of the corresponding solution path. To adapt RestBench into a graph-structured dataset, \cite{wu2024can} treats each API as a unique task node, modeling their relationships through two key dimensions: (1) categorical association and (2) resource dependencies. For instance, APIs offering movie-related functionalities, such as retrieving movie details or recommending films, are grouped under the 'movie' category, while APIs focused on person-related tasks, like searching for actors, are classified under the 'person' category. Additionally, if two APIs share a common parameter (e.g., movie-id), a link is established to represent resource dependency. To further enhance semantic distinction, GPT-4 is prompted to assign a unique name and detailed functional description to each API.

    \item \textbf{AsyncHow}\footnote{\href{https://github.com/fangru-lin/graph-llm-asynchow-plan}{https://github.com/fangru-lin/graph-llm-asynchow-plan}}~\cite{lin2024graph}: A curated dataset consisting of 1.6K data points for asynchronous planning. Each data point comprises a user instruction specifying tasks with their basic execution constraints, represented by a directed acyclic graph (DAG). In these DAGs, nodes denote actions, and edges represent ordering constraints. Each edge carries a weight indicating the time required to complete the preceding action and transition to the next. Additionally, edges also signify causal links, meaning an action can only proceed if all preceding linked actions are completed. To construct this dataset, we first collected planning tasks from WikiHow~\cite{koupaee2018wikihow}. LLMs were then used for preprocessing, time annotation, and step dependency annotation. Specifically, plans containing optional steps, irrelevant tasks, or those lacking quantifiable duration were filtered out. The GPT-3.5 is used to estimate the time duration for each step, and the GPT -4 is used to annotate step dependencies using the DOT language. To generate natural language questions with execution constraints, ten trivially different but plausible templates were used to paraphrase the graph-structured dot language into human-understandable texts. The optimal time duration for a plan was calculated by determining the longest path within the DAG representation of the workflow.

    \item \textbf{EXPLAGRAPHS}\footnote{\href{https://github.com/swarnaHub/ExplaGraphs}{https://github.com/swarnaHub/ExplaGraphs}}~\cite{saha2021explagraphs}: Give the initially collected triplets consisting of belief, argument, and stance, the commonsense explanation graphs are constructed through a generic create-verify-refine iterative framework. Firstly, annotators construct a commonsense-augmented explanation graph that explicitly explains the stance. Each graph comprises 3-8 facts, each of which is a triplet with two concepts as entities connected by their relations. The graphical representation allows us to automatically perform in-browser checks for structural constraints, thereby guaranteeing the structural correctness of the graph. To further ensure the semantic correctness of the constructed graph, annotators reason through the graph to infer the stance based solely on the belief and the explanation graph. Finally, for incorrect graphs, another annotator refines them either by adding a new fact, removing an existing one, or replacing an existing fact.

    \item \textbf{GSM8K}\footnote{\href{https://github.com/openai/grade-school-math}{https://github.com/openai/grade-school-math}}~\cite{cobbe2021training}: The GSM8K dataset is a collection of 8.5K high-quality, linguistically diverse math word problems designed to assess multi-step reasoning for question answering. These grade school-level problems, solvable by a bright middle school student, require between 2 and 8 steps, primarily using basic arithmetic operations ($+, -, \times, \div$) without needing concepts beyond early Algebra. Solutions are presented in natural language to facilitate general applicability, offering insight into the reasoning processes of large language models. This format highlights how models handle structured, step-by-step reasoning in response to real-world math problems.

    \item \textbf{PrOntoQA}\footnote{\href{https://github.com/asaparov/prontoqa}{https://github.com/asaparov/prontoqa}}~\cite{saparov2022language}: PrOntoQA is a question-answering dataset that provides examples featuring chains-of-thought to outline the reasoning needed for correct answers. The sentences are syntactically simple and well-suited for semantic parsing, making it valuable for formal analysis of predicted reasoning chains from large language models like GPT-3.
    
\end{itemize}

\subsubsection{Tools}
\begin{itemize}[leftmargin=*]
    \item \textbf{ToolBench}\footnote{\href{https://github.com/sambanova/toolbench}{https://github.com/sambanova/toolbench}}~\cite{shen2023taskbench}: Recent studies on software tool manipulation with LLMs primarily depend on closed model APIs (e.g., OpenAI), as there remains a significant performance gap between these proprietary models and available open-source LLMs. To investigate the underlying causes of this discrepancy and to advance the capabilities of open-source LLMs—particularly in tool manipulation—a benchmark named ToolBench has been developed. ToolBench includes a range of diverse software tools designed for real-world tasks and provides an accessible infrastructure for directly evaluating each model’s execution success rate.
\end{itemize}

\section{Tabular Graph \texorpdfstring{\emojitabtitle}{Tabular Graph}}
\label{sec-tab}

Tabular data is another type of structured data that is widely used in real-world applications~\citep{sahakyan2021explainable} and is typically stored in relational databases~\citep{codd2007relational}. Tabular data may consist of a single table containing samples and their attributes, or multiple tables that share primary and foreign keys. LLMs have been explored to process and solve tasks involving tabular data, primarily by transforming the tabular data into text through serialization~\citep{fang2024large, singha2023tabular}. However, such serialization may lead to some issues: (1) In the case of a single table, the feature columns should exhibit permutation invariance, but serializing the table data may disrupt this invariance; (2) For multiple tables, one table may be connected to another through primary/foreign keys, and leveraging these relationships is crucial for various tasks. Therefore, graph structures can be a suitable representation for tabular data. Additionally, when tables contain too many rows to fit within the context window of LLMs, graphs can facilitate efficient retrieval.

\subsection{Application Tasks}
Tables are widely used in real-world scenarios to store features and relationships between data. Understanding the structure of tabular data is crucial. As a result, there are numerous tasks that can benefit from the use of tabular graphs and below we list a few representative ones. 

\begin{itemize}[leftmargin=*]
    \item \textbf{Node-level tasks}: Node-level tasks include node classification and node regression, applied to tasks like cell type prediction~\citep{jin2023tabprompt}, fraud detection~\citep{rao2020xfraud, singh2023graphfc}, outlier detection~\citep{goodge2022lunar}, and click-through rate (CTR) prediction~\citep{li2019fi, du2022learning}.

    \item \textbf{Link-level tasks}: Link-level tasks involve link prediction, edge classification, and edge regression. Many tabular data tasks can be modeled as link-level tasks, such as data imputation~\citep{zhong2023data, you2020handling} and recommendation~\citep{robinson2024relbench}.

    \item \textbf{Graph-level tasks}: Graph-level tasks aim to predict the properties of the entire graph, such as table type classification and table similarity prediction~\citep{wang2021tuta, jia2023getpt, jin2023tabprompt}.

    \item \textbf{Table question answering}: Table QA involves generating answers by understanding and reasoning over tabular data. This task requires comprehension of both the content and their relationships within tables, making graph structures suitable for encoding such information. For example, \citet{zhang2020cfgnn, zhang2020graph} utilize graphs to enhance Table QA.

    \item \textbf{Table retrieval}: Table retrieval focuses on retrieving semantically relevant tables based on natural language queries~\citep{wang2021retrieving, cheng2021hitab}.
\end{itemize}

\subsection{Tabular Graph Construction}
Graphs are used in tabular data learning to model high-order feature interactions, high-order instance relationships, and relationships between instances across multiple tables. There are typically two types of nodes: instance nodes, which represent each row of a table, and feature nodes, which represent individual features. Generally, the following graphs are constructed:

\begin{itemize}[leftmargin=*]
    \item \textbf{Instance Graph}: An instance graph connects a table's rows (instances), modeling the relationships between instances. It is particularly useful for retrieving relevant instances within tabular data. There are mainly two approaches for constructing instance graphs:
    
    \begin{itemize}[leftmargin=*]
        \item \textit{Rule-based methods}: Instances are connected based on predefined rules. For example, heuristics derived from expert knowledge can be used to connect instances that share certain features~\citep{lu2021weighted}. Expert knowledge can be leveraged for the graph construction~\citep{parisot2018disease}. Another common approach is similarity-based methods, such as connecting instances through K-Nearest Neighbors (KNN)~\citep{gu2022structure, errica2024class} or based on similarity exceeding a threshold~\citep{spinelli2020missing, chen2023gedi}.
        \item \textit{Learnable methods}: In this approach, an instance graph is typically initialized using rules or heuristics. The edge weights are then adjusted during the learning process, allowing the model to dynamically refine the graph structure over time~\citep{kang2021k, liao2023tabgsl, cosmo2020latent, kazi2022differentiable}.
    \end{itemize}

    \item \textbf{Feature Graph:} A feature graph connects features, with edges representing correlations between pairs of features. Typically, feature relationships are modeled in a learnable manner~\citep{yan2023t2g, zhou2022table2graph}. Some works construct the feature graph by leveraging feature similarity~\citep{guo2021tabgnn, li2019fi}, while others use heuristics based on expert knowledge to establish connections~\citep{li2024graph}. Additionally, certain approaches link features if they belong to the same instance~\citep{rao2020xfraud}.

    \item \textbf{Instance-Feature graphs:} The instance-feature graph is a heterogeneous graph, which connects the instance with their corresponding features~\citep{guo2021dual, you2020handling, wang2021retrieving, wu2021towards}. 

    \item \textbf{Cell Graph:} Cell graphs treat each cell in the table as a node. \citet{xue2021tgrnet} build a cell graph, where each cell node contains both spatial and logical locations attributes while the adjacency matrix represents the neighbor relation or the same-row and same-column relation between two cells.

    \item \textbf{Tabular Hypergraph:} A hypergraph is a generalization of a graph in which an edge, known as a hyperedge, can connect multiple nodes. Since tables are often invariant to arbitrary row and column permutations, a hypergraph is a suitable structure for modeling tabular data~\citep{zahradnik2023deep}. Specifically, \citet{chen2024hytrel} model each row and column in a table as a hyperedge for pretraining purposes. Additionally, \citet{du2022learning} construct a hypergraph from tabular data to capture relationships between instances effectively.

    Additionally, \citet{cvetkov2023relational} model table as a knowledge graph, where the feature values or rows represent nodes, and the column names represent the relations. \citet{zhang2020graph} build a heterogeneous graph based on various predefined relations. \citet{wang2021tuta} construct trees for non-relational tables based on the hierarchical relations. 

    Previous approaches primarily focus on constructing graphs based on a single table. However, in relational databases, there are often multiple tables. A common method is to first merge these tables into a single table, and then apply the previously mentioned methods to the merged table~\citep{kanter2015deep}. However, this approach relies on manually joining the tables as part of a feature engineering step, which requires significant effort and substantial domain expertise. In the following, we will introduce cross-table graphs, which directly build a graph from multiple tables without the need for manual merging.

    \item \textbf{Cross-table Graphs:} Cross-table graphs for tabular data typically connect different instances across multiple tables, where nodes usually represent rows in each table and edges are defined by primary-foreign key relationships~\citep{fey2023relational, cvitkovic2020supervised, zhang2023gfs}. Additionally,\citet{wang20244dbinfer} propose Row2N/E: if a table has a primary key (PK), each row is treated as a node; however, if there are also two foreign key (FK) columns, each row is instead represented as an edge. These graphs model relationships across multiple tables, enabling the integration of information from different perspectives to facilitate various tasks.

    Additionally, \citet{bai2021atj} build a hypergraph based on multiple tables, where the primary or foreign key as nodes, and the nodes within the same table as hyperedge.

\end{itemize}

\subsection{Retriever}
Modeling tabular data into graphs is still in its early stage, and most works follow popular retrieval methods, as introduced in Sections~\ref{sec:retriver}. Additionally, \citet{fey2023relational} retrieve subgraphs based on timestamps to construct time-consistent computational graphs, while \citet{cvitkovic2020supervised} adopt a deterministic heuristic to select subgraphs.

\subsection{Generator}
Most existing methods that leverage tabular graphs still rely on GNNs or Graph Transformers as generators~\citep{li2024graph}. Besides, \citet{wang20244dbinfer} fuse both GNNs and tabular based predictors, such as DeepFM~\citep{guo2017deepfm}, FT-Transformer~\citep{gorishniy2021revisiting}, XGBoost~\citep{chen2016xgboost} and AutoGluon~\citep{erickson2020autogluon}. \citet{ivanov2021boost} jointly train gradient boosted decision tree (GBDT) and GNN. Recent advancements have seen the application of LLMs to tabular data tasks. This is typically achieved by converting tabular data into sequential formats suitable for LLM input, as explored by \citet{fang2024large}, \citet{dong2024large}, and \citet{sui2024table}. However, this transformation can lead to the loss of inherent structural information present in the original tabular format. With advancements in the application of LLMs to graph data, LLMs can be further enhanced to handle various tasks more effectively with the support of tabular graphs.

\subsection{Resources and Tools}
In this section, we list some data sources and tools for GraphRAG on tabular graphs.

\subsubsection{Data Resources}
Relational machine learning has recently gained popularity, leading to the development of several benchmarks and datasets for tabular data. For example:

\begin{itemize}[leftmargin=*]
    \item \textbf{RelBench}~\citep{robinson2024relbench, fey2023relational}\footnote{https://relbench.stanford.edu/} is a collection of realistic, large-scale, and diverse benchmark datasets for machine learning on relational databases. It includes various tasks, such as Node-level prediction tasks (e.g., multi-class classification, multi-label classification, regression), Link prediction tasks, Temporal and static prediction tasks. The benchmark features several datasets, including rel-amazon, rel-stack, rel-trial, rel-f1, rel-hm, rel-event, and rel-avito, providing a robust platform for evaluating models across multiple relational data scenarios and tasks.
    
    \item \textbf{TabGraphs}~\citep{bazhenov2024tabgraphs} evaluates various models — including simple baselines, tabular models, and GNNs — on graphs with tabular node features. This benchmark includes several TabGraphs datasets: tolokers-tab, questions-tab, city-reviews, browser-games, hm-categories, hm-prices, city-roads-M, city-roads-L, avazu-devices, web-fraud and web-traffic. 
    \item \textbf{Tabular-benchmark}~\citep{grinsztajn2022tree} \footnote{https://github.com/LeoGrin/tabular-benchmark} benchmarks 45 tabular datasets from diverse domains, comparing the performance of several tree-based and deep learning models. They evaluate tasks such as numerical classification, numerical regression, categorical classification, and categorical regression. 
    \item \citet{shwartz2022tabular} benchmark 11 tabular datasets, such as 
    MSLR~\citep{qin2013introducing},  Forest Cover Type, Higgs Boson, and Year Prediction~\citep{dua2017uci}.
    evaluating several tree-based models, deep learning models, and ensemble models.
    \item \textbf{DBInfer Benchmark}~\citep{wang20244dbinfer}\footnote{https://github.com/awslabs/multi-table-benchmark}  is a set of benchmarks for measuring machine learning solutions over data stored as multiple tables. It includes several large-scale relational database (RDB) datasets—such as AVS, OB, DN, RR, AB, SE, MAG, and SE—with tasks like retention, CTR, purchase, CVR, churn, rating, popularity, venue prediction, citation, charge, and prepay. The benchmark provides implementations of various baselines, including popular tabular models with and without auto-feature-engineering methods, as well as Graph Neural Networks, making it a robust resource for multi-table learning evaluations.
    \item \textbf{RTDL}\footnote{https://github.com/yandex-research/rtdl}: RTDL (Research on Tabular Deep Learning) is a collection of papers and packages on deep learning for tabular data. It provides several popular tabular deep learning models.
    \item \textbf{OpenML}~\footnote{https://www.openml.org/} is an open platform for sharing datasets, algorithms, and experiments. It provides a large collection of machine learning datasets across various domains, including many tabular datasets.
    \item \textbf{Kaggle}~\footnote{https://www.kaggle.com/} hosts a wide array of datasets for data science competitions, making it a valuable resource for benchmarking tabular machine learning models on diverse real-world data.
    \item \textbf{UC Irvine Machine Learning Repository}~\citep{dua2017uci}\footnote{https://archive.ics.uci.edu/} is a collection of databases, domain theories, and data generators that are used by the machine learning community for the empirical analysis of machine learning algorithms.
    \item \textbf{HiTab}~\citep{cheng2021hitab} is a hierarchical table dataset for question answering and natural language generation.
\end{itemize}

\subsubsection{Tools}
In this subsection, we list several tools that provide essential functionalities for data preparation, model training, and evaluation.  These tools can be effectively leveraged for GraphRAG applications on tabular graphs.

\begin{itemize}[leftmargin=*]
    \item \textbf{PyTorch Tabular}~\citep{joseph2021pytorch}\footnote{https://github.com/manujosephv/pytorch\_tabular}: PyTorch Tabular is a powerful library built on PyTorch designed to simplify the application of deep learning techniques to tabular data. It provides several data preprocessing functions, including normalization, standardization, encoding of categorical features, and dataloader preparation. Additionally, it includes a variety of tabular machine-learning models and evaluation functions, making it a comprehensive tool for handling tabular data.
    \item \textbf{DeepTables}~\citep{deeptables}\footnote{https://github.com/DataCanvasIO/DeepTables}: DeepTables is an easy-to-use toolkit that harnesses the power of deep learning for tabular data. Built on TensorFlow, it is designed to address classification and regression tasks on tabular data. DeepTables offers a variety of tabular models and supports AutoML.
    \item \textbf{PyTorch Frame}~\citep{hu2024pytorch}\footnote{https://github.com/pyg-team/pytorch-frame}: PyTorch Frame is a deep learning extension for PyTorch, designed for heterogeneous tabular data with different column types, including numerical, categorical, time, text, and images. t provides a wide range of tabular models and supports integration with diverse model architectures, including Large Language Models.
\end{itemize}

\section{Other Domains}\label{sec-other}
While substantial GraphRAG research has been investigated in previous domains, GraphRAG remains significantly underexplored in other domains such as infrastructure, biology, and scene. Therefore, we combine them into a single comprehensive section, focusing only on key studies in these fields.

\subsection{Infrastructure Graph \texorpdfstring{\emojiinfratitle}{Infrastructure Graph}}
\label{sec-infra}

%\hy{Strada-LLM: Graph LLM for traffic prediction}
%\yu{GenAI-powered Multi-Agent Paradigm for Smart Urban Mobility: Opportunities and Challenges for Integrating Large Language Models (LLMs) and Retrieval-Augmented Generation (RAG) with Intelligent Transportation Systems}

Infrastructure graphs, defined by Points of Presence (PoPs) interconnected through physical links, play a vital role in serving our daily activities across sectors such as power, water, gas, transportation, and communication. Due to the scarcity of GraphRAG research in infrastructure graphs, our review mainly focuses on graph construction methods and tasks alongside only a brief overview of recent (Graph)RAGs in this domain.

\begin{itemize}[leftmargin=*]
    \item \textbf{Power Networks}~\cite{zimmerman2010matpower, owerko2020optimal, liao2021review}: In power networks, nodes represent critical entities such as power plants, substations, transformers, and load points, while edges correspond to transmission and distribution lines facilitating the flow of electrical power across the network. Each node possesses features like power generation capacity, voltage level, load demand, geographical location, reliability metrics, and operational status, capturing the network's operational and spatial aspects. Edges are characterized by features such as transmission capacity, impedance, length, voltage level, and real-time power flow, all essential for analyzing line performance and flow efficiency. This graph-based representation enables a range of essential tasks, including transformer fault diagnosis, power outage prediction, power flow approximation, and power generation optimization.

    \item \textbf{Water Networks}~\cite{xing2022graph}: In water networks, nodes are junctions (pipe connections or users) and sources (reservoirs and tanks), and edges are pipes with water flow. Basic tasks include estimating node water heads and flow in all pipes given the water network layout, pipe characteristics, nodal demands, reservoir levels, and head measurements at limited locations~\cite{candelieri2014graph}.

    \item \textbf{Gas Networks}~\cite{wang2024multiscale, yang2023graph}: nodes represent connection points, gas sources, storage facilities, compressor stations and users, and edges edge respected natural gas pipelines, carrying user loads, diversion, and input. 

    \item \textbf{Transportation Networks}~\cite{wang2020traffic, jiang2022graph}: Road networks, where intersections serve as nodes and road connections as edges, are commonly used to model spatial dependencies for traffic flow and speed forecasting. Similarly, metro and bus networks use stations as nodes with routes as edges, capturing station topology. Region-level problems involve dividing cities into regular or irregular regions, represented as graph nodes, where spatial dependencies reflect land use patterns; regular regions often use grid partitions, while irregular ones use diverse methods like road- or zip-code-based partitions. At the road level, sensor, segment, intersection, and lane graphs capture different road elements, with nodes representing sensors, segments, intersections, or lanes. For station-level networks, nodes represent various transportation hubs (subway, bus, bike, or car-sharing stations) linked by natural connections like subway or road networks, creating an interconnected spatial dependency graph.

    \item \textbf{Communication Networks}~\cite{ferriol2023routenet, rusek2020routenet}: A computer network system typically comprises two interconnected graph layers: the logical and the physical. The logical layer represents the flow of data, where traffic is routed through multiple intermediate routers before reaching its destination. In this layer, nodes correspond to routers, and edges represent the logical paths that data packets traverse. Conversely, the physical topology layer refers to the underlying infrastructure, with Points of Presence (PoPs) from providers such as Comcast and Verizon serving as nodes and the physical fiber links between them forming the edges. Studies focused on the logical layer often model networks as graphs, where nodes represent devices (e.g., switches, routers) and edges denote links (e.g., fiber-optic cables) or traffic paths~\cite{ji2020traffic}. This graph-based representation captures both the forwarding behavior of the network (i.e., traffic interactions) and its connectivity (i.e., topological behavior), offering valuable insights for network management. For instance, optical topology-aware traffic engineering has proven effective in mitigating issues like fiber cuts and flash crowds. For the physical layer, much previous research has focused on inferring the complete as well as aligning the physical and logical layers.
\end{itemize}

To summarize, the tasks operated on infrastructure networks could be broadly encompassed as utility prediction (e.g., flow and node performance), flow simulation and generation, vulnerability analysis, and network maintenance and operation. Managing and understanding the complex physical relationships within these graphs is essential for improving infrastructure management in key areas like service delivery, function forecasting, and network optimization. In computer networks, for instance, the state of a logical traffic routing path is directly influenced by the underlying physical fiber links. Predicting critical metrics like traffic delay and jitter depends on assessing the status of all involved fiber links~\cite{ferriol2023routenet, rusek2020routenet}, as these metrics are highly interdependent. The intricate physical connections in infrastructure graphs create valuable opportunities for employing GraphRAG techniques to enhance infrastructure management and optimization. Although no existing works have explored GraphRAG in infrastructure, few works have leveraged RAG, which we briefly review in the next.

\citet{hussien2024rag} explore the use of RAG to empower automated vehicles with the ability to anticipate pedestrian and driver behaviors, including pedestrian road-crossing actions and driver lane changes. RAG retrieves explanatory documents from the JAAD and PSI datasets to provide insights into pedestrian behavior. Additionally, \citet{qian2024netbench, wang2024lens, liu2024large, bariah2024large} build on recent advances in generative models for vision and language, proposing to harness the capabilities of LLMs within computer networking. A systematic approach has been proposed to develop a foundation model for traffic tasks, such as traffic classification and generation, treating packet hexadecimal bytes as tokens. Furthermore, the potential of generative AI in advancing telecom and networking is reviewed. Specifically, \citet{kotaru2023adapting} introduces an "operator’s copilot," a natural language interface that leverages LLMs for efficient data retrieval. This interface helps manage thousands of counters and metrics, minimizing the need for specialists and accelerating issue resolution through data intelligence. These pioneering works in leveraging generative AI for computer networking lay a good foundation for rebuilding (Graph)RAG for infrastructure networks.

\subsection{Single-cell Graph \texorpdfstring{\emojibiotitle}{Biological Graph}}

Single-cell sequencing allows for the detailed analysis of molecular traits at the level of individual cells. For example, single-cell RNA sequencing (scRNA-seq)~\cite{kolodziejczyk2015technology} quantifies RNA transcript levels, offering valuable information about cell identity, developmental stages, and functional characteristics. 
Single-cell assay for transposase-accessible chromatin sequencing (scATAC-seq) ~\cite{buenrostro2013transposition} records the number of reads per accessible chromatin region, resulting in highly dimensional data matrices containing hundreds of thousands of genomic regions. 
With the explosion of the number of single-cell data, a variety of deep learning approaches ~\cite{molho2024deep, ding2024dance, tang2023single, wang2023mem, wen2023single} have been developed in recent years to tackle single-cell analysis, especially GNN methods ~\cite{lu2024graph, ding2024spatialctd, wen2022graph, shao2021scdeepsort, wangSinglecellClassificationUsing2021} have been applied to various downstream tasks. In the task of cell type annotation,
sigGCN ~\cite{wangSinglecellClassificationUsing2021} builds a gene-wise weighted adjacency matrix using data from the STRING database \cite{szklarczykSTRINGV11Protein2019} to establish a gene interaction network, with the node features representing corresponding gene expression levels. This graph is fed into a GCN-based autoencoder, which includes a convolutional layer and a max-pooling layer, followed by a flattened layer and a fully connected (FC) layer, to perform cell type annotation. 
scDeepSort \cite{shao2021scdeepsort} constructs a weighted bipartite graph where both cells and genes serve as nodes, with the edge weights representing the gene expression values for each cell-gene pair. Gene node features are derived through principal component analysis (PCA), while cell node features are obtained by aggregating the weighted features of the connected gene nodes.
The most common way to construct a single-cell graph is to build a KNN graph from single-cell data~\cite{wang2021scgnn}. To be specific, the data is first normalized to make it comparable across different cells. Next, dimensionality reduction methods, such as principal component analysis (PCA)~\cite{mackiewicz1993principal}, are employed to reduce the dataset’s high dimensionality while retaining key information. In this lower-dimensional space, distances between cells are calculated, often using Euclidean distance. Each cell’s K nearest neighbors are then determined, and a graph is formed where cells are represented as nodes, and edges connect each cell to its $K$ closest neighbors.

\textbf{Multi-omics single-cell graphs}:
% Multiome ATAC + Gene Expression; CITE-seq
Multi-omics single-cell technologies ~\cite{baysoy2023technological} integrate multiple layers of biological information—such as gene expression (RNA), chromatin accessibility (ATAC), and protein data—at the resolution of individual cells. Multiome~\cite{stuart2019integrative} refers to the simultaneous measurement of multiple molecular modalities, such as gene expression (RNA) and chromatin accessibility (ATAC), from the same single cell while CITE-seq (Cellular Indexing of Transcriptomes and Epitopes by Sequencing) ~\cite{mimitou2019multiplexed} is a multi-omics technology that allows for the simultaneous measurement of gene expression and protein surface markers at the single-cell level. 
Taking multiome dataset as an example, scMoGNN~\cite{wen2022graph} builds a heterogeneous graph consisting of cell nodes, gene nodes, and peak nodes. The edge between a cell node and a gene node indicates the expression level of that gene in the cell, while the edge between a cell node and a peak node reflects the expression level of that peak in the cell. Additionally, a threshold can be applied to filter and select the connecting edges. 
Additionally, pathways and gene activity can be incorporated as prior knowledge to establish connections between gene nodes, and connections between peaks and gene nodes respectively.
This graph representation captures both intra-modality (within each modality, like gene expression or chromatin accessibility) and cross-modality (between gene expression and chromatin accessibility) relationships, providing a comprehensive view of regulatory dynamics at the single-cell level

\textbf{Spatial transcriptomics single-cell graphs}: 
Unlike traditional transcriptomics, spatial transcriptomics retains the precise location of spot representing a cell or small group of cells within the tissue. 
Spatial transcriptomics primarily consists of three key data sources: gene expression, which captures transcript levels in specific spots; spot location, providing the spatial coordinates of each spot within the tissue; and spot images, offering visual context and morphology of the tissue at each spot.
In the task of cell type deconvolution, when building the KNN graph, GNNDeconvolver~\cite{ding2024spatialctd} integrates both cell position information and gene expression data to compute the similarity between cells. SpaGCN~\cite{hu2021spagcn} additionally incorporate spot morphology to calculate similarity between spots.
Together, constructing KNN graph with such three data sources provides a richer, more holistic view of the tissue, enabling deeper insights into cellular function, interactions, and spatial organization.

\subsection{Scene Graph \texorpdfstring{\emojiscenetitle}{Scene Graph}}

Scene graphs are a data structure designed to capture spatial and semantic relationships between objects within a scene. They are widely used in computer vision, graphics, and robotics to organize scenes, particularly when multiple objects interact, or the scene contains complex interactions. For example, SceneGraphs~\cite{he2024g} is a dataset for visual question answering, containing 100,000 scene graphs that describe the objects, attributes, and relationships within images. It presents tasks that test spatial reasoning and multi-step inference by asking users to answer open-ended questions based on textual descriptions derived from scene graphs. 

G-Retriever~\cite{he2024g} processes Scene Graphs by first parsing JSON data to index objects and attributes, allowing for efficient retrieval of relevant nodes and edges based on the query. It then constructs a subgraph with the necessary scene information, filtering out unrelated details. Finally, G-Retriever uses this subgraph along with an LLM to generate answers, leveraging spatial and semantic relationships for accurate reasoning and response generation.

P-RAG~\cite{xu2024p} engages with the environment through agents, building and updating a database of historical trajectories that guide the agent’s actions. Initially, the task’s goal instruction is provided. Before each action, the agent captures an observation image reflecting the current state, which is then converted into a scene graph format to facilitate processing by LLMs. By creating and retrieving scene graph context, P-RAG enhances the LLM's ability to accurately interpret and navigate complex scenes, making this approach especially useful in planning embodied everyday tasks.

\subsection{Random Graph \texorpdfstring{\emojirandomtitle}{Random Graph}}

Random graphs are a foundational concept in network theory~\citep{newman2018networks, janson2011random} and are widely used for modeling and studying complex networks in fields such as computer science, physics, biology, and social sciences. They are constructed based on probabilistic rules, leading to various possible structures that reflect the diversity seen in real-world networks. These random graphs can be used to analyze the retrieval, organization, and generation processes within GraphRAG.

There are many models to generate random graphs, such as Erdős-Rényi Model (ER Model)~\citep{erdds1959random}, Watts–Strogatz model~\citep{watts1998collective}, Barabási–Albert model~\cite{albert2002statistical}, Random geometric graph~\cite{cannings2005random}, scale-free networks~\citep{barabasi1999emergence} and stochastic block model~\cite{holland1983stochastic}. Additionally, various graph structures, such as Path Graphs, Complete Graphs, Star Graphs, and Barbell Graphs, can be generated to meet specific analytical needs.

Random graph generation models, such as Contextual stochastic block models (CSBM)~\citep{deshpande2018contextual} has been widely leveraged in GNNs analysis~\citep{baranwal2021graph, ma2021homophily, mao2024demystifying, han2024node}. Additionally, random graphs are increasingly applied to examine the behavior of LLMs. For instance, \citet{fatemi2023talk} leverage Erdős-Rényi graphs, scale-free networks, Barabási–Albert, stochastic block model, star, path and complete graph generators to test LLM performance on various graph reasoning tasks, such as edge existence, node degree, node/edge count, connected nodes and cycle check.
GraphLLM~\citep{chai2023graphllm} generates random graphs in various input formats to assess LLM performance on graph reasoning tasks, including substructure counting, maximum triplet sum, shortest path calculation, and bipartite graph matching. In this approach, a graph transformer encodes the graph, and the embedding-fusion method is used for generation. \citet{bachmann2024pitfalls} use star graphs to investigate limitations within the next-token prediction paradigm. Other studies, such as \citet{dai2024revisiting}, \citet{wang2024can}, \citet{guo2023gpt4graph}, and \citet{luo2024graphinstruct}, explore the graph reasoning abilities of LLMs across various tasks.

\section{Challenges and Future Work}\label{sec-challenge}
After proposing the holistic framework of GraphRAG and reviewing it in each domain, we begin in this section by outlining the challenges and opportunities associated with each key component of GraphRAG, including graph construction, retriever, organizer, and generator. Then, we discuss the challenges and opportunities for GraphRAG as a holistic system with its evaluation and application. 

 \subsection{Graph Construction} 

 \begin{itemize}[leftmargin=*]
     \item \textbf{How to construct graphs?} There are numerous ways to construct graphs, yet different tasks or domains may require different graph structures. For example, deciding on the granularity of nodes and edges, as well as which entities or relationships to extract, is critical. This process may also need to address challenges such as entity disambiguation, entity alignment, and coreference resolution. Understanding when a graph structure is necessary, whether to use single or multiple graphs and how to construct them in the most appropriate way for a specific application is essential but often complex. \\
    
    \item \textbf{The format of graphs.} Graphs can be represented in various forms—are these representations equivalent, or do they offer unique advantages? Choosing the most effective representation for a given task can significantly influence performance. \\
    
    \item \textbf{Multi-modal Graphs.} Despite building text-based graphs, the retrieved resources, such as images, audio, or video, can be multi-modal. Constructing a cohesive graph from multi-modal data presents a significant challenge, as it requires integrating diverse data types while preserving meaningful relationships. \\
    
    \item \textbf{Dynamic graphs.} Many real-world scenarios involve dynamic data that evolve over time, which are essential for downstream tasks. Developing strategies to construct, update, and store graphs dynamically while maintaining both efficiency and effectiveness presents a further challenge worthy of exploration.
\end{itemize}
  %  \item \textbf{Challenge 2: Query} 
   % \yu{Deciper the intention}
    %\yu{Entity Ambiguity}

\subsection{Retriever}
\begin{itemize}[leftmargin=*]
    \item \textbf{Differentiating Neural and Symbolic Knowledge}: Graph-structured data usually involves two distinct types of knowledge: symbolic-formatted knowledge, such as relations in knowledge graphs, and neural-formatted knowledge, such as entity names. Developing techniques to differentiate these two types of knowledge and designing corresponding retriever strategies for retrieving these two types of knowledge is worthy of further investigation.

    \item \textbf{Harmonization between internal and external knowledge}: Since GraphRAG is often applied when internal knowledge alone is insufficient to address the task, assessing any overlap between internal and external knowledge using a robust calibration method is crucial. In addition, external knowledge may sometimes conflict with internal knowledge. To handle this, it is essential to design an effective knowledge validation and reconciliation mechanism, allowing selective retrieval and updating of external content as needed.

    \item \textbf{Trade-off among Accuracy, Diversity, and Novelty}: Real-world user retrieval often involves complex intentions. Beyond delivering accurate content to ensure high utility—such as achieving high question-answering accuracy—there may also be a demand for diversity and novelty in the retrieved information. Balancing retrieved content's accuracy, diversity, and novelty remains an open-ended challenge.

    \item \textbf{Reasoning, planning, and thinking along the way}: A real-world retriever may need to dynamically and adaptively update its retrieve process in response to both the initial query and the content retrieved along the way. The question of how to equip the retriever with these adaptive thinking, reasoning, and planning abilities remains a big problem.

\end{itemize}
    
\subsection{Organizer} 
\begin{itemize}[leftmargin=*]
    \item \textbf{Balancing Completeness and Conciseness.} The retrieved graphs may be large, potentially containing significant information that is not related to the query. The Organizer should balance the need for complete information with the risk of overwhelming the model. This requires techniques for pruning irrelevant nodes and edges while retaining essential context, which is especially challenging in large or noisy graphs. Additionally, some knowledge may already be captured by LLMs, raising the question of how to identify and remove redundant information to further improve efficiency. \\ 
    
    \item \textbf{Optimal Data Structuring:} Determining the most effective way to structure the retrieved content is a challenge. For example, deciding how to convert a structured graph into a format that the generator can leverage, how to arrange the order of retrieved content, and how to preserve the original data structure for structure-sensitive tasks are all important considerations. Different tasks may benefit from different structuring methods. \\
    
    \item \textbf{Aligning Different Resources:} Retrieved content may come from various sources and in diverse formats, such as text, graphs, and images. Aligning these components effectively to help the generator poses a significant challenge, especially when integrating multiple data modalities. \\
    
    \item \textbf{Data Augmentation: } Incorporating data augmentation within the Organizer involves enriching the retrieved graph content with nodes, edges, or features to improve the robustness of the model and improve downstream tasks. However, this process requires balancing real and augmented data to avoid introducing irrelevant or redundant information.
\end{itemize}  
    
\subsection{Generator}
\begin{itemize}[leftmargin=*]
    \item \textbf{Correct Format for Prompting.} The content retrieved after organization varies significantly in format — such as texts, triplets, or graphs — while current LLMs can only process text inputs. Exploring the most effective format for optimal LLM performance on specific tasks and designing more flexible generators that go beyond text input could be worthwhile avenues for research.

    \item \textbf{Structural Encoding.} When the retrieved content is a subgraph and the downstream generator is an LLM, ensuring that the LLM can interpret the graph's structural information is essential. Designing effective structural encodings and integrating them into token embeddings pose a key challenge. Although some previous works have incorporated various graph encodings into the text decoding process, no systematic study has yet demonstrated whether LLMs can distinguish between these encodings and accurately recognize their corresponding geometric structures.
\end{itemize}

\subsection{GraphRAG as a System} Rather than individual components, GraphRAG is a system. Designing an efficient and cohesive GraphRAG system presents additional challenges: \\
\begin{itemize}[leftmargin=*]
    \item \textbf{Integration Across Components:} Ensuring seamless interaction among the Graph Construction, Retriever, Organizer, and Generator components is essential. Each part must operate harmoniously to maintain efficiency and accuracy, which can be challenging to optimize the system globally. \\
    \item \textbf{Scalability:} As data volumes and query demands increase, each component of the GraphRAG system, such as Graph Construction, Retriever, Organizer, and Generator, must efficiently handle larger, more complex graphs without compromising performance.For example, efficient graph storage, optimized querying (e.g., subgraph sampling and pathfinding), streamlined organization of retrieved components, and responsive generation are all essential. Additionally, training, serving, fine-tuning, and evaluating within GraphRAG demand sophisticated engineering on modern hardware and software stacks, including efficient training, data-efficient fine-tuning, communication-efficient algorithms, implementation of reinforcement learning with human feedback, GPU acceleration and other specialized hardware, model compression for deployment, and online maintenance. \\
   % \textbf{3) Security:} As a system-level concern, security involves ensuring that GraphRAG is protected from potential misuse.  Key security challenges include safeguarding against attempts to recover sensitive training data embedded within models and defending against adversarial inputs designed to manipulate models into producing target outputs. Developing robust defenses against these vulnerabilities is essential for maintaining the integrity and confidentiality of the GraphRAG system. \\
   % \textbf{4) Trustworthiness and Explainability:} For users to trust GraphRAG outputs, the system should provide insights into how answers were derived. This is especially challenging in complex graph-based retrieval and generation systems, where multiple relationships and entities contribute to a response. Ensuring transparency and explainability is crucial, particularly for applications in sensitive domains like healthcare, finance, and law. \hy{May move this part to the Trustworthiness}
    \item \textbf{Trustworthiness:}
    Real-world GraphRAG systems often operate in high-stakes domains such as education~\cite{Thway2023Battling}, healthcare~\cite{Xiong2023Benchmarking, Zakka2023Almanac}, and law~\cite{Wiratunga2023CBR}, which imposes multiple desires when deploying GrapRAG systems. Therefore, deeply understanding the broader desires beyond merely optimizing utility is essential to ensure effective deployment. As motivated by trustworthy and safety research in other fields~\citep{chen2024fairness, huang2024position, wang2024safety}, these key goals beyond utility include reliability, robustness, fairness, and privacy. Although previous works have initiated the exploration of this multi-purpose optimization of RAG~\cite{wagle2023empirical, zhou2024trustworthiness}, they mostly focus on RAG without dedicated investigation of how the relational information captured by GraphRAG could cause additional trustworthy and safety concerns. 
    \begin{itemize}
        \item \textbf{Reliability}~\cite{kumar2023conformalpredictionlargelanguage, quach2024conformallanguagemodeling, su2024apienoughconformalprediction, ye2024benchmarkingllmsuncertaintyquantification, li2023traq, ni2024towards}: Reliability requires the system to deliver consistently low error rates across different scenarios. In RAGs, uncertainty quantification presents two primary challenges: Firstly, during the generation phase, uncertainty stems from the inherent probabilistic generation of LLMs. Standard techniques such as conformal prediction have been applied here with few adaptations~\cite{kumar2023conformalpredictionlargelanguage, quach2024conformallanguagemodeling, su2024apienoughconformalprediction, ye2024benchmarkingllmsuncertaintyquantification}. The general idea is to estimate the non-conformal score based on the calibration set and filter out those high-risk answers during the testing phase. Secondly, the retriever and its interaction with LLMs (e.g., treating LLMs as the reasoning agent to perform adaptive retrieval) also possess uncertainty~\cite{li2023traq, ni2024towards}. Unlike RAGs, retrievers and generators in GraphRAG demonstrate the multi-hop nature, which raises new requirements for multi-stage uncertainty quantification. For example, the error rate of one-step generation might accumulate after multi-step generation, and how to calibrate this accumulated error at the global level is the key challenge. The very first work~\cite{ni2024towards} proposed the learn-then-test framework to guarantee the global error rate. However, this solution may not consider the uncertainty of human-LLM interactions. Moreover, the additional learn-then-test component takes additional time and computational load for calibration. Future research aims to develop methods that can quantify and calibrate the uncertainty of (Graph)RAG as a system holistically.

        \item \textbf{Robustness}~\cite{fang2024enhancing, xiang2024certifiably, xu2024unveil, yoran2023making, dziri2024faith, zhang2024darg, zhu2024dyval}: Robustness aims the system to deliver equal-quality response under extreme scenarios such as the presence of noisy and irrelevant content. Existing research on RAGs has highlighted the significant degradation in LLMs' performance when retrieved content includes noise or irrelevant information~\cite{yoran2023making, fang2024enhancing}. Some studies address this challenge by adversarially training LLMs in noisy environments. However, there has been limited exploration of LLMs' robustness from the structural perspective. For example, would the LLM reasoning performance change when the underlying reasoning graph gets perturbed~\cite{dziri2024faith, zhang2024darg, zhu2024dyval}? Addressing these issues requires a deeper focus on the interplay between structural integrity and the robustness of GraphRAG systems, paving the way for robust performance in tasks requiring complex reasoning and planning.

        % the out-of-distribution not only happens at the instance level but also at the relation level. For example, \citet{zhang2024darg} has shown that the mathematical reasoning capability of LLMs gradually decreases as the reasoning graph structure (e.g., graph depth and width) becomes more complex. Beyond generalizability as one aspect of robustness, this relational information provides malicious actors additional channels to design adversarial attacks, such as injecting malicious passage nodes to dismantle graph-structured data sources and misguide graph traversal. 

        \item \textbf{Safety}~\cite{zou2024poisonedragknowledgecorruptionattacks, xue2024badrag, wang2024poisonedlangchainjailbreakllms, deng2024pandorajailbreakgptsretrieval, xiang2024certifiably}: Recent studies have highlighted that LLMs are susceptible to various adversarial attacks. Techniques such as prompt manipulation, hint injection, and input perturbation enable attackers to bypass safety mechanisms and exploit vulnerabilities, posing substantial social risks. RAGs that combine the power of LLMs with external databases introduce unique safety challenges. Existing works in RAGs have developed several threats (e.g., adversarial passage injection~\cite{zou2024poisonedragknowledgecorruptionattacks}, group-query targeting~\cite{xue2024badrag}, and jailbreak attacks~\cite{wang2024poisonedlangchainjailbreakllms, deng2024pandorajailbreakgptsretrieval}) and defense strategies (such as isolate-then-aggregate strategy~\cite{xiang2024certifiably}). However, all of them have overlooked the structural vulnerabilities inherent in graph-structured data. For instance, attackers can exploit the network structure of graph-based databases by leveraging principles from network science to design highly impactful attacks. Attackers can maximize their influence on the system's behavior by strategically targeting nodes with specific structural properties, such as high degree or centrality. This highlights the need for robust defenses that account for both content-based and structural threats in RAG systems.

        \item \textbf{Privacy}~\citep{chen2024survey, zafar2023building, qian2017social}: Privacy is a critical concern in RAG systems~\citep{zeng2024good, zhou2024trustworthiness}, especially when operating in domains involving sensitive or personal data, such as healthcare, education, or finance~\citep{yao2024survey, das2024security, murali2023towards}. Unlike traditional RAG systems, GraphRAG introduces additional privacy risks due to the relational nature of graphs, which may inadvertently reveal private information through connections or patterns. For instance, even if a sensitive node is protected, it may still be retrieved indirectly via its neighbors. Additionally, in graphs exhibiting homophily—where connected nodes tend to share similar attributes—sensitive information can be inferred from neighboring nodes. The use of GNNs for graph encoding further exacerbates these risks. The message-passing mechanism in GNNs propagates features across nodes and edges, potentially leading to the leakage of sensitive or confidential information during the propagation process~\citep{zhang2024survey}.  Addressing these challenges requires advanced privacy-preserving techniques that account for the interconnected nature of graphs, rather than treating the data as independent and identically distributed (iid).  Privacy protection must be designed across the entire GraphRAG pipeline, from graph construction to generation, ensuring that sensitive information remains secure while preserving the graph's utility for downstream tasks.

        \item \textbf{Explainability}: Explainability is essential for fostering trust in RAG systems, especially in high-stakes domains like law, healthcare, and finance~\citep{zhao2024explainability, cambria2024xai, yu2023temporal, elsborg2023using}. Compared to traditional RAG, GraphRAG offers enhanced explainability through the explicit relationships encoded between nodes in the graph~\citep{abu2024supporting}. These explicit connections allow the system to generate clear and interpretable explanations tailored to specific tasks, making it more transparent and trustworthy for end users. For example, in the mutli-hop question answering, the GraphRAG can be leveraged to generate reasoning paths to the question~\citep{luo2023reasoning, keqing, sunthink, jiang2024hykge}. These paths provide step-by-step explanations, allowing users to understand how intermediate conclusions were derived and verify the relevance and correctness of the system’s logic. Similarly, in molecular property prediction, specific subgraphs representing functional groups or structural motifs can be utilized to explain predictions, linking molecular features to observed properties or behaviors. However, achieving such explainability requires GraphRAG to retrieve reasonable and contextually relevant subgraphs, which remains a challenge. Ensuring that explanations remain faithful to the underlying model logic while balancing comprehensiveness and simplicity will be critical challenges for future research. 

        % \item \textbf{Accountability}
    \end{itemize}
\end{itemize}

\subsection{ Evaluation of GraphRAG.} Evaluating the performance of a GraphRAG system is challenging due to its complex, multi-component nature. Standard benchmarks may not fully capture the nuances of graph-based construction, retrieval, organization, and generation, so tailored benchmarks are essential to assess each component and their overall impact on the system. \\ 

\begin{itemize}[leftmargin=*]
    \item \textbf{Component-Level Optimality:} Each component’s performance directly influences the GraphRAG system as a whole. Evaluating each component, such as Graph Construction, Retriever, Organizer, and Generator, requires specific designs suited to their unique roles. This may involve constructing different datasets and evaluation metrics that align with the intended function of each component.
    
    \item \textbf{End-to-End Benchmarks:}  To assess the system’s overall effectiveness, comprehensive end-to-end benchmarks are essential. These should evaluate the quality of generated outputs, system response time, efficiency, and resource utilization, providing a holistic view of GraphRAG’s performance in real-world applications.
    
    \item \textbf{Task and Domain-Specific Evaluation:} Different tasks and domains may impose unique requirements on the GraphRAG system, necessitating specialized designs for each component. Methods that perform well in one domain may not generalize effectively to others, highlighting the need for diverse benchmarks. Additionally, Multi-task and multi-domain evaluations help determine the system’s adaptability and effectiveness across varied contexts.
    
    \item \textbf{Trustworthiness benchmark.} Ensuring the trustworthiness of the GraphRAG system is critical, especially in applications where decisions rely heavily on accurate and unbiased information. A trustworthiness benchmark should evaluate aspects such as robustness against adversarial inputs, data privacy protection, and transparency in how answers are derived, ensuring that outputs are explainable and reliable.
\end{itemize}

\subsection{New Applications} While we have introduced applications of GraphRAG in several domains and tasks, there are still many domains that can leverage GraphRAG, such as Code generation~\cite{liu2024codexgraph} and robust cyber defense~\cite{rahman2024retrieval}. Extending its use to other domains is promising but presents unique challenges. Determining how to adapt GraphRAG effectively for new fields requires understanding the specific requirements, data structures, and graph configurations unique to each application area. Developing tailored strategies to construct, retrieve, organize, and generate data for diverse domains is essential for maximizing GraphRAG’s adaptability and effectiveness.

\section{Conclusion}\label{sec-conclusion}

In this survey, we first introduced the demand and rationale behind GraphRAG, highlighting its ability to enhance retrieval-augmented generation by integrating graph-structured information. We then unified the architectural designs of existing GraphRAG approaches into a holistic framework, comprising five key components: graph construction, retriever, organizer, generator, and data source. For each component, we reviewed representative techniques. Given the diversity of graph structures and their applications across different domains, we also explored GraphRAG designs tailored to specific domains. By reviewing GraphRAG applications across varied domains—from knowledge graphs and document graphs to scientific and social graphs, we illustrated how its flexibility allows it to meet unique demands and address a wide range of tasks. Finally we discuss challenges and opportunities that have the potential to push the boundaries for GraphRAG. 

%However, implementing an efficient and effective GraphRAG system presents unique challenges, as discussed across its different components: graph construction, retrieval optimization, organization, and generation. Each component introduces distinct demands, and the system’s overall effectiveness relies on seamless integration and adaptability to diverse data types, domains, and tasks. We outlined critical evaluation considerations, including component-level and end-to-end benchmarks, as well as domain-specific assessments to ensure performance across various contexts. Additionally, we discussed challenges related to trustworthiness and the need for expanding GraphRAG’s applications, highlighting areas for future research and development.

%The potential applications of GraphRAG span multiple domains, from question answering and recommendation systems to complex planning and reasoning tasks and knowledge-based fields like healthcare and finance. As this technology evolves, developing domain-adaptive and efficient GraphRAG systems will be key to expanding its use in real-world applications. Future research may focus on overcoming the highlighted challenges, refining evaluation frameworks, and enhancing scalability, efficiency, and trustworthiness. By addressing these areas, GraphRAG can further improve information retrieval and generation, providing a valuable tool for advancing AI capabilities across diverse fields.

\bibliographystyle{plainnat}
\bibliography{reference}
\end{document}